\documentclass[notoc]{JHEP3}

\usepackage{epsfig}

\newcommand\fverb{\setbox\pippobox=\hbox\bgroup\verb}
\newcommand\fverbdo{\egroup\medskip\noindent%
                        \Fbox{\unhbox\pippobox}\ }
\newcommand\fverbit{\egroup\item[\fbox{\unhbox\pippobox}]}
\newbox\pippobox

\def\JHEP{\it JHEP}

\def\app#1#2#3{{\it Astropart. Phys. }{\bf #1} (#2) {#3}}

\newcommand{\newc}{\newcommand}
\newc\eg{{\it {e.g.}}}  \newc\etal{{\it {et\ al.}}} \newc\ie{{\it i.e.}}
\newc\etc{{\it {etc}}}

\newc{\mplanck}{M_{\rm P}}      \newc{\mpl}{M_{\rm Pl}}
\newc{\msusy}{M_{\rm SUSY}}      \newc{\ms}{M_{\rm S}}

\newc{\jxf}{J({\xf})}
\newc{\jxfexact}{J_{\rm exact}({\xf})}  \newc{\jxfexp}{J_{\rm exp}({\xf})}
\newc{\VEV}[1]{\langle #1 \rangle}

\newc{\xf}{x_f}
\newc\vrel{v_{\rm rel}}
\newcommand\mchi{m_{\chi}}              
\newcommand\mchii{m_{\chi_{i}^{0}}}  \newcommand\mchij{m_{\chi_{j}^{0}}}
\newcommand\mchargk{m_{\chi_{k}^{\pm}}}  \newcommand\mchargl{m_{\chi_{l}^{\pm}}}

\newcommand\hl{h}           \newcommand\ha{A}
\newcommand\mh{m_{h}}          \newcommand\mH{m_{H}}
\newcommand\mA{m_{A}}          \newcommand\mcH{m_{H^{\pm}}}

\newc\hpm{H^\pm} \newc\hp{H^+} \newc\hm{H^-}
\newc\sfermion{\widetilde f}  \newc\msfermion{m_{\sfermion}}
\newc\stoponetwo{{\widetilde t}_{1,2}}
\newc\sbotonetwo{{\widetilde b}_{1,2}}
\newc\stauonetwo{{\widetilde \tau}_{1,2}}

\newcommand\msfa{m_{\widetilde f_a}}
\newcommand\msfb{m_{\widetilde f_b}}

\newc\second{{\rm sec}}

\newc\alphas{\alpha_s}

\newc{\gstar}{g_\ast}           \newc{\gsstar}{g_{s\ast}}
\newc{\geff}{g_{\rm eff}}

\newc{\sthw}{\sin\theta_W}              \newc{\cthw}{\cos\theta_W}
\newc{\bino}{\widetilde B}              \newc{\wino}{\widetilde W_3}
\newc{\higgsinob}{{\widetilde H}^0_b}   \newc{\higgsinot}{{\widetilde H}^0_t}

\newc{\abund}{\Omega h^2}
\newc{\abundchi}{\Omega_\chi h^2}
\newc{\rhocrit}{\rho_{crit}}
\newc{\rhochi}{\rho_{\chi}}

\newcommand\gev{\,\mbox{GeV}}

\newc{\ra}{\rightarrow}
\newc{\beq}{\begin{equation}}
\newc{\eeq}{\end{equation}}
\newc{\bea}{\begin{eqnarray}}
\newc{\eea}{\end{eqnarray}}

\def\nn              {\nonumber}

\newcommand\lsim{\mathrel{\rlap{\lower4pt\hbox{\hskip1pt$\sim$}}
    \raise1pt\hbox{$<$}}}
\newcommand\gsim{\mathrel{\rlap{\lower4pt\hbox{\hskip1pt$\sim$}}
    \raise1pt\hbox{$>$}}}

\long\def\begincomment#1\endcomment{%
        \begingroup\sf\baselineskip12pt#1\endgroup}

\newcommand\mZ{m_{Z}}        \newcommand\mW{m_{W}}
\newcommand\T{\mathcal{T}}       \newcommand\Y{\mathcal{Y}}
\newcommand\F{\mathcal{F}}
\newcommand\sw{\sin\theta_W}
\newcommand\cw{\cos\theta_W}
\newcommand\tw{\tan\theta_W}
\newcommand{\swsq}{\sin^2 \theta_W}
\newcommand{\delmux}{\stackrel{\leftrightarrow}{\partial^{\,\mu}}}

\title{Exact Cross Sections \\ for the Neutralino WIMP Pair-Annihilation}

\author{Takeshi Nihei\\
    Department of Physics, College of Science and Technology,
        Nihon University, \\
        \hspace{0.4in}
        1-8-14, Kanda-Surugadai, Chiyoda-ku, Tokyo, 101-8308, Japan \\
        E-mail: \email{nihei@phys.cst.nihon-u.ac.jp}}

\author{Leszek Roszkowski
\\
        Department of Physics, Lancaster University,
        Lancaster LA1 4YB, England\\
        E-mail: \email{L.Roszkowski@lancaster.ac.uk}}

\author{Roberto Ruiz de Austri\\
        Physics Division, School of Technology, 
        Aristotle University of Thessaloniki, \\
        \hspace{1.1in}
        GR - 540 06 Thessaloniki, Greece \\
        E-mail: \email{rruiz@gen.auth.gr}}
\abstract{
We derive a full set of exact, analytic expressions for the annihilation of
        the lightest neutralino pairs into all two-body tree-level final
        states in the framework of minimal supersymmetry. We make no
        simplifying assumptions about the neutralino nor about
        sfermion masses and mixings other than the absence of explicit
        CP--violating terms. The expressions should be particularly
        useful in computing the neutralino WIMP relic abundance
        without the usual approximation of partial wave expansion.
}


\begin{document}


\section{Introduction}\label{intro:sec}
The quest for identifying the nature of the dark matter (DM) in the
Universe continues~\cite{kt90,jkg96}. It is generally believed that
most of the DM is made of some hypothetical weakly-interacting massive
particles (WIMPs).  From the particle theory point of view, a commonly
considered candidate for the WIMP is the lightest neutralino under the
assumption that it is the lightest supersymmetric particle (LSP).  In
most approaches the LSP is stable due to an additional
R-parity~\cite{susyreview}. The 
neutralino, being massive, often provides a sizeable contribution to
the relic density. In addition, the requirement that the neutralino,
or some other stable particle relic, does not ``overclose'' the
Universe, often provides a strong constraints on a supersymmetric
model. The two other robust candidates for the LSP and cold dark
matter (CDM) are the axino (superpartner of the axion)~\cite{ckr} and
the gravitino~\cite{gravitino}.

Continuing improvements in determining the abundance of cold dark
matter, and other components of the Universe, have now reached an
unprecedented precision of a few per cent~\cite{cmbrreview}. In light
of this, one needs to be able to perform an accurate enough
computation of the WIMP relic abundance, which would allow for a
reliable comparison between theory and observation.

The literature on the relic abundance of the neutralino is vast and
still growing. (For a comprehensive review, see Ref.~\cite{jkg96}.)  A
brief, and by no means complete, account of major developments, can be
summarized as follows.  The original paper by
Goldberg~\cite{goldberg83} considered the neutralino in the photino
limit and pointed out the strong constraints from its relic
abundance. This was soon followed the first analysis by Ellis,
\etal~\cite{ehnos} and Krauss~\cite{krauss83}, of the general
neutralino case. Several other early papers
subsequently appeared with more detailed and elaborate analyses.  In
particular, Griest~\cite{griest88} was first to compute in detail the
annihilation into the ordinary fermion-pair ($f\bar f$) final states
through the $Z$--exchange, and later Griest, Kamionkowski and
Turner~\cite{gkt90} conducted the first more complete analysis of the
general neutralino case into $WW$, $ZZ$ and Higgs-pair final
states. The Higgs contribution to $f\bar f$ was first computed in
Refs.~\cite{erl90,gkt90}.  Olive and Srednicki~\cite{os91} considered
all the annhilation channels but only in the limit of the pure gaugino
and higgsino cases where several important resonances and final states
are absent. Drees and Nojiri~\cite{dn93} computed a first complete set
of expressions for the product of the cross section times velocity
using the helicity amplitude technique. When expanded in the
nonrelativistic limit, these give expressions for the first two coefficients
of the partial wave expansion.

In the early papers the partial wave expansion of the
thermally-averaged product of the neutralino pair-annihilation cross
section and their relative velocity, $\langle \sigma v \rangle\approx
a+b x$ was used in most cases.  The method
is normally expected to give an accurate enough approximation (about
few per cent) but only far enough from $s$--channel resonances and
thresholds for new final states, as was first pointed out by Griest
and Seckel~\cite{gs91} and further emphasized in
Refs.~\cite{gg91,an93,lny93,nrr1}.  In particular, it was shown in 
Ref.~\cite{an93} that, because of the very narrow
width of the lightest supersymmetric Higgs $\hl$, in the vicinity of
its $s$--channel exchange the error can be as large as a few orders of
magnitude. Partial remedies were suggested, for example in~\cite{an93,dy96},
by numerically integrating the full cross section near resonances
only and by matching this with the expansion-based calculations
further away from poles. Such methods are not fully satisfactory
since they do not include interference terms. A
recent detailed analysis~\cite{nrr1} showed that in the case of the
often wide $s$--channel resonance caused by the exchange of the
pseudoscalar $\ha$, the expansion produces a significant error over
the range of neutralino mass which can be as big as several tens of
$\gev$. Furthermore subdominant channels and often neglected
interference terms can also sometimes play a sizeable role. 

A formalism for computing the relic abundance also became refined. In
particular, the effect of replacing the usually assumed common heat
bath for both annihilating particles~\cite{kt90} by a more accurate
treatment of involving two separate thermal distributions was
considered by Gondolo and Gelmini~\cite{gg91} and by Srednicki,
\etal~\cite{swo88}. In practical terms the numerical difference is
usually negligible when one does the usual partial wave expansion. The first
coefficient $a$ is universal while the second ones $b$ differ by
$3/2 a$, where usually $a\ll b$.
Gondolo and Gelmini~\cite{gg91} further derived a very useful compact
expression for the thermally-averaged product of the neutralino
pair-annihilation cross section and their relative velocity as a
single integral over the cross section, as we will see below. Gondolo
and collaborators
next developed a Fortran code DarkSusy~\cite{darksusy00} where the
relic density of neutralinos is numerically computed without using the
partial wave expansion approximation. 

An additional effect of reducing the relic abundance of WIMPs through
co-annihilation was first pointed out by Griest and Seckel~\cite{gs91}.
In some cases there may exist some other states which
are not much heavier than the stable WIMP and may therefore be still
present in the thermal plasma around the WIMP decoupling. In the
framework of minimal supersymmetry with the lightest neutralino LSP
with a significant higgsino component, the co-annihilation with the
next-to-lightest neutralino and the lightest chargino is often
important~\cite{mizuta92,edsjo97}.
Other cases of interest 
involve neutralino annihilation with the lighter superpartner of the
$\tau$--lepton~\cite{efos98} and with lighter
stop~\cite{bdd00}. 

In this paper, we will present a full set of exact analytic
expressions for the cross section of the neutralino pair-annihilation
in the general Minimal Supersymmetric Standard Model (MSSM) for a
general neutralino case.
From the point of view of low-energy supersymmetry, the most natural
choice for the LSP and CDM is a nearly pure gaugino (bino) as was
first shown by Roszkowski~\cite{chiasdm}. Remarkably, just
such a case of the [electrically neutral] LSP naturally emerges in
most case in the Constrained MSSM~\cite{rr93,an93,kkrw94}. Nevertheless, in
our analysis we will
make {\em no} simplifying assumptions about the neutralino, nor will
we assume 
the degeneracy of the left-- and right--sfermion masses. We will include
all tree-level final states and all intermediate states. We will also
keep finite widths in $s$--channel resonances. We will only neglect
possible CP-violating phases in the SUSY sector. We will also not
consider the effect of co-annihilation here but will address it in a
subsequent publication. A complete set of
expressions presented here does not rely on the partial wave
expansion, includes all the terms and is valid both near and further
away from resonances and thresholds for new final states.

Some of the results presented here are not new but we include them
here nevertheless in order to provide a complete and self-consistent
reference containing the full set of exact expressions. 
In particular, the cross sections for the neutralino
pair-annihilation into the SM fermion-pair ($f\bar f$) final states
were first computed in Ref~\cite{griest88} and for the $WW$, $ZZ$ and
Higgs boson-pair final states in Ref.~\cite{gkt90}. 

Given the complexity of analytic expressions presented here and in
other papers and due to often different conventions used, it is not doable to
perform a real comparison with the literature. Instead, we have
performed a numerical check of our cross section with the results
obtained by using 
DarkSusy~\cite{darksusy00}. We have found, for the same values of
input parameters, an impressive agreement, at the level of a few per
cent, for all the annihilation channels, which we find
reassuring. 
(Recently another numerical code has been derived~\cite{annecy01} and
also numerically agreed with DarkSusy.)

While the exact analytic expressions presented here are applicable
both near and away from special cases where the method of partial wave
expansion fails, sometimes one may find it less CPU-time consuming to
use the latter one. Starting from our exact cross sections we have
therefore derived a complete set of expressions for the usual first
two terms of expansion for all the dominant channels. We will present
them here as well.

The annihilation into $f\bar f$ is often dominant.  However, other
final states can also play important role, depending on the case.  In
a previous paper~\cite{nrr1} we performed a detailed numerical
comparison of the relic abundance computed using the exact formulae
with the one obtained using the expansion formulae, for all the
channels, including subdominant ones. Our analysis confirmed that the
expansion gives highly inaccurate results near resonances and new
thresholds. We also showed that very far from such cases the error is
typically rather small, of the order of a few per cent.  However, we
found that, because of the existence of several resonances ($Z$ and
the Higgs bosons), the expansion produces large errors, compared to an
exact treatment, over a sizeable range of the neutralino mass $\mchi$,
even of a several tens of~$\gev$. In other words, the widely used
method of expansion may lead to significant errors in a sizeable
fraction of the neutralino mass.

The plan of the paper is as follows. In Sect.~\ref{relicdensity:sec} we
review the formalism for computing the relic density that we
employ. In Sect.~\ref{mssm:sec} we introduce the relevant ingredients
of the MSSM and list all the neutralino pair-annihilation
channels. Explicit expressions for the annihilation cross secion are
given in Sect.~\ref{exact:sec}. In Sect.~\ref{expansion:sec} we discuss
expansion and provide a list of formulae for the first two
coefficients in the case of equal-mass final states. In
Sect.~\ref{summary:sec} we summarize our work. Appendix~A contains a
complete list of Lagrangian terms and couplings which are used in the
paper while in Appendix~B we provide expressions for several auxiliary
functions used in the text.

%
\section{Calculation of the Relic Density}\label{relicdensity:sec}
The relic abundance of some stable species $\chi$ is defined as
$\Omega_{\chi} \equiv \rho_\chi/{\rho_{crit}}$, where $\rho_\chi=\mchi
n_\chi$ is the relic's mass density, $n_\chi$ is its number density,
$\rho_{crit}\equiv 3 H_0^2/8\pi G_N = 1.9\times 10^{-29}\,
(h^2)\,g/cm^3$ is the critical density and $G_N$ is the gravitational
constant.
(For a review of relic density calculations, see,
\eg, Refs.~\cite{kt90,jkg96}.)
The time evolution and subsequent
freeze-out of $n_\chi$ in an expanding Universe are described by the
Boltzmann equation
\begin{eqnarray}
\frac{d n_\chi}{dt} & = & - 3 H n_\chi
 - \langle\sigma v_{\rm M{\o}l}\rangle
\left[n_\chi^2 - (n_\chi^{\rm eq})^2\right], \label{Boltzmann-eq:eq}
\end{eqnarray}
where $n_\chi^{\rm eq}$ is the number density that the species would
have in thermal equilibrium,
$H(T)$ is the Hubble expansion rate,
$\sigma(\chi \chi
\rightarrow {\rm all})$ denotes the cross section of the species
annihilation into ordinary particles,
$v_{\rm M\o l}$ is a so-called M{\o}ller velocity~\cite{gg91} which is
the relative velocity
of the annihilating particles, and
$\langle\sigma v_{\rm M\o l}\rangle$ represents the thermal average of
$\sigma v_{\rm M\o l}$ which will be given below.
In the early Universe, the species $\chi$ were initially in thermal
equilibrium, $n_\chi$ $=n_\chi^{\rm eq}$. When their typical
interaction rate $\Gamma_\chi$ became less than the Hubble parameter,
$\Gamma_\chi\lsim H$, the annihilation
process froze out. Since then their number density in a co-moving volume
has remained basically constant.

The thermally-averaged product of the neutralino pair-annihilation
cross section and their relative velocity $\langle \sigma v_{\rm M\o
l} \rangle$ is most properly defined in terms of separate thermal
baths for both annihilating particles~\cite{gg91,swo88}
\begin{equation}
\langle \sigma v_{\rm M\o l} \rangle(T)= 
\frac{\int d^3p_1 d^3p_2\, \sigma v_{\rm
M\o l}\, e^{-E_1/T} e^{-E_2/T}} {\int d^3p_1 d^3p_2\,  e^{-E_1/T} e^{-E_2/T} }
\label{sigmavdef:eq}
\end{equation}
where $p_1= (E_1, {\bf p}_1)$ and  $p_2= (E_2, {\bf p}_2)$
are the 4-momenta of the two colliding particles, and $T$
is the
temperature of the bath.
(Note that one often uses another definition of
$\langle \sigma v_{\rm M\o l} \rangle$
which involves a single thermal bath for both
neutralinos. Compare, \eg, Refs.~\cite{kt90,dn93,jkg96}. The numerical
difference between the two formulae is usually rather small.)
The above expression can be reduced to a one-dimensional
integral which can be written
in a Lorentz-invariant form as~\cite{gg91}
\begin{eqnarray}
\langle\sigma v_{\rm M\o l}\rangle(T) 
& = & \frac{1}{8 m_\chi^4 T K_2^2(m_\chi/T)}
\int_{4 m_\chi^2}^\infty ds \, \sigma(s) (s-4m_\chi^2)\sqrt{s}
K_1\left(\frac{\sqrt{s}}{T}\right), \label{thermal-average:eq}
\end{eqnarray}
where  $s=(p_1+p_2)^2$ is a usual Mandelstam variable and $K_i$ denotes the
modified Bessel function of order $i$. 
In computing the relic abundance one first evaluates 
eq.~(\ref{thermal-average:eq}) and then uses this to solve the
Boltzmann eq.~(\ref{Boltzmann-eq:eq}).

There are a number of methods of solving eq.~(\ref{Boltzmann-eq:eq}). 
One often used, approximate, although in general quite accurate (for a
recent discussion see Ref.~\cite{nrr1}),
solution to the Boltzmann equation
is based on solving iteratively the equation
\begin{eqnarray}
x_f^{-1} & = & \ln \left( \frac{m_\chi}{2 \pi^3} \sqrt{\frac{45}{2g_* G_N}}
\langle\sigma v_{\rm M\o l}\rangle({x_f})\, x_f^{1/2} \right),
\label{freeze-out-temperature:eq}
\end{eqnarray}
where $g_*$ represents the effective
number of degrees of freedom at freeze-out ($\sqrt{g_*}\simeq 9$).
Typically one finds that the freeze-out point
$x_f\equiv T_f/\mchi$ is roughly given by $1/25$--$1/20$.
One usually introduces $\jxf$ defined as
\beq
\jxf\equiv \int_0^{x_f}dx \langle\sigma v_{\rm M\o l}\rangle(x),
\label{jxfdef:eq}
\eeq
where
$x=T/m_\chi$. 

The relic density at present is given by
\begin{eqnarray}
\rho_\chi & = & \frac{1.66}{M_{\rm Pl}} \left(\frac{T_\chi}{T_\gamma}\right)^3
T_\gamma^3 \sqrt{g_*} \frac{1}{ \jxf}, \label{relic-density:eq}
\end{eqnarray}
where $M_{\rm Pl}=1/{\sqrt{G_N}}$ denotes the Planck mass,
$T_\chi$ and $T_\gamma$ are the present temperatures of the neutralino
and the photon, respectively. The suppression factor
$(T_\chi/T_\gamma)^3$ $\approx$ $1/20$ follows from entropy
conservation in a comoving volume~\cite{reheatfactor}.

%
\section{WIMP Annihilation in the MSSM}\label{mssm:sec}
In this Section we introduce the relevant parameters and
definitions. We will be working in the framework of the general
MSSM.
(For a review, see, \eg, Ref.~\cite{susyreview}. We
follow the conventions of Ref.~\cite{gh}.)
The lightest neutralino is
a mass eigenstate given by a linear combination of the bino
$\widetilde{B}$, the neutral wino $\widetilde{W}^0_3$ and the two neutral
higgsinos $\widetilde{H}^0_b$ and $\widetilde{H}^0_t$
\begin{eqnarray}
\chi &\equiv \chi^0_1 = & N_{11} \widetilde{B} + N_{12} \widetilde{W}^0_3 + 
           N_{13} \widetilde{H}^0_b + N_{14} \widetilde{H}^0_t. 
\label{chi-1:eq}
\end{eqnarray}
The neutralino mass matrix is determined by
the $U(1)_Y$ and $SU(2)_L$ gaugino mass parameters $M_1$
and $M_2$, respectively (and we impose the usual GUT relation
$M_1=\frac{5}{3}\tan^2\theta_W  M_2 $),
the Higgs/higgsino mass parameter $\mu$,  the usual weak angle
$\theta_W$ and
$\tan \beta=v_t/v_b$ -- the ratio of the vacuum expectation values of
the two neutral Higgs fields.

The neutralino mass matrix is given by
\begin{eqnarray}
{\cal M}_{\chi^0} &=&
\left( \begin{array}{cccc}
M_1  &   0   &  -m_Z \sw \cos\beta  &  m_Z \sw \sin\beta \\
 0   &  M_2  &   m_Z \cw \cos\beta  &  -m_Z \cw \sin\beta \\
-m_Z \sw \cos\beta  &   m_Z \cw \cos\beta  &   0    &  -\mu  \\
 m_Z \sw \sin\beta  &  -m_Z \cw \sin\beta  &  -\mu  &   0
\end{array}
\right). \nonumber \\
& &
\end{eqnarray}
The neutralino mass matrix is diagonalized by
a unitary matrix $N$
\begin{eqnarray}
\label{neutdiag}
N^* {\cal M}_{\chi^0} N^{-1} & = &
{\rm diag}(m_{\chi^0_1},m_{\chi^0_2},m_{\chi^0_3},m_{\chi^0_4}).
\end{eqnarray}
In the absence of possible CP violating phases, one
can choose a basis such that the mixing matrix $N$ is real, in which case
some of the neutralino masses will in general be negative.

The chargino mass matrix is given by
\begin{eqnarray}
{\cal M}_{\chi^\pm} &=&
\left( \begin{array}{cc}
          M_2           & \sqrt{2}m_W \sin\beta \\
 \sqrt{2}m_W \cos\beta  &           \mu
\end{array} \right).
\end{eqnarray}
The chargino mass matrix is diagonalized by
two unitary matrices $U$ and $V$
\begin{eqnarray}
\label{chardiag}
U^* {\cal M}_{\chi^\pm} V^{-1} & = & {\rm diag}(m_{\chi^\pm_1},m_{\chi^\pm_2}).
\end{eqnarray}

There are two neutral scalar Higgs bosons $h$ and $H$, a pseudoscalar
$A$ plus a pair of charged Higgs $H^\pm$. (We will typically suppress
the Higgs charge assignment except where this may lead to
ambiguities.)

Other relevant parameters which determine the masses of scalars and
various couplings are the squark soft mass parameters $m_Q$, $m_U$ and
$m_D$, the slepton soft mass parameters $m_L$ and $m_E$, and the
pseudoscalar mass $m_A$. We also include the trilinear terms $A_i$
($i=t,b,\tau$) of the third generation which are important in
determining the masses and couplings for the stop $\stoponetwo$,
sbottom $\sbotonetwo$ and stau $\stauonetwo$ states,
respectively. 

In general the flavor-violating 
sfermion mass--squared ($6\times 6$) matrices are given by 
\begin{eqnarray}
{\cal M}^2_{\widetilde{f}} & = &
\left(
\begin{array}{cc}
m^2_{LL}(\widetilde{f}) & m^2_{LR}(\widetilde{f}) \\
m^2_{RL}(\widetilde{f}) & m^2_{RR}(\widetilde{f}) \\
\end{array}
\right),
\label{eqn:squarkmass}
\end{eqnarray}
with 
\begin{eqnarray}
\label{eqn:ll}
m^2_{LL}(\widetilde{f}) & = & M^2_{\widetilde{f}L} + M_f^\dagger M_f 
+ m_Z^2 \cos 2\beta \left( T_{3f}-Q_f \sin^2 \theta_W \right) {\bf 1},  \\
m^2_{RR}(\widetilde{f}) & = & M^2_{\widetilde{f}R} + M_f M_f^\dagger 
+ m_Z^2 \cos 2\beta\,Q_f \sin^2 \theta_W {\bf 1}, \\
\label{eqn:lr}
m^2_{LR}(\widetilde{f}) & = & 
\left\{
\begin{array}{c}
M_f^{\dagger}\left(A_f^{\dagger} -\mu\cot\beta \right) 
            \hspace{5mm}{\rm for} \hspace{4mm} T_{3f}=+1/2 \vspace{1mm}\\
M_f^{\dagger}\left(A_f^{\dagger} -\mu\tan\beta \right) 
            \hspace{5mm}{\rm for} \hspace{4mm} T_{3f}=-1/2
\end{array}
\right., \\
\label{eqn:rr}
m^2_{RL}(\widetilde{f}) & = & m^{2 \,\dagger}_{LR}(\widetilde{f}), 
\end{eqnarray}
where $\widetilde{f}$ and $f$
denote here different types of (s)fermions (up-- and down--type
(s)quarks, charged and neutral (s)leptons). 
For (s)neutrinos only the $LL$ part of 
eq.~(\ref{eqn:squarkmass}) should be taken.
$M^2_{\widetilde{f}L}$ and $M^2_{\widetilde{f}R}$ denote
$3\times 3$ soft SUSY--breaking sfermion mass matrices,
and $M_f$ denotes here a $3\times 3$ fermion mass matrix. Finally, 
$A_f$ is a scalar trilinear coupling matrix of the same
dimension while $Q_f$ and $T_{3f}$ are the respective
electric and isospin charges.
All the interaction terms and couplings that we will need below are
summarized in Appendix~A.

In the MSSM, the neutralino LSP's can pair-annihilate into a number of
final states, if kinematically allowed. A complete list of all
tree-level two-body final states is given in Table~\ref{tableone}.
We only neglect two-body loop processes into final state photon pairs
and gluon pairs because they are always subdominant in computing the
relic density~\cite{jkg96}. We also neglect three-body final
states since they are unlikely to be competitive with two-body
final states. They were shown to dominate in the higgsino case just
below the $WW$ and $t\bar t$ final states~\cite{ck98} but in such
regions neutralino co-annihilation with the lightest chargino
and next-to-lightest neutralino reduce the relic density to very small
values anyway. 

The channels $WW$, $ZZ$, ${\bar t}t$, $W^\pm H^\mp$, $Zh$, $ZH$, $Ah$
and $AH$, are not $s$-wave suppressed, and, once kinematically
allowed, can give dominant contributions. But even the $s$-wave
suppressed channels ${\bar f}f$ ($f\neq t$), 
$hh$, $Hh$, $HH$, $AA$, $H^+ H^-$ and $ZA$ can
play some role, especially if the other channels are not yet
kinematically allowed. This in particular is the case with the light
fermion-pair final states for which the cross section is suppressed by
the square of the corresponding fermion mass but which are always kinematically
allowed and often dominant.

\TABLE[h!] {\label{tableone} 
\begin{tabular}{|p{1.5in}|p{1.1in}|p{1.6in}|} \hline
& \multicolumn{2}{c|}{ Exchanged particles } \\
\cline{2-3}
\multicolumn{1}{|c|}{Process} &
\hspace{.2in}{$s$--channel} &
\hspace{.2in}{$t$-- and $u$--channel} \\ \hline
\hline
\hspace{0.2in} $\chi\chi\rightarrow hh$ &
\hspace{.2in} $h,H$ &
\hspace{.6in} $\chi^0_{i}$ \\ \hline
\hspace{0.2in} $\chi\chi\rightarrow HH$ &
\hspace{.2in} $h,H$ &
\hspace{.6in} $\chi^0_{i}$ \\ \hline
\hspace{0.2in} $\chi\chi\rightarrow hH$ &
\hspace{.2in} $h,H$ &
\hspace{.6in} $\chi^0_{i}$ \\ \hline
\hspace{0.2in} $\chi\chi\rightarrow AA$ &
\hspace{.2in} $h,H$ &
\hspace{.6in} $\chi^0_{i}$ \\ \hline
\hspace{0.2in} $\chi\chi\rightarrow hA$ &
\hspace{.2in} $A,Z$ &
\hspace{.6in} $\chi^0_{i}$ \\ \hline
\hspace{0.2in} $\chi\chi\rightarrow HA$ &
\hspace{.2in} $h,H$ &
\hspace{.6in} $\chi^0_{i}$ \\ \hline
\hspace{.2in} $\chi\chi\rightarrow H^{+}H^{-}$ &
\hspace{.2in} $h,H,Z$ &
\hspace{.6in} $\chi^\pm_{k}$ \\ \hline
\hspace{0.2in} $\chi\chi\rightarrow W^{\pm}H^{\mp}$ &
\hspace{.2in} $h,H,A$ &
\hspace{.6in} $\chi^\pm_{k}$ \\ \hline
\hspace{0.2in} $\chi\chi\rightarrow Zh$ &
\hspace{.2in} $A,Z$ &
\hspace{.6in} $\chi^0_{i}$ \\ \hline
\hspace{0.2in} $\chi\chi\rightarrow ZH$ &
\hspace{.2in} $A,Z$ &
\hspace{.6in} $\chi^0_{i}$ \\ \hline
\hspace{0.2in} $\chi\chi\rightarrow ZA$ &
\hspace{.2in} $h,H$ &
\hspace{.6in} $\chi^0_{i}$ \\ \hline
\hspace{0.2in} $\chi\chi\rightarrow W^{+}W^{-}$ &
\hspace{.2in} $h,H,Z$ &
\hspace{.6in} $\chi^\pm_{k}$\\ \hline
\hspace{0.2in} $\chi\chi\rightarrow ZZ$ &
\hspace{.2in} $h,H$ &
\hspace{.6in} $\chi^0_{i}$ \\ \hline
\hspace{0.2in} $\chi\chi\rightarrow f\bar f$ &
\hspace{.2in} $h,H,A,Z$ &
\hspace{.6in} $\widetilde{f}_a$\\ \hline
\end{tabular}
\caption{A complete set of neutralino pair-annihilation channels into 
tree-level two-body final states in the MSSM. 
The indices $i,k,a$ run as follows:
$i=1,\ldots,4$, $k=1,2$ and $a=1,\ldots,6$.}
}

%
\section{Exact Expressions}\label{exact:sec}
We now proceed to present a full set of exact, analytic expressions
for the total cross section $\sigma(\chi \chi \rightarrow {\rm all})$
for the neutralino pair-annihilation processes into all allowed
(tree-level) two-body final states in the general MSSM. We have
included all contributing diagrams as well as all interference terms
and kept finite widths of all $s$--channel resonances. 
We have made no simplifying
assumptions about sfermion masses although we assumed that there are
no CP violating phases in SUSY parameters.

To start with, it is convenient to introduce a Lorentz-invariant
function $w(s)$~\cite{swo88}
\beq
w(s)= \frac{1}{4} \int d\, {\rm LIPS}\, |{\cal A} (\chi \chi
\rightarrow {\rm all})|^2
\label{wdef:eq}
\eeq
where $|{\cal A} (\chi \chi \rightarrow {\rm all})|^2$
denotes the absolute square of the reduced matrix
element for the annihilation of two $\chi$ particles, averaged over
initial spins and summed over final spins. The function $w(s)$ is
related to the annihilation cross section $\sigma(s)$ in
eq.~(\ref{thermal-average:eq}) via~\cite{lny93}
\beq
w(s)=  \frac{1}{2} \sqrt{ s (s-4\mchi^2)}\, \sigma(s).
\label{wtosigma:eq}
\eeq

Since $w(s)$ receives contributions from all the kinematically allowed
annihilation process $\chi\chi\rightarrow f_1 f_2$, it can be written as
\begin{eqnarray}
w(s)&=&\frac{1}{32\,\pi}\sum_{f_1 f_2} \bigg[
c \, \theta\left(s-(m_{f_1}+m_{f_2})^2 \right)\,
\beta_f(s,m_{f_1},m_{f_2})\,\widetilde{w}_{f_1f_2}(s)\bigg],
\label{wtowtilde:eq}
\end{eqnarray}
where
the summation extends over all possible two-body final states
$f_1f_2$, $m_{f_1}$ and $m_{f_2}$ denote their respective masses, and
\begin{eqnarray}
\label{color:eq}
c &=& \left\{ \begin{array}{ll}
c_{f}  & \mbox{if $f_{1 (2)}=f(\bar f)$} \\
 1     & \mbox{otherwise,}  \,
\end{array}
      \right.
\end{eqnarray}
where $c_{f}$ is the color factor of SM fermions
($c_{f}=3$ for quarks and $c_{f}=1$ for leptons).
The kinematic factor $\beta_f$ is defined as
\begin{eqnarray}
\beta_f(s,m_{f_1},m_{f_2})\equiv
\left[1-\frac{(m_{f_1}+m_{f_2})^2}{s}\right]^{1/2}
\left[1-\frac{(m_{f_1}-m_{f_2})^2}{s}\right]^{1/2}.
\label{kdef:eq}
\end{eqnarray}
In the CM frame, which we choose for convenience, the function
$\widetilde{w}_{f_1 f_2}(s)$ can be expressed as
\begin{eqnarray}
\widetilde{w}_{f_1 f_2}(s)\equiv \frac{1}{2}\int_{-1}^{+1} \!d \cos \theta_{CM}
\,|{\cal A}(\chi \chi \rightarrow f_1 f_2)|^2, \label{wtildedef:eq}
\end{eqnarray}
where $\theta_{CM}$ denotes the scattering angle in the CM frame. In other
words, we write
$|{\cal A}(\chi \chi \rightarrow f_1 f_2)|^2$ as a function of
$s$ and $\cos \theta_{CM}$, which greatly simplifies the computation.

We will follow Table~\ref{tableone} in presenting explicit expressions
for $\widetilde{w}_{f_1 f_2}(s)$ for all the two-body final
states. All the couplings are defined in Appendix~A. All other
auxiliary functions, are listed in Appendix B. Feynman diagrams
corresponding to all the annihilation channels are given in Chapter~6
of Ref.~\cite{jkg96}.

\vspace{0.3cm}
\begin{center}
\fbox{\boldmath ${1.\:\:\chi\chi\rightarrow hH}$}
\end{center}
This process involves the $s$-channel CP--even Higgs boson ($h$ and
$H$) exchange and the $t$- and $u$-channel neutralino ($\chi_{i}^{0}$,
$i=1,\ldots,4$) exchange
\begin{eqnarray}
 \widetilde{w}_{hH}=
 \widetilde{w}_{hH}^{(h,H)}
+\widetilde{w}_{hH}^{(\chi^{0})} +\widetilde{w}_{hH}^{(h,H-\chi^{0})}:
\end{eqnarray}
\vspace{0.25cm}
$\bullet$ \underline{CP--even Higgs--boson ($h,H$) exchange:}
\begin{eqnarray}
 \widetilde{w}_{hH}^{(h,H)}& = & \frac{1}{2}\: \left
  | \sum_{r=h,H} \frac{C^{hHr}\: C_{S}^{\chi\chi r}}
  {s-m_{r}^{2}+i\, \Gamma_{r}\, m_{r}} \right |^{2} (s-4\,\mchi^{2});
\end{eqnarray}
\vspace{0.25cm}
$\bullet$
 \underline{neutralino\ ($\chi_{i}^{0}$) exchange:}
\begin{eqnarray}
\widetilde{w}_{hH}^{(\chi^{0})}  = \sum_{i,j=1}^{4}
  C_{S}^{\chi_{i}^{0} \chi h } \,(C_{S}^{\chi_{j}^{0} \chi h})^{*} \,
  C_{S}^{\chi_{i}^{0} \chi H }\, (C_{S}^{\chi_{j}^{0} \chi H})^{*}
  \bigg[\mchii\,\mchij I^{hH}_{ij}
  +\mchi \mchii J^{hH}_{ij} + K^{hH}_{ij} \bigg], 
\end{eqnarray}
where
\begin{eqnarray}
 I^{hH}_{ij} & = &
  (s-4\,\mchi^{2})\Big[\T_{0}-\Y_{0}\Big]
  (s,\mchi^{2},\mh^{2},\mH^{2},\mchii^{2},\mchij^{2}),\\
 J^{hH}_{ij} & = &
  \Big[-4\,\T_{1}+2(-2\,\mchi^{2}+\mh^{2}+\mH^{2})\T_{0}-2\,\Y_{1} 
\nonumber \\
  & & \hspace{0.1in} -2(s-4\,\mchi^{2})\Y_{0}\Big]
  (s,\mchi^{2},\mh^{2},\mH^{2},\mchii^{2},\mchij^{2}),\\
 K^{hH}_{ij} & = &
  \Big\{-T_{2}-(s+2\,\mchi^{2}-\mh^{2}-\mH^{2})\T_{1}
  -(\mchi^{2}-\mh^{2})(\mchi^{2}-\mH^{2})\T_{0} \nonumber \\
  & & \hspace{0.1in} -\Y_{2}+[-(\mchi^{2}+\mh^{2})(\mchi^{2}+\mH^{2})
  +4\,\mchi^{4}]\Y_{0}\Big\}
  (s,\mchi^{2},\mh^{2},\mH^{2},\mchii^{2},\mchij^{2}); \nn \\
\end{eqnarray}
\vspace{0.25cm}
$\bullet$
 \underline{Higgs ($h,H$)--neutralino\ ($\chi_{i}^{0}$) interference term:}
\begin{eqnarray}
 \widetilde{w}_{hH}^{(h,H-\chi^{0})}& = &  2 \:
  \sum_{i=1}^{4} Re \left[\sum_{r=h,H} \left(\frac{C^{hHr}\:
  C_{S}^{\chi\chi r}}{s-m_{r}^{2}+i\, \Gamma_{r}\, m_{r}}\right)^{*}\,
  C_{S}^{\chi_{i}^{0} \chi h }
  C_{S}^{\chi_{i}^{0} \chi H } \right] \nonumber \\
  & & \times\left\{-2\,\mchi+[\mchi\,(\mH^{2}+\mh^{2}-2\,\mchi^{2}
  -2\,\mchii^{2}) \right.  \nonumber \\
  & & \left.  \hspace{0.2in} +\mchii\,(s-4\,\mchi^{2})]
  \F(s,\mchi^{2},\mH^{2},\mh^{2},\mchii^{2}) \right\}.
\end{eqnarray}
As mentioned above, the couplings $C_S^{\chi^0_i\chi^0_j r}$ and
$C^{hHr}$ ($r$ $=$ $h,H$), as well as all the other couplings
appearing in this Section, are defined in Appendix A. The functions
$\F$, ${\mathcal T}_k$ and ${\mathcal Y}_k$ ($k$ $=$ $0,\ldots,4$),
and all other auxiliary functions, are listed in Appendix B. By
$\Gamma_h$ and $\Gamma_H$ we denote the widths of $h$ and $H$,
respectively.

The expressions for $hh$ final state are obtained from the above by
replacing $m_H$, $C^{hHr}$, $C_{S}^{\chi_{i}^{0} \chi H }$,
$C_{S}^{\chi_{j}^{0} \chi H }$ with $m_h$, $C^{hhr}$,
$C_{S}^{\chi_{i}^{0} \chi h }$, $C_{S}^{\chi_{j}^{0} \chi h }$,
respectively, and multiplying $\omega$ by a factor of 1/2 for
identical particles in the final state. The contributions for $HH$
final state are obtained in an analogous way.

\vspace{0.3cm}
\begin{center}
\fbox{\boldmath ${2.\:\:\chi\chi\rightarrow AA}$}
\end{center}
Similarly to the final state $hH$, this process proceeds via the
$s$-channel CP--even Higgs boson ($h$ and $H$) exchange and the $t$-
and $u$-channel neutralino ($\chi_{i}^{0}$, $i=1,\ldots,4$) exchange
\begin{eqnarray}
 \widetilde{w}_{AA}=
 \widetilde{w}_{AA}^{(h,H)}
+\widetilde{w}_{AA}^{(\chi^{0})} +\widetilde{w}_{AA}^{(h,H-\chi^{0})}:
\end{eqnarray}
\vspace{0.25cm}
$\bullet$
 \underline{CP--even Higgs--boson ($h,H$) exchange:}
\begin{eqnarray}
 \widetilde{w}_{AA}^{(h,H)}& = & \frac{1}{4}\:
  \left | \sum_{r=h,H} \frac{C^{AAr}\: C_{S}^{\chi\chi r}}
  {s-m_{r}^{2}+i\, \Gamma_{r}\, m_{r}} \right |^{2} (s-4\,\mchi^{2});
\end{eqnarray}
\vspace{0.25cm}
$\bullet$
 \underline{neutralino\ ($\chi_{i}^{0}$) exchange:}
\begin{eqnarray}
 \widetilde{w}_{AA}^{(\chi^{0})} & = & \frac{1}{2}
  \sum_{i,j=1}^{4} \Big(C_{P}^{\chi_{i}^{0} \chi A }\Big)^2 \,
  \Big(C_{P}^{\chi_{j}^{0} \chi A *}\Big)^2
  \bigg[ \mchii\,\mchij I^{AA}_{ij}
  +\mchi \mchii J^{AA}_{ij} + K^{AA}_{ij} \bigg],
\end{eqnarray}
where
\begin{eqnarray}
 I^{AA}_{ij} & = &
  (s-4\,\mchi^{2})\Big[\T_{0}-Y_{0}\Big]
  (s,\mchi^{2},\mA^{2},\mA^{2},\mchii^{2},\mchij^{2}),\\
 J^{AA}_{ij} & = &
  \Big[4\,\T_{1}+4(\mchi^{2}-\mA^{2})\T_{0}+2\,\Y_{1}
  +2(s-4\,\mchi^{2})\Y_{0}\Big]
  (s,\mchi^{2},\mA^{2},\mA^{2},\mchii^{2},\mchij^{2}), \nonumber \\
  & & \\
 K^{AA}_{ij} & = &
  \Big\{-T_{2}-[s+2\,(\mchi^{2}-\mA^{2})]\T_{1}
  -(\mchi^{2}-\mA^{2})^{2}\T_{0} \nonumber \\
  & & \hspace{0.2in} - \Y_{2}
     +[4\,\mchi^{4}-(\mchi^{2}+\mA^{2})^{2}]\Y_{0}\Big\}
  (s,\mchi^{2},\mA^{2},\mA^{2},\mchii^{2},\mchij^{2});
\end{eqnarray}
\vspace{0.25cm}
$\bullet$
 \underline{Higgs ($h,H$)--neutralino\ ($\chi_{i}^{0}$) interference term:}
\begin{eqnarray}
 \widetilde{w}_{AA}^{(h,H-\chi^{0})}& = &  - \: \sum_{i=1}^{4}
  Re \left[\sum_{r=h,H} \left(\frac{C^{AAr}\: C_{S}^{\chi\chi r}}
  {s-m_{r}^{2}+i\, \Gamma_{r}\, m_{r}}\right)^{*}\,
  \big(C_{P}^{\chi_{i}^{0} \chi A}\big)^2  \right] \nonumber \\
  & & \times \Big\{-2\,\mchi +[2\,\mchi \,(\mA^{2}-\mchi^{2}
  -\mchii^{2}) \nonumber \\
  & & \hspace{0.2in} -\mchii\,(s-4\,\mchi^{2})]
  \F(s,\mchi^{2},\mA^{2},\mA^{2},\mchii^{2})\Big\}.
\end{eqnarray}

\vspace{0.3cm}
\begin{center}
\fbox{\boldmath ${3.\:\:\chi\chi\rightarrow hA}$}
\end{center}
This process proceeds via the $s$-channel $Z$ and CP--odd Higgs boson ($A$)
exchange as well as the $t$- and $u$-channel neutralino ($\chi_{i}^{0}$,
$i=1,\ldots,4$) exchange
\begin{eqnarray}
\widetilde{w}_{hA} &=&
\widetilde{w}_{hA}^{(A)} 
+\widetilde{w}_{hA}^{(Z)} +\widetilde{w}_{hA}^{(\chi^{0})}
+\widetilde{w}_{hA}^{(A-Z)} +\widetilde{w}_{hA}^{(A-\chi^{0})}
+\widetilde{w}_{hA}^{(Z-\chi^{0})}:
\end{eqnarray}
\vspace{0.25cm}
$\bullet$
 \underline{CP--odd Higgs-boson ($A$) exchange:}
\begin{eqnarray}
 \widetilde{w}_{hA}^{(A)}& = & \frac{1}{2}\:
  \left | \frac{C^{hAA}\: C_{P}^{\chi\chi A}}
  {s-\mA^{2}+i\,\Gamma_{A}\,\mA} \right  |^{2}s\,;
\end{eqnarray}
\vspace{0.25cm}
$\bullet$
 \underline{$Z$--boson exchange:}
\begin{eqnarray}
 \widetilde{w}_{hA}^{(Z)}& = & \frac{1}{3}\:
  \left | \frac{C^{hAZ}\: C_{A}^{\chi\chi Z}}
  {s-\mZ^{2}+i\, \Gamma_{Z}\, \mZ} \right |^{2}
  \frac{1}{\mZ^{4}\,s} \nonumber \\
  & & \times \Big\{\mZ^{4}\,s^{3}+s^{2}[6\,\mchi^{2}\,
  (\mA^{2}-\mh^{2})^{2}-4\,\mchi^{2}\,\mZ^{4}
  -2\,\mZ^{4}\,(\mA^{2}+\mh^{2})] \nonumber \\
  & & \hspace{0.2in}+s\,\mZ^{2}[-12\,\mchi^{2}\,
  (\mA^{2}-\mh^{2})^{2}+8\,\mZ^{2}\,\mchi^{2}\,(\mA^{2}+\mh^{2}) \nonumber \\
  & & \hspace{0.8in}+\mZ^{2}\,(\mA^{2}-\mh^{2})^{2}]
  +2\,\mZ^{4}\,\mchi^{2}\,(\mA^{2}-\mh^{2})^{2}\Big\};
\end{eqnarray}
\vspace{0.25cm}
$\bullet$
 \underline{neutralino\ ($\chi_{i}^{0})$ exchange:}
\begin{eqnarray}
 \widetilde{w}_{hA}^{(\chi^{0})} & = &
  \sum_{i,j=1}^{4}   C_{S}^{\chi_{i}^{0} \chi h } \,
  C_{P}^{\chi_{i}^{0} \chi A }\, \Big(C_{S}^{\chi_{j}^{0} \chi h}\Big)^{*}\,
  \Big(C_{P}^{\chi_{j}^{0} \chi A}\Big)^{*}
  \bigg[ \mchii\,\mchij I^{hA}_{ij}
  + \mchi \mchii J^{hA}_{ij} + K^{hA}_{ij} \bigg], \nonumber \\
\end{eqnarray}
where
\begin{eqnarray}
 I^{hA}_{ij} & = &
  s \Big[\T_{0}-\Y_{0}\Big]
     (s,\mchi^{2},\mh^{2},\mA^{2},\mchii^{2},\mchij^{2}) ,\\
 J^{hA}_{ij} & = &
  2(\mA^{2}-\mh^{2})\Big[\T_{0}-\Y_{0}\Big]
     (s,\mchi^{2},\mh^{2},\mA^{2},\mchii^{2},\mchij^{2}), \\
 K^{hA}_{ij} & = &
  \Big\{-T_{2}+(2\,\mchi^{2}+\mA^{2}+\mh^{2}-s)\T_{1}+\Y_{2}
\\
  & & \hspace{0.09in}+\{\mchi^{2}(\mA^{2}+\mh^{2}-\mchi^{2})
  -\mA^{2}\mh^{2}\}\Big[\T_{0}-\Y_{0}\Big]\Big\}
     (s,\mchi^{2},\mh^{2},\mA^{2},\mchii^{2},\mchij^{2}); \nonumber 
\end{eqnarray}
\vspace{0.25cm}
$\bullet$
 \underline{Higgs ($A$)--$Z$ interference term:}
\begin{eqnarray}
 \widetilde{w}_{hA}^{(A-Z)}& =& 2\,
  Re \left [ \left(\frac{C^{hAA}\: C_{P}^{\chi\chi A}}
  { s-\mA^{2}+i\, \Gamma_{A}\, \mA}\right)^{*}\,
  \frac{C^{hAZ}\: C_{A}^{\chi\chi Z}}{ s-\mZ^{2}+i\, \Gamma_{Z}\, \mZ}\right]
    \nonumber \\
  & & \times \frac{\mchi\,(s-\mZ^{2})\,(\mA^{2}-\mh^{2})}{\mZ^{2}};
\end{eqnarray}
\vspace{0.25cm}
$\bullet$
 \underline{Higgs ($A$)--neutralino\ ($\chi_{i}^{0}$) interference term:}
 \begin{eqnarray}
  \widetilde{w}_{hA}^{(A-\chi^{0})}& = &
   2\,\sum_{i=1}^{4} Re \left [ \left(\frac{C^{hAA}\:
   C_{P}^{\chi\chi A}}{ s-\mA^{2}+i\, \Gamma_{A}\, \mA}\right)^{*}\,
   C_{S}^{\chi_{i}^{0}\chi h}\: C_{P}^{\chi_{i}^{0}\chi A}\right] \nonumber \\
   & & 
   \times \Big[s\,\mchii+\mchi\,(\mA^{2}-\mh^{2})\Big]\,
                 \F(s,\mchi^{2},\mh^   {2},\mA^{2},\mchii^{2});
 \end{eqnarray}
\vspace{0.25cm}
$\bullet$
 \underline{$Z$--neutralino\ ($\chi_{i}^{0}$) interference term:}
\begin{eqnarray}
 \widetilde{w}_{hA}^{(Z-\chi^{0})}& = &
  2\,\sum_{i=1}^{4} Re \left [ \left(\frac{C^{hAZ}\:
  C_{A}^{\chi\chi Z}}{ s-\mZ^{2}+i\, \Gamma_{Z}\, \mZ}\right)^{*}\,
  C_{S}^{\chi_{i}^{0}\chi h}\: C_{P}^{\chi_{i}^{0}\chi A}\right] 
       \frac{1}{\mZ^2}\nonumber \\
  & & \times \bigg[(s+2\,\mchii^{2}-2\,\mchi^{2}-\mA^{2}-\mh^{2})\,\mZ^{2}
      \nonumber \\
  & & \hspace{0.2in}+2\,\Big\{ \mchii \, s \,[\mZ^{2}\,\mchii
  +\mchi \,(\mA^{2}-\mh^{2})]
  -\mZ^{2}\,\mchi\,\mchii\,(\mA^{2}-\mh^{2}) \nonumber \\
   & & \hspace{0.5in}+\mZ^{2}\,\mA^{2}\,\mh^{2}
   -\mZ^{2}\,(\mA^{2}+\mh^{2})\,
  (\mchii^{2}+\mchi^{2})+\mZ^{2}\,(\mchii^{2}-\mchi^{2})^{2} \nonumber \\
  & & \hspace{0.5in} +\mchi^{2}\,(\mA^{2}-\mh^{2})^{2}\Big\}
   \F(s,\mchi^{2},\mh^{2},\mA^{2},\mchii^{2})\bigg].
\end{eqnarray}
The expressions for $HA$ final state are obtained by replacing $m_h$,
$C_{S}^{\chi_{i}^{0} \chi h }$ and $C^{hAZ}$ in the above with $m_H$,
$C_{S}^{\chi_{i}^{0} \chi H }$ and $C^{HAZ}$, respectively.

\vspace{0.3cm}
\begin{center}
\fbox{\boldmath ${4.\:\:\chi\chi\rightarrow H^{+}H^{-}}$}
\end{center}
This process proceeds via the $s$-channel $Z$ and CP--even Higgs boson
($h,H$) exchange as well as the $t$- and $u$-channel chargino
($\chi_{k}^{\pm}$, $k=1,2$) exchange
\begin{eqnarray}
 \widetilde{w}_{H^{+}H^{-}} &=&
\widetilde{w}_{H^{+}H^{-}}^{(h,H)} +\widetilde{w}_{H^{+}H^{-}}^{(Z)}
+\widetilde{w}_{H^{+}H^{-}}^{(\chi^{\pm})}
+\widetilde{w}_{H^{+}H^{-}}^{(h,H-\chi^{\pm})}
+\widetilde{w}_{H^{+}H^{-}}^{(Z-\chi^{\pm})}:
\end{eqnarray}
\vspace{0.25cm}
$\bullet$
 \underline{CP--even Higgs--boson ($h,H$) exchange:}
\begin{eqnarray}
 \widetilde{w}_{H^{+}H^{-}}^{(h,H)}& = &
  \frac{1}{2}\: \left |\sum_{r=h,H} \frac{C^{H^{+}H^{-}r}\:
  C_{S}^{\chi\chi r}}
  { s-m_{r}^{2}+i\, \Gamma_{r}\, m_{r}}\right |^{2} (s-4\,\mchi^{2});
\end{eqnarray}
\vspace{0.25cm} \\
$\bullet$
 \underline{$Z$--boson exchange:}
\begin{eqnarray}
 \widetilde{w}_{H^{+}H^{-}}^{(Z)}& = &
  \frac{1}{3}\: \left | \frac{C^{H^{+}H^{-}Z}\:
  C_{A}^{\chi\chi Z}}{ s-\mZ^{2}+i\, \Gamma_{Z}\, \mZ}
  \right |^{2} (s-4\,\mchi^{2})\,(s-4\,\mcH^{2})\,;
\end{eqnarray}
\vspace{0.25cm}
$\bullet$
 \underline{chargino ($\chi_{k}^{\pm})$ exchange:}
\begin{eqnarray}
 \widetilde{w}_{H^{+}H^{-}}^{(\chi^{\pm})} & = &
  \sum_{k,l=1}^{2} \bigg[ \mchargk\,\mchargl I^{H^{+}H^{-}}_{kl}
  +m_{\chi} \mchargk J^{H^{+}H^{-}}_{kl} + K^{H^{+}H^{-}}_{kl} \bigg],
\end{eqnarray}
where
\begin{eqnarray}
 I^{H^{+}H^{-}}_{kl} & = &
  \Big\{(C_{-k}^{H}\,C_{-l}^{H}+D_{-k}^{H}\,D_{-l}^{H})
  \,(-2 \mchi^{2}) \Big[\T_{0}-\Y_{0}\Big] \nonumber \\
  & & +(C_{-k}^{H}\,C_{-l}^{H}-D_{-k}^{H}\,D_{-l}^{H})
  (s-2 \mchi^{2})\Big[\T_{0}-\Y_{0}\Big]\Big\}
  (s,\mchi^{2},\mcH^{2},\mcH^{2},\mchargk^{2},\mchargl^{2}), \nonumber \\
  & & \\
 J^{H^{+}H^{-}}_{kl} & = &
 C_{-k}^{H}\,C_{+l}^{H}\Big[-4\,\T_{1}
 -4(\mchi^{2}-\mcH^{2})\T_{0} \nonumber \\
 & & \hspace{0.6in} -2\, \Y_{1}-2(s-4\,\mchi^{2})\,\Y_{0}\Big]
 (s,\mchi^{2},\mcH^{2},\mcH^{2},\mchargk^{2},\mchargl^{2}), \\
 K^{H^{+}H^{-}}_{kl} & = &
 \bigg\{(C_{+k}^{H}\,C_{+l}^{H}+D_{+k}^{H}\,D_{+l}^{H})
  \Big[-{\mathcal T}_2-(s-2\,\mcH^2)\,{\mathcal T}_1
  -(\mchi^{2}-\mcH^{2})^{2}\,{\mathcal T}_0 \nonumber \\
  & & \hspace{1.6in}
+\,2 \mchi^{2}(\mchi^{2}-\mcH^{2})\, \Y_{0}\Big] \nonumber \\
  & & +\,(C_{+k}^{H}\,C_{+l}^{H}-D_{+k}^{H}\,D_{+l}^{H})
  \Big[-2\,\mchi^{2}\,\T_{1}-\Y_{2} \nonumber \\
  & & \hspace{1.6in}
+\,(\mchi^{4}-\mcH^{4})\Y_{0}\Big]\bigg\}
  (s,\mchi^{2},\mcH^{2},\mcH^{2},\mchargk^{2},\mchargl^{2}),  \nonumber \\
  & &
\end{eqnarray}
and
\begin{eqnarray}
\label{chch-coupl1:eq}
  C_{\pm k}^{H} &\equiv&
    |C_{S}^{\chi_{k}^{+}\chi H^{-}}|^{2}\pm
    |C_{P}^{\chi_{k}^{+}\chi H^{-}}|^{2}, \\
\label{chch-coupl2:eq}
  D_{\pm k}^{H}& \equiv& C_{S}^{\chi_{k}^{+}\chi H^{-}}
    \Big(C_{P}^{\chi_{k}^{+}\chi H^{-}}\Big)^{*} \pm
    \:C_{P}^{\chi_{k}^{+}\chi H^{-}}
    \Big(C_{S}^{\chi_{k}^{+}\chi H^{-}}\Big)^{*};
\end{eqnarray}
\vspace{0.25cm}
$\bullet$
 \underline{Higgs ($h,H$)--chargino\ ($\chi_{k}^{\pm}$) interference term:}
\begin{eqnarray}
 \widetilde{w}_{H^{+}H^{-}}^{(h,H-\chi^{\pm})}& = &
 2\: \sum_{k=1}^{2} Re \left[ \sum_{r=h,H}
 \left(\frac{C^{H^{+}H^{-}r}\:
 C_{S}^{\chi\chi r}}{ s-m_{r}^{2}+i\, \Gamma_{r}\,
 m_{r}}\right)^{*}\right] 
\nonumber \\
 & & \times \bigg\{ C_{+k}^{H}\,(-2\,\mchi)
 +\Big[C_{+k}^{H}\,2\,\mchi\,(m_{H^{\pm}}^{2}-\mchi^{2}-\mchargk^{2}) 
\nonumber \\
 & & \hspace{0.2in} +C_{-k}^{H}\,\mchargk\,(s-4\,\mchi^{2})\Big]\,
 \F(s,\mchi^{2},\mcH^{2},\mcH^{2},\mchargk^{2}) \bigg\};
\end{eqnarray}
\vspace{0.25cm} \\
$\bullet$
 \underline{$Z$--chargino\ ($\chi_{k}^{\pm}$) interference term:}
\begin{eqnarray}
 \widetilde{w}_{H^{+}H^{-}}^{(Z-\chi^{\pm})}& = &
  2\:\sum_{k=1}^{2} Re \left[\left(\frac{C^{H^{+}H^{-}Z}\:
  C_{A}^{\chi\chi Z}}{ s-\mZ^{2}+i\, \Gamma_{Z}\, \mZ}\right)^{*}\right]
  D_{+k}^{H} \nonumber \\
  & & \times \bigg\{s-2\,\mcH^{2}-2\,\mchi^{2}+2\,\mchargk^{2}  \nonumber \\
  & & \hspace{0.2in} +2\,\Big[s\,\mchargk^{2}+\mcH^{4}
  -2\,(\mchi^{2}+\mchargk^{2})\,\mcH^{2} \nonumber \\
  & & \hspace{0.5in} +(\mchi^{2}-\mchargk^{2})^{2}\Big]\,
  \F(s,\mchi^{2},\mcH^{2},\mcH^{2},\mchargk^{2}) \bigg\}.
\end{eqnarray}

\vspace{0.3cm}
\begin{center}
\fbox{\boldmath ${5.\:\:\chi\chi\rightarrow W^{+}H^{-}}$}
\end{center}
This process involves the $s$-channel CP--even ($h,H$) and odd ($A$)
Higgs boson exchange as well as the $t$- and $u$-channel chargino
($\chi_{k}^{\pm}$, $k=1,2$) exchange
\begin{eqnarray}
 \widetilde{w}_{WH} &=&
 \widetilde{w}_{WH}^{(h,H)}
+\widetilde{w}_{WH}^{(A)} +\widetilde{w}_{WH}^{(\chi^{\pm})}
+\widetilde{w}_{WH}^{(h,H-\chi^{\pm})} +\widetilde{w}_{WH}^{(A-\chi^{\pm})}:
\label{eq:WH}
\end{eqnarray}
\vspace{0.25cm}
$\bullet$
 \underline{CP--even Higgs--boson ($h,H$) exchange:}
\begin{eqnarray}
 \widetilde{w}_{WH}^{(h,H)}& = &
  \frac{1}{2}\: \left |\sum_{r=h,H} \frac{C^{W^{-}H^{+}r}\:
  C_{S}^{\chi\chi i}}{ s-m_{r}^{2}+i\, \Gamma_{r}\, m_{r}} \right |^{2}
  (s-4\,\mchi^{2}) \nonumber \\
  & & \times \frac{s^{2}-2\,(\mcH^{2}+\mW^{2})\,s+(\mcH^{2}-\mW^{2})^{2}}
  {\mW^{2}};
\end{eqnarray}
\vspace{0.25cm}
$\bullet$
 \underline{CP--odd Higgs--boson ($A$) exchange:}
\begin{eqnarray}
 \widetilde{w}_{WH}^{(A)}& = &
  \frac{1}{2}\: \left | \frac{C^{W^-H^+A}\:
   C_{P}^{\chi\chi A}}{ s-\mA^{2}+i\, \Gamma_{A}\, \mA} \right |^{2}
   \,s\, \nonumber \\
   & & \times \frac{s^{2}-2\,(\mcH^{2}+\mW^{2})\,s
   +(\mcH^{2}-\mW^{2})^{2}}{\mW^{2}};
\end{eqnarray}
\vspace{0.25cm}
$\bullet$
 \underline{chargino ($\chi_{k}^{\pm})$ exchange:}
\begin{eqnarray}
 \widetilde{w}_{WH}^{(\chi^{\pm})} & = &
  \frac{1}{\mW^{2}}\sum_{k, l=1}^{2}
  \bigg[ \mchargk\,\mchargl I^{WH}_{kl}
      + \mchi \mchargk J^{WH}_{kl} + K^{WH}_{kl} \bigg],
\end{eqnarray}
where
\begin{eqnarray}
 I^{WH}_{kl} & = &
  \bigg\{(C_{+k}^{HW\!*}C_{+l}^{HW}+D_{+k}^{HW\!*}D_{+l}^{HW}) \nonumber \\
  & & \hspace{3mm} \times \Big[ - \T_2
  -(s-m_W^2-\mcH^2-2\,\mchi^2)\,\T_1
  + G_{WH}^{T(1)}\,\T_0
  + 6\,m_W^2 \mchi^2\,\Y_0 \,\Big] \nonumber \\
  & & \hspace{0.1in} + \ (C_{+k}^{HW\!*}C_{+l}^{HW}-D_{+k}^{HW\!*}D_{+l}^{HW})
 \Big[ \,6\,m_W^2 \mchi^2\,\T_0
  - \Y_2 \nonumber \\
  & & \hspace{1in} +G_{WH}^{Y(1)}\,\Y_0 \,\Big]\bigg\}
    (s,\mchi^{2},\mcH^{2},m_{W}^{2},\mchargk^{2},\mchargl^{2}), \\
%
 J^{WH}_{kl} & = &
  \bigg\{ Re(C_{+k}^{HW\!*}C_{-l}^{HW}+D_{+k}^{HW\!*}D_{-l}^{HW})
  \Big[ -2\,\T_2
        + 2\,(2\,\mchi^2-m_W^2)\,\T_1 \nonumber \\
  & & \hspace{1in} - 2(\mchi^2-m_W^2)(\mchi^2+2\,m_W^2)\T_0
        -3\,m_W^2\,\Y_1
  - G_{WH}^{Y(2)}\,\Y_0 \,\Big] \nonumber \\
  & & \hspace{0.1in} + 
\ Re(C_{+k}^{HW\!*}C_{-l}^{HW}-D_{+k}^{HW\!*}D_{-l}^{HW})
  \Big[ \,6\,m_W^2\,\T_1
        + 6\,m_W^2(\mchi^2-\mcH^2)\T_0 \nonumber \\
  & & \hspace{1in} - 2\,\Y_2
                    - (s-\mcH^2-2\,m_W^2)\Y_1 \nonumber \\
  & & \hspace{1in} - G_{WH}^{Y(3)}\,\Y_0 \,\Big]\bigg\}
  (s,\mchi^{2},\mcH^{2},m_{W}^{2},\mchargk^{2},\mchargl^{2}), \\
%
 K^{WH}_{kl} & = &
  \bigg\{(C_{-k}^{HW\!*}C_{-l}^{HW}+D_{-k}^{HW\!*}D_{-l}^{HW})
  \Big[ \,(s-\mchi^2-2\,m_W^2)\,\T_2
       - G_{WH}^{T(2)}\,\T_1  \nonumber \\
  & &   \hspace{1in} - \ G_{WH}^{T(3)}\,\T_0
       + (s-2\,m_W^2)\,\Y_2
       + G_{WH}^{Y(4)}\,\Y_0 \,\Big] \nonumber \\
  & & + \ (C_{-k}^{HW\!*}C_{-l}^{HW}-D_{-k}^{HW\!*}D_{-l}^{HW})
  \Big[ \,6 \,m_W^2 \mchi^2\,\T_1
       - \mchi^2\,\Y_2 \nonumber \\
   & & \hspace{1in} + G_{WH}^{Y(5)}\,\Y_0 \,\Big]\bigg\}
   (s,\mchi^{2},\mcH^{2},m_{W}^{2},\mchargk^{2},\mchargl^{2}),
\end{eqnarray}
and
\begin{eqnarray}
G_{WH}^{T(1)} \! &=& \! s (\mchi^2+2\mW^2) - \mchi^4
- \mchi^2 (\mcH^2+3\mW^2) - \mW^2\mcH^2, \nonumber \\
G_{WH}^{T(2)} \! &=& \! s (\mchi^2+2m_W^2)-2\mchi^4
+\mchi^2 (\mcH^2-m_W^2)-2m_W^2 (\mcH^2+m_W^2), \nonumber \\
G_{WH}^{T(3)} \! &=& \! (\mchi^2-\mcH^2)\,(\mchi^2-\mW^2)
\,(\mchi^2+2\,\mW^2), \nonumber \\
G_{WH}^{Y(1)} \! &=& \! G_{WH}^{T(1)} \, , \nonumber \\
G_{WH}^{Y(2)} \! &=& \! 3\,s\,m_W^2 - 12\,m_W^2 \mchi^2
 + 3\,m_W^2 \mcH^2 - 3\,m_W^4, \nonumber \\
G_{WH}^{Y(3)} \! &=& \! s^2 - s(2\mchi^2+2\mcH^2+3m_W^2)
 + 2\mchi^4 + (2\mchi^2+\mcH^2)(\mcH^2+3m_W^2)
 - 2m_W^4, \nonumber \\
G_{WH}^{Y(4)}  \! &=& \! s \,(\mchi^2-2 m_W^2)(\mchi^2-\mcH^2)
+ 4 m_W^2 \mchi^4   \nonumber \\
\! & & \! + \ \mchi^2 (\mcH^4 - 5 m_W^2 \mcH^2 + 2 m_W^4 )
- 2 m_W^4 \mcH^2,  \nonumber \\
G_{WH}^{Y(5)} \! &=& \! s \,\mchi^2 (\mchi^2-m_W^2) - \mchi^6 -
\mchi^4 (\mcH^2 + 3 m_W^2) + \mchi^2 m_W^2 (2 \mcH^2 + 3 m_W^2),
\nonumber
\end{eqnarray}
and
\begin{eqnarray}
 C_{\pm k}^{HW} & \equiv &
   C_{S}^{\chi_{k}^{+}\chi H^{-}}
   \Big(C_{V}^{\chi_{k}^{+}\chi W^{-}}\Big)^{*} \pm
   \:C_{P}^{\chi_{k}^{+}\chi H^{-}}
   \Big(C_{A}^{\chi_{k}^{+}\chi W^{-}}\Big)^{*}, \\
 D_{\pm k}^{HW} & \equiv &
   C_{S}^{\chi_{k}^{+}\chi H^{-}}
   \Big(C_{A}^{\chi_{k}^{+}\chi W^{-}}\Big)^{*}\pm
   \:C_{P}^{\chi_{k}^{+}\chi H^{-}}
   \Big(C_{V}^{\chi_{k}^{+}\chi W^{-}}\Big)^{*};
\end{eqnarray}
\vspace{0.25cm}
$\bullet$
 \underline{Higgs ($h,H$)--chargino ($\chi_{k}^{\pm}$) interference term:}
\begin{eqnarray}
 \widetilde{w}_{WH}^{(h,H-\chi^{\pm})}& = &
 -2 \: Re \sum_{k=1}^{2} \sum_{r=h,H} \left(\frac{C^{W^-H^+r}\:
 C_{S}^{\chi\chi r}}{ s-m_{r}^{2}+i\, \Gamma_{r}\, m_{r}} \right )^{*} \,
 \frac{1}{\mW^{2}} \nonumber \\
  & & \hspace{0.2in} \times \Big[C_{+k}^{HW\!*} H^{(+)WH}_{k}
 +C_{-k}^{HW\!*} H^{(-)WH}_{k}\Big],
\end{eqnarray}
where
\begin{eqnarray}
 H^{(+)WH}_{k} & = &
  2\,\mchi\,\mchargk\,(s+\mW^{2}-\mcH^{2})+\mchi\,\mchargk\,
  \Big[s^{2}+2\,s\,(\mchargk-\mchi^{2}-\mcH^{2}) \nonumber \\
  & & \hspace{0.1in}+\mcH^{4}-\mW^{4}+2\,(\mchargk^{2}-\mchi^{2})\,
  (\mW^{2}-\mcH^{2})\Big]
  \F(s,\mchi^{2},\mcH^{2},m_{W}^{2},\mchargk^{2}), \nonumber \\
  & &       \\
 H^{(-)WH}_{k} & = &
  - s^{2}+s\,(\mW^{2}+\mcH^{2} +2\,\mchi^{2})+2\,\mchi^{2}\,
  (\mW^{2}-\mcH^{2}) \nonumber \\
  & & \hspace{0.2in}+\Big\{-s^{2}\,\mchargk^{2}+s[2\,\mchi^{2}\,
  (\mchargk^{2}-\mchi^{2})
  + \mchi^{2}\,(\mcH^{2}+3\,\mW^{2}) \nonumber \\
  & & \hspace{1.4in} +\mchargk^{2}(\mcH^{2}+\mW^{2})
     - 2\,\mW^{2}\,\mcH^{2}] \nonumber \\
  & & \hspace{0.5in}+\mchi^{2}\,
  (\mW^{2}-\mcH^{2})\,(2\,\mchargk^{2}-2\,\mchi^{2}+\mcH^{2}
   - 3\,\mW^{2}) \Big\} \nonumber \\
  & & \hspace{0.3in} \times \F(s,\mchi^{2},\mcH^{2},m_{W}^{2},\mchargk^{2});
\end{eqnarray}
\vspace{0.25cm}
 $\bullet$
 \underline{Higgs ($A$)--chargino ($\chi_{k}^{\pm}$) interference term:}
\begin{eqnarray}
 \widetilde{w}_{WH}^{(A-\chi^{\pm})} & = &
  -2 \: Re \sum_{k=1}^{2} \left(\frac{C^{W^-H^+A}\:
  C_{P}^{\chi\chi A}}{ s-\mA^{2}+i\, \Gamma_{A}\, \mA} \right )^{*}\,
  \frac{1}{\mW^{2}} \nonumber \\
  & & \hspace{0.2in} \times \Big[D_{+k}^{HW\!*}A^{(+)WH}_{k}
  + D_{-k}^{HW\!*} A^{(-)WH}_{k}\Big],
\end{eqnarray}
where
\begin{eqnarray}
 A^{(+)WH}_{k} & = &
  \mchi\,\mchargk\Big[s^{2}-2\,s\,(\mcH^{2}+\mW^{2}) \nonumber \\
   & & \hspace{0.1in} +(\mcH^{2}-\mW^{2})^{2}\Big]
   \F(s,\mchi^{2},\mcH^{2},m_{W}^{2},\mchargk^{2}), \\
 A^{(-)WH}_{k} & = &
   s\,(s-\mW^{2}-\mcH^{2})+
   \Big\{ s^{2}\,\mchargk^{2}
   -s\,[(\mchi^{2}+\mchargk^{2})\,
   (\mcH^{2}+\mW^{2}) \nonumber \\
   & & \hspace{0.1in} -2\,\mW^{2}\,\mcH^{2}]
   +\mchi^{2}\,(\mcH^{2}-\mW^{2})^{2}\Big\}
   \F(s,\mchi^{2},\mcH^{2},m_{W}^{2},\mchargk^{2}).
\end{eqnarray}
Note that $\widetilde{w}_{WH}$ in eq.(\ref{eq:WH}) does not include
the contribution for $W^- H^+$ final state. 
The contributions to the $W^- H^+$ and $W^+ H^-$ final states are
obviously identical and so the total contribution for
$W^\pm H^\mp$ is twice that of $W^+ H^-$.

\vspace{0.3cm}
\begin{center}
\fbox{\boldmath ${6.\:\:\chi\chi\rightarrow Zh}$}
\end{center}
This process proceeds via the $s$-channel $Z$ and CP--odd Higgs boson ($A$)
exchange as well as the $t$- and $u$-channel neutralino ($\chi_{i}^{0}$,
$i=1,\ldots,4$) exchange
\begin{eqnarray}
 \widetilde{w}_{Zh} &=& \widetilde{w}_{Zh}^{(A)}
+\widetilde{w}_{Zh}^{(Z)} +\widetilde{w}_{Zh}^{(\chi^{0})}
+\widetilde{w}_{Zh}^{(A-Z)} +\widetilde{w}_{Zh}^{(A-\chi^{0})}
+\widetilde{w}_{Zh}^{(Z-\chi^{0})}:
\end{eqnarray}
\vspace{0.25cm} \\
$\bullet$
 \underline{CP--odd Higgs--boson ($A$) exchange:}
\begin{eqnarray}
  \widetilde{w}_{Zh}^{(A)}& = & \frac{1}{2}\:
   \left | \frac{C^{ZhA}\: C_{P}^{\chi\chi A}}
    {s-\mA^{2}+i\, \Gamma_{A}\, \mA} \right |^{2} \,s\,
      \frac{s^{2}-2\,(\mh^{2}+\mZ^{2})\,s+(\mh^{2}-\mZ^{2})^{2}}
   {\mZ^{2}};
\end{eqnarray}
\vspace{0.25cm}
$\bullet$
 \underline{$Z$--boson exchange:}
\begin{eqnarray}
 \widetilde{w}_{Zh}^{(Z)}& = & \frac{1}{12}\:
  \left | \frac{C^{ZZh}\: C_{A}^{\chi\chi Z}}
  {s-\mZ^{2}+i\, \Gamma_{Z}\, \mZ} \right |^{2}
  \frac{1}{\mZ^{6}\,s} \nonumber \\
  & & \times\left\{m_{\chi}^{2}[6\,s^{4}-12\,(\mh^{2}+2\,\mZ^{2})\,s^{3}
  +(32\,\mZ^{4}+12\,\mZ^{2}\,\mh^{2}+6\,\mh^{4})\,s^{2} \right.\nonumber \\
  & & \hspace{0.5in}-(64\,\mZ^{4}-20\,\mZ^{2}\,\mh^{2}
  +12\,\mh^{4})\,s\,\mZ^{2}+2\,(\mh^{2}-\mZ^{2})^{2}\,\mZ^{4}] \nonumber \\
  & & \left.\hspace{0.2in}+\mZ^{4}[s^{3}+(10\,\mZ^{2}-2\,\mh^{2})\,s^{2}
  +(\mh^{2}-\mZ^{2})^2\,s]\right\};
\end{eqnarray}
\vspace{0.25cm}
$\bullet$
 \underline{neutralino\ ($\chi_{i}^{0})$ exchange:}
\begin{eqnarray}
 \widetilde{w}_{Zh}^{(\chi^{0})} & = &
  \frac{1}{\mZ^{2}}\sum_{i,j=1}^{4}
C_{S}^{\chi_{i}^{0}\chi h} C_{S}^{\chi_{j}^{0}\chi h *}
  \bigg[ \mchii\,\mchij I^{Zh}_{ij}
   +\mchi \mchii J^{Zh}_{ij} + K^{Zh}_{ij} \bigg],
\end{eqnarray}
where
\begin{eqnarray}
 I^{Zh}_{ij} & = &
  \bigg\{(C_{Vi}^{\chi} C_{Vj}^{\chi *}
  +C_{Ai}^{\chi} C_{Aj}^{\chi *}) \nonumber \\
  & & \hspace{0.2in} \times \Big[ - \T_2
       - (s-m_Z^2-\mh^2-2\,\mchi^2)\,\T_1
       + G_{Zh}^{T(1)}\,\T_0
       + 6\,m_Z^2 \mchi^2\,\Y_0 \,\Big] \nonumber \\
  & & + (C_{Vi}^{\chi} C_{Vj}^{\chi *}
  -C_{Ai}^{\chi} C_{Aj}^{\chi *})
  \Big[ \,6\,m_Z^2 \mchi^2\,\T_0
        - \Y_2 \nonumber \\
    & & \hspace{1.5in} + G_{Zh}^{Y(1)}\,\Y_0 \,\Big]\bigg\}
     (s,\mchi^{2},\mh^{2},\mZ^{2},\mchii^{2},\mchij^{2}), \\
 J^{Zh}_{ij} & = &
  \bigg\{(C_{Vi}^{\chi} C_{Vj}^{\chi *}
  +C_{Ai}^{\chi} C_{Aj}^{\chi *})
  \Big[ -2\,\T_2
       + 2\,(2\,\mchi^2-m_Z^2)\,\T_1 \nonumber \\
  & & \hspace{1.5in} - \ 2(\mchi^2-m_Z^2)(\mchi^2+2\,m_Z^2)\T_0
                    - 3\,m_Z^2\,\Y_1
                    - \ G_{Zh}^{Y(2)}\,\Y_0 \,\Big] \nonumber \\
  & & + (C_{Vi}^{\chi} C_{Vj}^{\chi *}
         - C_{Ai}^{\chi} C_{Aj}^{\chi *})
  \Big[ \,6\,m_Z^2\,\T_1
       + 6\,m_Z^2(\mchi^2-\mh^2)\T_0 \nonumber \\
  & & \hspace{1.5in} - \ 2\,\Y_2
                    - (s-\mh^2-2\,m_Z^2)\Y_1 \nonumber \\
  & & \hspace{1.5in}                  - G_{Zh}^{Y(3)}\,\Y_0 \,\Big]\bigg\}
       (s,\mchi^{2},\mh^{2},\mZ^{2},\mchii^{2},\mchij^{2}), \\
 K^{Zh}_{ij} & = &
  \bigg\{(C_{Vi}^{\chi} C_{Vj}^{\chi *}
   + C_{Ai}^{\chi} C_{Aj}^{\chi *})
  \Big[ \,(s-\mchi^2-2\,m_Z^2)\,\T_2
       - G_{Zh}^{T(2)}\,\T_1  \nonumber \\
  & &   \hspace{1.5in} - \ G_{Zh}^{T(3)}\,\T_0
       + (s-2\,m_Z^2)\,\Y_2
       + G_{Zh}^{Y(4)}\,\Y_0 \,\Big] \nonumber \\
  & & + (C_{Vi}^{\chi} C_{Vj}^{\chi *}
        - C_{Ai}^{\chi} C_{Aj}^{\chi *})
 \Big[ \,6 \,m_Z^2 \mchi^2\,\T_1
       - \mchi^2\,\Y_2 \nonumber \\
  & &  \hspace{1.5in}  +G_{ZH}^{Y(5)}\,\Y_0 \,\Big]\bigg\}
       (s,\mchi^{2},\mh^{2},\mZ^{2},\mchii^{2},\mchij^{2}),  
%
\end{eqnarray}
and
\begin{eqnarray}
 \!\! G_{Zh}^{T(1)} \! &=& \!
s (\mchi^2+2\mZ^2) - \mchi^4
- \mchi^2 (\mh^2+3\mZ^2) - \mZ^2\mh^2, \nonumber \\
 \!\! G_{Zh}^{T(2)} \! &=& \! s (\mchi^2+2m_Z^2)-2\mchi^4
+\mchi^2 (\mh^2-m_Z^2)-2m_Z^2 (\mh^2+m_Z^2), \nonumber \\
 \!\! G_{Zh}^{T(3)} \! &=& \!
(\mchi^2-\mh^2)\,(\mchi^2-\mZ^2)
\,(\mchi^2+2\,\mZ^2), \nonumber \\
 \!\! G_{Zh}^{Y(1)} \! &=& \! G_{Zh}^{T(1)} \, , \nonumber \\
 \!\! G_{Zh}^{Y(2)} \! &=& \!
3\,s\,m_Z^2 - 12\,m_Z^2 \mchi^2
 + 3\,m_Z^2 \mh^2 - 3\,m_Z^4, \nonumber \\
 \!\! G_{Zh}^{Y(3)} \! &=& \!
s^2 - s(2\mchi^2+2\mh^2+3m_Z^2)
 + 2\mchi^4 + (2\mchi^2+\mh^2)(\mh^2+3m_Z^2)
 - 2m_Z^4, \nonumber \\
 \!\! G_{Zh}^{Y(4)} \! &=& \!
s \,(\mchi^2-2 m_Z^2)(\mchi^2-\mh^2)
+ 4 m_Z^2 \mchi^4   \nonumber \\
\! & & \! + \ \mchi^2 (\mh^4 - 5 m_Z^2 \mh^2 + 2 m_Z^4 )
- 2 m_Z^4 \mh^2,  \nonumber \\
 \!\! G_{Zh}^{Y(5)} \! &=& \!
s \,\mchi^2 (\mchi^2-m_Z^2) - \mchi^6 - \mchi^4 (\mh^2 + 3 m_Z^2) 
        + \mchi^2 m_Z^2 (2 \mh^2 + 3 m_Z^2), \nonumber
\end{eqnarray}
and
\begin{eqnarray}
\label{Zh-coupl1:eq}
     C_{Vi}^{\chi}& \equiv &  C_{V}^{\chi_{i}^{0}\chi Z}, \\
\label{Zh-coupl2:eq}
     C_{Ai}^{\chi}& \equiv&  C_{A}^{\chi_{i}^{0}\chi Z};
\end{eqnarray}
\vspace{0.25cm}
$\bullet$
 \underline{Higgs ($A$)--$Z$ interference term:}
\begin{eqnarray}
 \widetilde{w}_{Zh}^{(A-Z)}& =&
  Re \left [\left(\frac{C^{ZhA}\: C_{P}^{\chi\chi A}}
  {s-\mA^{2}+i\, \Gamma_{A}\, \mA}\right)^{*}\,
  \frac{C^{ZZh}\: C_{A}^{\chi\chi Z}}{s-\mZ^{2}+i\,\Gamma_{Z}\,\mZ}
      \right] \nonumber \\
  & & \times \frac{m_{\chi}\,(\mZ^{2}-s)}
  {\mZ^{4}}\,[s^{2}-2\,(\mh^{2}+\mZ^{2})\,s+(\mh^{2}-\mZ^{2})^{2}];
\end{eqnarray}
\vspace{0.25cm}
$\bullet$
 \underline{Higgs ($A$)--neutralino\ ($\chi_{i}^{0}$) interference term:}
\begin{eqnarray}
  \widetilde{w}_{Zh}^{(A-\chi^{0})} & = & 2\,\sum_{i=1}^{4}
   Re \left[\left(\frac{C^{ZhA}\: C_{P}^{\chi\chi A}}
   {s-\mA^{2}+i\, \Gamma_{A}\, \mA}\right)^{*}\,
   C_{S}^{\chi_{i}^{0}\chi h}\: C_{A}^{\chi_{i}^{0}\chi Z}\right] 
 \frac{1}{\mZ^2} \nonumber \\
   & & \times  \bigg[-s^{2}+(\mh^{2}+\mZ^{2})\,s
   -\Big\{s^{2}\,\mchii\,(\mchi+\mchii) \nonumber \\
   & & \hspace{0.2in} +s\,[2\,\mh^{2}\,\mZ^{2}
   -(\mh^{2}+\mZ^{2})\,(\mchi+\mchii)^{2}] \nonumber  \\
   & & \hspace{0.2in}+\mchi\,(\mchi+\mchii)\,(\mh^{2}-\mZ^{2})^{2}\Big\}\,
   \F(s,\mchi^{2},\mh^{2},\mZ^{2},\mchii^{2})\bigg];
\end{eqnarray}
\vspace{0.25cm}
$\bullet$
 \underline{$Z$--neutralino\ ($\chi_{i}^{0}$) interference term:}
\begin{eqnarray}
 \widetilde{w}_{Zh}^{(Z-\chi^{0})} & = &
  -\sum_{i=1}^{4} Re \left [ \left(\frac{C^{ZZh}\: C_{A}^{\chi\chi Z}}
  { s-\mZ^{2}+i\, \Gamma_{Z}\, \mZ}\right)^{*}\,
  C_{S}^{\chi_{i}^{0}\chi h}\: C_{A}^{\chi_{i}^{0}\chi Z}\right]\,
  \frac{1}{\mZ^{4}} \nonumber \\
  & & \times \bigg\{ \mchi[-2\,s^{2}+s\,(2\,\mh^{2}
  +3\,\mZ^{2})+\mZ^{2}\,(2\,\mchi^{2}-9\,\mZ^{2}-\mh^{2}-2\,\mchii^{2})]
      \nonumber \\
  & & \hspace{0.2in}+\mchii\,\mZ^{2}
  [-s+(2\,\mchi^{2}+\mZ^{2}+\mh^{2}-2\,\mchii^{2})] \nonumber \\
  & &\hspace{0.2in} -2\,\Big[\mchi\big\{s^{2}\,\mchii^{2}
  +s\,[-(\mchi^{2}+\mchii^{2})\,(\mZ^{2}+\mh^{2})+2\,\mZ^{2}\,\mh^{2}]
      \nonumber \\
  & & \hspace{0.9in}-2\,\mZ^{6}+5\,\mchi^{2}\,\mZ^{4}
  +4\,\mchii^{2}\,\mZ^{4}+\mZ^{2}\,(\mchi^{2}-\mchii^{2})^{2} \nonumber \\
  & & \hspace{0.9in}-\mh^{2}\,\mZ^{2}\,(3\,\mZ^{2}+2\,\mchi^{2})
  +\mchi^{2}\,\mh^{4}\big\} \nonumber \\
  & & \hspace{0.5in}+\mchii[s^{2}\,\mchi^{2}
  +s\,(-3\,\mchi^{2}\,\mZ^{2}+\mZ^{2}\,\mchii^{2}
  -2\,\mchi^{2}\,\mh^{2}-2\,\mZ^{4}) \nonumber \\
  & & \hspace{0.8in}+10\,\mchi^{2}\,\mZ^{4}-\mchii^{2}\,\mZ^{4}
  +\mZ^{2}\,(\mchi^{2}-\mchii^{2})^{2} \nonumber \\
  & & \!\hspace{0.8in}+\mh^{2}\,\mZ^{2}\,(\mZ^{2}-\mchi^{2}-\mchii^{2})
  +\mchi^{2}\,\mh^{4}\big]\Big] 
  \F(s,\mchi^{2},\mh^{2},\mZ^{2},\mchii^{2})\bigg\}. \nonumber \\
  \end{eqnarray}
The expressions for $ZH$ final state are found by replacing
$h$ with $H$ in the above. \\
\vspace{0.25cm}

\vspace{0.3cm}
\begin{center}
\fbox{\boldmath ${7.\:\:\chi\chi\rightarrow ZA}$}
\end{center}
This process involves the $s$-channel CP--even Higgs boson ($h$ and
$H$) exchange and the $t$- and $u$-channel neutralino ($\chi_{i}^{0}$,
$i=1,\ldots,4$) exchange
\begin{eqnarray}
 \widetilde{w}_{ZA} &=&
 \widetilde{w}_{ZA}^{(h,H)}
+\widetilde{w}_{ZA}^{(\chi^{0})} +\widetilde{w}_{ZA}^{(h,H-\chi^{0})}:
\end{eqnarray}
\vspace{0.25cm}
$\bullet$
 \underline{CP-even Higgs-boson $(h,H)$ exchange:}
\begin{eqnarray}
  \widetilde{w}_{ZA}^{(h,H)}& = &
   \frac{1}{2}\: \left |\sum_{r=h,H} \frac{C^{ZAr}\:
   C_{S}^{\chi\chi r}}{ s-m_{r}^{2}+i\, \Gamma_{r}\, m_{r}}
   \right |^{2} \,(s-4\,\mchi^{2})\, \nonumber \\
   & & \times \frac{s^{2}-2\,(\mA^{2}+\mZ^{2})\,s
   +(\mA^{2}-\mZ^{2})^{2}}{\mZ^{2}}\,;
\end{eqnarray}
\vspace{0.25cm}
$\bullet$
 \underline{neutralino ($\chi_{i}^{0}$) exchange:}
\begin{eqnarray}
 \widetilde{w}_{ZA}^{(\chi^{0})} & = &
  \frac{1}{\mZ^{2}}\sum_{i,j=1}^{4}
C_{P}^{\chi_{i}^{0}\chi A} C_{P}^{\chi_{j}^{0}\chi A *}
  \bigg[ \mchii\,\mchij I^{ZA}_{ij}
      +m_{\chi} \mchii J^{ZA}_{ij} + K^{ZA}_{ij} \bigg],
\end{eqnarray}
where
\begin{eqnarray}
 I^{ZA}_{ij} & = &
  \bigg\{ (C_{Vi}^{\chi} C_{Vj}^{\chi *}
  +C_{Ai}^{\chi} C_{Aj}^{\chi *}) \nonumber \\
  & & \hspace{2mm} \times \Big[ - \T_2
       - (s-m_Z^2-\mA^2-2\,\mchi^2)\,\T_1
       + G_{ZA}^{T(1)}\,\T_0
       + 6\,m_Z^2 \mchi^2\,\Y_0 \,\Big] \nonumber \\
  & & + (C_{Vi}^{\chi} C_{Vj}^{\chi *}
      - C_{Ai}^{\chi} C_{Aj}^{\chi *})
 \Big[- \,6\,m_Z^2 \mchi^2\,\T_0
       + \Y_2 \nonumber \\
  & &  \hspace{1.5in} - G_{ZA}^{Y(1)}\,\Y_0 \,\Big] \bigg\}
     (s,\mchi^{2},\mA^{2},\mZ^{2},\mchii^{2},\mchij^{2}),
\\
J^{ZA}_{ij} & = &
 \bigg\{ (C_{Vi}^{\chi} C_{Vj}^{\chi *}
 + C_{Ai}^{\chi} C_{Aj}^{\chi *})
 \Big[ 2\,\T_2
       - 2\,(2\,\mchi^2-m_Z^2)\,\T_1 \nonumber \\
 & & \hspace{2.3cm} + 2(\mchi^2-m_Z^2)(\mchi^2+2\,m_Z^2)\T_0
                    + 3\,m_Z^2\,\Y_1
                    + G_{ZA}^{Y(2)}\,\Y_0 \,\Big] \nonumber \\
 & & + (C_{Vi}^{\chi} C_{Vj}^{\chi *}
       -C_{Ai}^{\chi} C_{Aj}^{\chi *})
 \Big[ \,6\,m_Z^2\,\T_1
       + 6\,m_Z^2(\mchi^2-\mA^2)\T_0 \nonumber \\
 & & \hspace{1.5in} - 2\,\Y_2
                    - (s-\mA^2-2\,m_Z^2)\Y_1 \nonumber \\
 & & \hspace{1.5in} - G_{ZA}^{Y(3)}\,\Y_0 \,\Big] \bigg\}
   (s,\mchi^{2},\mA^{2},\mZ^{2},\mchii^{2},\mchij^{2}),
\\
 K^{ZA}_{ij} & = &
  \bigg\{ (C_{Vi}^{\chi} C_{Vj}^{\chi *}
          +C_{Ai}^{\chi} C_{Aj}^{\chi *})
  \Big[ \,(s-\mchi^2-2\,m_Z^2)\,\T_2
       - G_{ZA}^{T(2)}\,\T_1  \nonumber \\
 & &   \hspace{1.5in} - G_{ZA}^{T(3)}\,\T_0
       + (s-2\,m_Z^2)\,\Y_2
       + G_{ZA}^{Y(4)}\,\Y_0 \,\Big] \nonumber \\
 & & + (C_{Vi}^{\chi} C_{Vj}^{\chi *}
       -C_{Ai}^{\chi} C_{Aj}^{\chi *})
 \Big[ -6 \,m_Z^2 \mchi^2\,\T_1
       + \mchi^2\,\Y_2 \nonumber \\
 & &  \hspace{1.5in} - G_{ZA}^{Y(5)}\,\Y_0 \,\Big]\bigg\}
   (s,\mchi^{2},\mA^{2},\mZ^{2},\mchii^{2},\mchij^{2}),
%
\end{eqnarray}
and
\begin{eqnarray}
G_{ZA}^{T(1)} \! &=& \! s (\mchi^2+2\mZ^2) - \mchi^4
- \mchi^2 (\mA^2+3\mZ^2) - \mZ^2\mA^2, \nonumber \\
G_{ZA}^{T(2)} \! &=& \! s (\mchi^2+2m_Z^2)-2\mchi^4
+\mchi^2 (\mA^2-m_Z^2)-2m_Z^2 (\mA^2+m_Z^2), \nonumber \\
G_{ZA}^{T(3)} \! &=& \! (\mchi^2-\mA^2)\,(\mchi^2-\mZ^2)
\,(\mchi^2+2\,\mZ^2), \nonumber \\
G_{ZA}^{Y(1)} \! &=& \! G_{ZA}^{T(1)} , \nonumber \\
G_{ZA}^{Y(2)} \! &=& \! 3\,s\,m_Z^2 - 12\,m_Z^2 \mchi^2
 + 3\,m_Z^2 \mA^2 - 3\,m_Z^4, \nonumber \\
G_{ZA}^{Y(3)} \! &=& \! s^2 - s(2\mchi^2+2\mA^2+3m_Z^2)
 + 2\mchi^4 + (2\mchi^2+\mA^2)(\mA^2+3m_Z^2)
 - 2m_Z^4, \nonumber \\
G_{ZA}^{Y(4)} \! &=& \! s \,(\mchi^2-2 m_Z^2)(\mchi^2-\mA^2)
+ 4 m_Z^2 \mchi^4   \nonumber \\
\! & & \! + \ \mchi^2 (\mA^4 - 5 m_Z^2 \mA^2 + 2 m_Z^4 )
- 2 m_Z^4 \mA^2,  \nonumber \\
G_{ZA}^{Y(5)} \! &=& \! s \,\mchi^2 (\mchi^2-m_Z^2) - \mchi^6 - \mchi^4 (\mA^2 + 3
m_Z^2) + \mchi^2 m_Z^2 (2 \mA^2 + 3 m_Z^2). \nonumber
\end{eqnarray}
$G_{ZA}^{T(1-3)}$ and $G_{ZA}^{Y(1-5)}$ are obtained from
$G_{Zh}^{T(1-3)}$ and $G_{Zh}^{Y(1-5)}$ by replacing $\mh$ with
$\mA$. \vspace{0.25cm} $C_{Vi}^{\chi}$ and $C_{Ai}^{\chi}$ have
already been defined for the $Zh$ final state in
eqs.~(\ref{Zh-coupl1:eq}) and~(\ref{Zh-coupl2:eq}). \\
$\bullet$
 \underline{Higgs $(h,H)$--neutralino ($\chi_{i}^{0}$) interference term:}
\begin{eqnarray}
  \widetilde{w}_{ZA}^{(h,H-\chi^0)}& =&
   -2\,\sum_{i=1}^{4} Re \left [\sum_{r=h,H}
   \left(\frac{C^{ZAr}\: C_{S}^{\chi\chi r}}
   { s-m_{r}^{2}+i\, \Gamma_{r}\, m_{r}}\right)^{*}\,
   C_{A}^{\chi_{i}^{0}\chi Z}\: C_{P}^{\chi_{i}^{0}\chi A}\right]
 \frac{1}{\mZ^2} \nonumber\\
   & & \times  \bigg[-s^{2}+(\mA^{2}+\mZ^{2})\,s+2\,\mchi\,
   (\mchii-\mchi)\,(-s-\mZ^{2}+\mA^{2}) \nonumber \\
   & & \hspace{0.2in}+\Big\{-(\mchii+\mchi)\,(s-\mA^{2})
   [\mchii\,s+2\,\mchi\,(\mchii-\mchi)^{2}-\mchi\,\mA^{2}] \nonumber \\
   & & \hspace{0.4in}+\mZ^{2}\,\mchi[-2\,(\mchii-\mchi)^{2}\,
   (\mchii+\mchi)+4\,\mchi\,\mA^{2} \nonumber  \\
   & & \hspace{0.6in}+\mZ^{2}\,(\mchii-3\,\mchi)+3\,s\,\mchi]
   +s\,\mZ^{2}\,(\mchii^{2}-2\,\mA^{2})\Big\} \nonumber \\
   & & \hspace{0.3in}\:\times
               \F(s,\mchi^{2},\mA^{2},\mZ^{2},\mchii^{2})\bigg].
\end{eqnarray}

\vspace{0.3cm}
\begin{center}
\fbox{\boldmath ${8.\:\:\chi\chi\rightarrow WW}$}
\end{center}
This process involves the $s$-channel CP--even Higgs boson ($h$ and
$H$) and $Z$ exchange, and the $t$- and $u$-channel chargino
($\chi_{k}^{\pm}$, $k=1,2$) exchange
\begin{eqnarray}
 \widetilde{w}_{WW} &=&
 \widetilde{w}_{WW}^{(h,H)}
+\widetilde{w}_{WW}^{(Z)} +\widetilde{w}_{WW}^{(\chi^{\pm})}
+\widetilde{w}_{WW}^{(h,H-\chi^{\pm})} +\widetilde{w}_{WW}^{(Z-\chi^{\pm})}:
\end{eqnarray}
\vspace{0.25cm}
%
$\bullet$
 \underline{CP--even Higgs--boson $(h,H)$ exchange:}
\begin{eqnarray}
 \widetilde{w}_{WW}^{(h,H)}& = &
  \left |\sum_{r=h,H} \frac{C^{WWr}\: C_{S}^{\chi\chi r}}
  { s-m_{r}^{2}+i\, \Gamma_{r}\, m_{r}} \right |^{2} \,(s-4\,\mchi^{2})\,
   \frac{s^{2}-4\,\mW^{2}\,s+12\,\mW^{4}}{8\,\mW^{4}};
\end{eqnarray}
\vspace{0.25cm}
$\bullet$
 \underline{$Z$-boson exchange:}
\begin{eqnarray}
 \widetilde{w}_{WW}^{(Z)}& = & \left | \frac{C^{WWZ}\: C_{A}^{\chi\chi
  Z}} {s-\mZ^{2}+i\, \Gamma_{Z}\, \mZ} \right |^{2} (s-4\,\mchi^{2})\,
  \nonumber \\ 
& & \times \frac{s^{3}
  +16\,\mW^{2}\,s^{2}-68\,\mW^{4}\,s-48\,\mW^{6}}{12\,\mW^{4}};
\end{eqnarray}
\vspace{0.25cm}
$\bullet$
 \underline{chargino\ ($\chi_{k}^{\pm}$) exchange:}
\begin{eqnarray}
 \widetilde{w}_{WW}^{(\chi^{\pm})} & = &
  \frac{1}{\mW^{4}}\sum_{k,l=1}^{2}
  \bigg[ \mchargk\,\mchargl I^{WW}_{kl}
      + \mchi \mchargk J^{WW}_{kl} + K^{WW}_{kl} \bigg],
\end{eqnarray}
where
\begin{eqnarray}
 I^{WW}_{kl} & = &
  \bigg\{ (C_{-k}^{W\!*}C_{-l}^{W}+D_{-k}^{W\!*}D_{-l}^{W}) \nonumber \\
  & & \hspace{5mm} \times \Big[ (s-4\,\mW^2)\, \T_2
       - G_{WW}^{T(1)} \,\T_1
       + G_{WW}^{T(2)}\,\T_0
       + (s-4\,\mW^2)\, \Y_2
       - G_{WW}^{Y(1)} \,\Y_0 \,\Big] \nonumber \\
 & & + (C_{-k}^{W\!*}C_{-l}^{W}-D_{-k}^{W\!*}D_{-l}^{W})
       \Big[ - 18\,m_W^4 \mchi^2\,\T_0 \nonumber \\
 & &   \hspace{5mm} - \mchi^2 (-s^{2}+4\,\mW^2\,s+6\,m_W^4)\,
        \Y_0 \,\Big] \bigg\}
       (s,\mchi^{2},m_{W}^{2},m_{W}^{2},\mchargk^{2},\mchargl^{2}),
\\
 J^{WW}_{kl} & = &
  \bigg\{ C_{-k}^{W\!*}C_{+l}^{W}
 \Big[ 12\,\mW^{2}\,\T_2
       + 12\,\mW^2\,(\mW^2-2\,\mchi^2)\,\T_1 \nonumber \\
 & & \hspace{0.2in}
+ 12\mW^2\,(\mchi^4+\mW^2\mchi^2-2\mW^4)\T_0
       - 4\,(s-m_W^2)\Y_2 \nonumber \\
 & & \hspace{0.2in} + 2\,m_W^2\,(2\,s-3\,m_W^2)\,\Y_1
       + \ G_{WW}^{Y(2)}\,\Y_0 \,\Big] \bigg\}
            (s,\mchi^{2},m_{W}^{2},m_{W}^{2},\mchargk^{2},\mchargl^{2}),
\\
 K^{WW}_{kl}  & = &
  \bigg\{ (C_{+k}^{W\!*}C_{+l}^{W}+D_{+k}^{W\!*}D_{+l}^{W})
 \Big[-\T_4-(s-2\,m_W^2-4\,\mchi^2)\,\T_3
       -G_{WW}^{T(3)}\,\T_2   \nonumber \\
 & & \hspace{0.2in} - G_{WW}^{T(4)}\,\T_1
       -(m_W^2-\mchi^2)^2(2\,m_W^2+\mchi^2)^2\,\T_0
       +\mchi^2(s-4\,m_W^2)\,\Y_2
       - G_{WW}^{Y(3)}\,\Y_0 \,\Big] \nonumber \\
 & & +  (C_{+k}^{W\!*}C_{+l}^{W}-D_{+k}^{W\!*}D_{+l}^{W})
       \Big[ -\,18 \,m_W^4 \mchi^2\,\T_1
  +\hspace{0.1in} \Y_4+ (2\,\mchi^4-4\,s\,m_W^2+ m_W^4)\,\Y_2 \nonumber \\
 & & \hspace{1.7in}  - G_{WW}^{Y(4)}\,\Y_0 \,\Big]\bigg\}
      (s,\mchi^{2},m_{W}^{2},m_{W}^{2},\mchargk^{2},\mchargl^{2}),
\end{eqnarray}
and
\begin{eqnarray}
G_{WW}^{T(1)} \! &=& \!
2[s (\mchi^2+2\,\mW^2) - 4\,\mW^2 (\mchi^2+\mW^2)], \nonumber \\
G_{WW}^{T(2)} \! &=& \! s (\mchi^2+2\,m_W^2)^2
+\mW^2 (-4\,\mchi^4-10\,\mchi^2\,\mW^2-4\,m_W^4), \nonumber \\
G_{WW}^{T(3)} \! &=& \!
 -2\,s\,(\mchi^{2}+2\,\mW^{2})+6\,\mchi^{4}
+6\,\mchi^{2}\,\mW^{2}+5\,\mW^4, \nonumber \\
G_{WW}^{T(4)} \! &=& \! s\,(\mchi^{2}+2\,\mW^{2})^{2}-8\,\mW^{6}
-6\,\mchi^{4}\,\mW^{2}-4\,\mchi^{6}, \nonumber \\
G_{WW}^{Y(1)} \! &=& \! -s\,\mchi^{2}\,(\mchi^{2}+2\,\mW^{2})
+6\,\mchi^{2}\,\mW^{4}+4\,\mW^{2}\,(\mchi^{2}-m_W^{2})^{2}, \nonumber \\
G_{WW}^{Y(2)} \! &=& \! 2\,[2\,\mW^{2}\,s^{2}+s\,(2\,\mchi^{2}\,\mW^{2}
-2\,\mchi^{4}-11\,\mW^{4}) \nonumber \\
& & +2\,\mW^{2}(\mchi^{4}+4\,\mchi^{2}\,\mW^{2}+\mW^{4})], \nonumber \\
G_{WW}^{Y(3)} \! &=& \! \mchi^2(\mW^{2}-\mchi^{2})
[s(\mchi^{2}+3\,\mW^{2})-2\,\mW^{2}(2\,\mchi^{2}+\mW^2)], \nonumber \\
G_{WW}^{Y(4)} \! &=& \! (\mW^{2}-\mchi^{2}) 
[s\,(8\,\mW^{4}-4\,\mchi^{2}\,\mW^{2})
-\mchi^{2}(8\,\mW^{4}-\mchi^4-\mchi^{2}\,\mW^{2})], \nonumber
\end{eqnarray}
and
\begin{eqnarray}
\label{WW-coupl1:eq}
   C_{\pm k}^{W}& \equiv& |C_{V}^{\chi_{k}^{+}\chi W^{-}}|^{2} \pm
                          |C_{A}^{\chi_{k}^{+}\chi W^{-}}|^{2},  \\
\label{WW-coupl2:eq}
   D_{\pm k}^{W}& \equiv&
      C_{V}^{\chi_{k}^{+}\chi W^{-}}
       \Big(C_{A}^{\chi_{k}^{+}\chi W^{-}}\Big)^{*}\pm
     \:C_{A}^{\chi_{k}^{+}\chi W^{-}}
       \Big(C_{V}^{\chi_{k}^{+}\chi W^{-}}\Big)^{*};
\end{eqnarray}
\vspace{0.25cm}
$\bullet$
 \underline{Higgs ($h,H$)--chargino\ ($\chi_{k}^{\pm}$) interference term:}
\begin{eqnarray}
 \widetilde{w}_{WW}^{(h,H-\chi^{\pm})} & = &
  \sum_{k=1}^{2} Re \left[\left(\sum_{r=h,H} \frac{C^{WWr}\:
   C_{S}^{\chi\chi r}}{ s-m_{r}^{2}+i\, \Gamma_{r}\, m_{r}} \right )^{*}
        \right]   \frac{1}{\mW^{4}} \nonumber \\
   & & \hspace{0.2in}\times \Big[C_{+k}^{W} H^{(+)WW}_{k}
   + C_{-k}^{W} H^{(-)WW}_{k}\Big],
\end{eqnarray}
where
\begin{eqnarray}
 H^{(+)WW}_{k} & = &
  \mchi[s^{2}+2\,s\,(\mchargk^{2}-\mW^{2}-\mchi^{2})
   +8\,\mW^{4}+4\,\mW^{2}\,\mchargk^{2}-4\,\mW^{2}\,\mchi^{2}] \nonumber \\
  & & +2\,\mchi\Big\{ s^{2}\,\mchargk^{2}
  +s\,[2\,\mW^{4}-3\,\mW^{2}\,\mchi^{2}-\mchargk^{2}\,\mW^{2}
  +(\mchi^{2}-\mchargk^{2})^{2}] \nonumber \\
  & & \hspace{0.4in}-4\,\mW^{6}
  +2\,\mW^{4}\,(\mchi^{2}+\mchargk^{2}) \nonumber \\
  & &\hspace{0.4in} +2\,\mW^{2}\,(\mchi^{2}-\mchargk^{2})^{2}\Big\}
  \F(s,\mchi^{2},m_{W}^{2},m_{W}^{2},\mchargk^{2}), \\
 H^{(-)WW}_{k} & = &
  -\mchargk\,(s^{2}-2\,\mW^{2}\,s)
  + \mchargk \Big[-(\mchi^{2}+\mchargk^{2})\,s^{2}
  + 2\,s\,(2\,\mW^{4}+\mW^{2}\,\mchargk^{2} \nonumber \\
  & & \hspace{0.4in} +3\,\mW^{2}\,\mchi^{2})
  -24\,\mchi^{2}\,\mW^{4}\Big]
  \F(s,\mchi^{2},m_{W}^{2},m_{W}^{2},\mchargk^{2});
\end{eqnarray}
\vspace{0.25cm}
$\bullet$
 \underline{$Z$--chargino\ ($\chi_{k}^{\pm}$) interference term:}
\begin{eqnarray}
  \widetilde{w}_{WW}^{(Z-\chi^{\pm})}& = &
   \frac{1}{3\,\mW^4}\sum_{k=1}^{2}
   Re \left[\left( \frac{C^{WWZ}\: C_{A}^{\chi\chi Z}}
    { s-\mZ^{2}+i\, \Gamma_{Z}\, \mZ} \right )^{*}\right] \,
    D_{+k}^{W} \nonumber \\
    & & \times \bigg\{s^{3}+s^{2}\,(18\,\mW^{2}-\mchi^{2}-3\,\mchargk^{2})
            \nonumber \\
    & & \hspace{0.2in}+ 2 \,s\,[-14\,\mW^{4}-18\,\mchi^{2}\,\mW^{2}
    + 6\,\mchargk^{2}\,\mW^{2} -3\,(\mchi^{2}-\mchargk^{2})^{2}] \nonumber \\
    & & \hspace{0.15in}+ 4 \,\mW^{2}\,[3\,(\mchi^{2}
    - \mchargk^{2})^{2}-\mW^{2}\,(11\,\mchi^{2}
    - 3\,\mchargk^{2}) - 6 \,\mW^{4}]\nonumber \\
    & & \hspace{0.15in} + 6 \,\Big[s^{2}\,\mchargk^{2}\,
    (\mchi^{2}-\mchargk^{2}+4\,\mW^{2}) \nonumber
    +\,s[8\,\mW^{6}-5\,\mW^{4}\,(\mchargk^{2}+3\,\mchi^{2}) \nonumber \\
    & & \hspace{0.3in}+2\,\mW^{2}\,(3\,\mchi^{4}
    -5\,\mchi^{2}\,\mchargk^{2}+2\,\mchargk^{4})
    +(\mchi^{2}-\mchargk^{2})^{3}] \nonumber \\
    & & \hspace{0.3in}+\mW^{2}[4\,\mW^{6}
    -2\,\mW^{4}\,(3\,\mchargk^{2}+5\,\mchi^{2}) \nonumber \\
    & & \hspace{0.3in}+8\,\mW^{2}\,\mchi^{2}\, (\mchi^{2}-\mchargk^{2})
     +2\,(\mchargk^{2}-\mchi^{2})^{3}]\Big]
    \F(s,\mchi^{2},m_{W}^{2},m_{W}^{2},\mchargk^{2})\bigg\}. \nonumber \\
    & &
\end{eqnarray}

\vspace{0.3cm}
\begin{center}
\fbox{\boldmath ${9.\:\:\chi\chi\rightarrow ZZ}$}
\end{center}
This process involves the $s$-channel CP--even Higgs boson ($h$ and
$H$) exchange and the $t$- and $u$-channel neutralino ($\chi_{i}^{0}$,
$i=1,\ldots,4$) exchange
\begin{eqnarray}
 \widetilde{w}_{ZZ} &=&
 \widetilde{w}_{ZZ}^{(h,H)}
+\widetilde{w}_{ZZ}^{(\chi^{0})} +\widetilde{w}_{ZZ}^{(h,H-\chi^{0})}:
\end{eqnarray}
\vspace{0.25cm}
$\bullet$
 \underline{CP--even Higgs--boson $(h,H)$ exchange:}
\begin{eqnarray}
  \widetilde{w}_{ZZ}^{(h,H)}& = &
    \left |\sum_{r=h,H} \frac{C^{ZZr}\: C_{S}^{\chi\chi r}}
    {s-m_{r}^{2}+i\, \Gamma_{r}\, m_{r}} \right |^{2} \,
    (s-4\,\mchi^{2})\,
      \frac{s^{2}-4\,\mZ^{2}\,s+12\,\mZ^{4}}{16\,\mZ^{4}};
\end{eqnarray}
\vspace{0.25cm}
$\bullet$
 \underline{neutralino\ ($\chi_{i}^{0}$) exchange:}
\begin{eqnarray}
 \widetilde{w}_{ZZ}^{(\chi^{0})} & = &
  \frac{1}{2\,\mZ^{4}}\sum_{i,j=1}^{4}
  \bigg[ \mchii\,\mchij I^{ZZ}_{ij}
      + \mchi \mchii J^{ZZ}_{ij} + K^{ZZ}_{ij} \bigg],
\end{eqnarray}
where
\begin{eqnarray}
I^{ZZ}_{ij} & = &
 \bigg\{(C_{-i}^{Z*}C_{-j}^{Z}+D_{-i}^{Z*}D_{-j}^{Z}) \nonumber \\
 & & \hspace{0.1in} \times \Big[ (s-4\,\mZ^2)\, \T_2
       - G_{ZZ}^{T(1)} \,\T_1
       + G_{ZZ}^{T(2)}\,\T_0
       + (s-4\,\mZ^2)\, \Y_2
       - G_{ZZ}^{Y(1)} \,\Y_0 \,\Big] \nonumber \\
 & & + \ (C_{-i}^{Z*}C_{-j}^{Z}-D_{-i}^{Z*}D_{-j}^{Z})
      \Big[- 18\,m_Z^4 \mchi^2\,\T_0 \nonumber \\
 & & \hspace{0.2in}- \mchi^2 (-s^{2}+4\,\mZ^2\,s+6\,m_Z^4) \,\Y_0 \,\Big]
      \bigg\} (s,\mchi^{2},\mZ^{2},\mZ^{2},\mchii^{2},\mchij^{2}),
\\
J^{ZZ}_{ij} & = & C_{-i}^{Z*}C_{+j}^{Z}
 \Big[ 12\,\mZ^{2}\,\T_2
       + 12\,\mZ^2\,(\mZ^2-2\,\mchi^2)\,\T_1 \nonumber \\
 & & \hspace{0.2in} + 12\,\mZ^2\,(\mchi^4+\mchi^2\,m_Z^2-2\,m_Z^4)\T_0
        - \ 4\,(s-m_Z^2) \Y_2 \nonumber \\
 & & \hspace{0.2in} + 2\,m_Z^2\,(2\,s-3\,m_Z^2)\,\Y_1
                  +  G_{ZZ}^{Y(2)}\,\Y_0\Big]
     (s,\mchi^{2},\mZ^{2},\mZ^{2},\mchii^{2},\mchij^{2}),
 \nonumber \\
 & & \\
K^{ZZ}_{ij} & = & 
      C_{+i}^{Z*}C_{+j}^{Z} \Big[-\T_4-(s-2\,m_Z^2-4\,\mchi^2)\,\T_3
       -G_{ZZ}^{T(3)}\,\T_2
       - G_{ZZ}^{T(4)}\,\T_1 \nonumber \\
 & & \hspace{0.8in} -(m_Z^2-\mchi^2)^2(2\,m_Z^2+\mchi^2)^2 \,\T_0
       +\Y_4 \nonumber \\
 & & \hspace{0.8in} + G_{ZZ}^{Y(3)} \,\Y_2
       - G_{ZZ}^{Y(4)}\,\Y_0 \,\Big]
       (s,\mchi^{2},\mZ^{2},\mZ^{2},\mchii^{2},\mchij^{2}),
\end{eqnarray}
and
\begin{eqnarray}
G_{ZZ}^{T(1)} \! &=& \!
2[s (\mchi^2+2\,\mZ^2) - 4\,\mZ^2 (\mchi^2+\mZ^2)], \nonumber \\
G_{ZZ}^{T(2)} \! &=& \! s (\mchi^2+2\,m_Z^2)^2
+\mZ^2 (-4\,\mchi^4-10\,\mchi^2\,\mZ^2-4\,m_Z^4), \nonumber \\
G_{ZZ}^{T(3)} \! &=& \!
 -2\,s\,(\mchi^{2}+2\,\mZ^{2})+6\,\mchi^{4}
+6\,\mchi^{2}\,\mZ^{2}+5\,\mZ^4, \nonumber \\
G_{ZZ}^{T(4)} \! &=& \!
s\,(\mchi^{2}+2\,\mZ^{2})^{2}-8\,\mZ^{6}+18\,\mZ^{4}\,\mchi^{2}
-6\,\mchi^{4}\,\mZ^{2}-4\,\mchi^{6}, \nonumber \\
G_{ZZ}^{Y(1)} \! &=& \! -s\,\mchi^{2}\,(\mchi^{2}+2\,\mZ^{2})
+6\,\mchi^{2}\,\mZ^{4}+4\,\mZ^{2}\,(\mchi^{2}-m_Z^{2})^{2}, \nonumber \\
G_{ZZ}^{Y(2)} \! &=& \! 2\,[2\,\mZ^{2}\,s^{2}+s\,(2\,\mchi^{2}\,\mZ^{2}
-2\,\mchi^{4}-11\,\mZ^{4}) \nonumber \\
 & &\hspace{0.2in} +2\,\mZ^{2}\,(\mchi^{4}
+4\,\mchi^{2}\,\mZ^{2}+\mZ^{4})], \nonumber \\
G_{ZZ}^{Y(3)} \! &=& \! s(\mchi^{2}-4\,\mZ^{2})+2\,\mchi^{4}
-4\,\mchi^{2}\mZ^{2}+\mZ^{4}, \nonumber \\
G_{ZZ}^{Y(4)} \! &=& \! (\mZ^{2}-\mchi^{2})
[s\,(8\,\mZ^{4}+\mchi^{4}-\mchi^{2}\,\mZ^{2})
+\mchi^{2}(-10\,\mZ^{4}+\mchi^4-3\,\mchi^{2}\,\mZ^{2})], \nonumber
\end{eqnarray}
and
\begin{eqnarray}
\label{ZZ-coupl1:eq}
   C_{\pm i}^{Z} & \equiv &
    |C_{V}^{\chi_{i}^{0}\chi Z}|^{2} \pm
    |C_{A}^{\chi_{i}^{0}\chi Z}|^{2}, \\
\label{ZZ-coupl2:eq}
   D_{\pm i}^{Z} & \equiv &
    C_{V}^{\chi_{i}^{0}\chi Z}
    \Big(C_{A}^{\chi_{i}^{0}\chi Z}\Big)^{*}\pm
    \:C_{A}^{\chi_{i}^{0}\chi Z} \Big(C_{V}^{\chi_{i}^{0}\chi Z}\Big)^{*};
\end{eqnarray}
\vspace{0.25cm}
$\bullet$
 \underline{Higgs $(h,H)$--neutralino\ ($\chi_{i}^{0}$) interference term:}
\begin{eqnarray}
  \widetilde{w}_{ZZ}^{(h,H-\chi^{0})} & = &
   \sum_{i=1}^{4} Re \left[\left(\sum_{r=h,H} \frac{C^{ZZr}\:
    C_{S}^{\chi\chi r}}
    { s-m_{r}^{2}+i\, \Gamma_{r}\, m_{r}} \right )^{*}\right] \,
    \frac{1}{2\,\mZ^{4}} \nonumber \\
    & & \hspace{0.2in} \times \Big[C_{+i}^{Z} H^{(+)ZZ}_{i}
    + C_{-i}^{Z} H^{(-)ZZ}_{i}\Big],
\end{eqnarray}
where
\begin{eqnarray}
H^{(+)ZZ}_{i} & = &
  \mchi\,\Big\{s^{2}-2\,s\,(\mchi^{2}+\mZ^{2}-\mchii^{2})
  -4\,\mZ^{2}\,(\mchi^{2}-2\,\mZ^{2}-\mchii^{2}) \nonumber \\
  & &  \hspace{0.2in}+2\,\Big[s^{2}\,\mchii^{2}
  +s\,\{(\mchi^{2}-m_{\chi_{i}^{0}}^{2})^{2}
   -\mZ^{2}\,(3\,\mchi^{2}-2\,\mZ^{2}+\mchii^{2})\} \nonumber \\
  & &  \hspace{0.5in}+2\,\mZ^{2}\,\{(\mchi^{2}-\mchii^{2})^{2}
  +\mZ^{2}\,(\mchi^{2}-2\,\mZ^{2}+\mchii^{2})\}\Big] \nonumber \\
  & & \hspace{1.3in}\times
       \F(s,\mchi^{2},\mZ^{2},\mZ^{2},m_{\chi_{i}^{0}}^{2}) \Big\}, \\
H^{(-)ZZ}_{i} & = &
   -\mchii \Big\{s^{2}-2\,\mZ^{2}\,s
   +\Big[s^{2}\,(\mchi^{2}+m_{\chi_{i}^{0}}^{2})
   -2\,\mZ^{2}\,(2\,\mZ^{2}+3\,\mchi^{2}+\mchii^{2})\,s \nonumber \\
   & & \hspace{1.7in} +24\,\mZ^{4}\,\mchi^{2}\Big]
   \F(s,\mchi^{2},\mZ^{2},\mZ^{2},m_{\chi_{i}^{0}}^{2})\Big\}.
\end{eqnarray}
Note that $D_{+ i}^{Z}$ $=$ 0, because $C_{V}^{\chi_{i}^{0}\chi_{j}^{0} Z}$ is pure
imaginary and $C_{A}^{\chi_{i}^{0}\chi_{j}^{0} Z}$ is real.
%

\vspace{0.3cm}
\begin{center}
\fbox{\boldmath ${10.\:\:\chi\chi\rightarrow \bar{f}f}$}
\end{center}
This process involves the $s$-channel Higgs boson ($h$, $H$ and $A$)
and $Z$ boson exchange and the $t$- and $u$-channel sfermion
($\widetilde{f}_{a}$) exchange
\begin{eqnarray}
 \widetilde{w}_{\bar{f}f} &=&
 \widetilde{w}_{\bar{f}f}^{(h,H)}
+\widetilde{w}_{\bar{f}f}^{(A)} +\widetilde{w}_{\bar{f}f}^{(Z)}
+\widetilde{w}_{\bar{f}f}^{(\widetilde{f})}
+\widetilde{w}_{\bar{f}f}^{(h,H-\widetilde{f})} 
+\widetilde{w}_{\bar{f}f}^{(A-Z)}
+\widetilde{w}_{\bar{f}f}^{(A-\widetilde{f})}
+\widetilde{w}_{\bar{f}f}^{(Z-\widetilde{f})}: \nn \\
\end{eqnarray}
\vspace{0.25cm}
%
$\bullet$
 \underline{CP--even Higgs--boson $(h,H)$ exchange:}
\begin{eqnarray}
  \widetilde{w}_{\bar{f}f}^{(h,H)}& =&
   \left |\sum_{r=h,H} \frac{C_{S}^{ffr}\:
    C_{S}^{\chi\chi r}}{ s-m_{r}^{2}+i\, \Gamma_{r}\, m_{r}}
    \right |^{2} \,(s-4\,\mchi^{2})\,(s-4\,m_{f}^{2});
\end{eqnarray}
\vspace{0.25cm}
$\bullet$
 \underline{CP--odd Higgs--boson ($A$) exchange:}
\begin{eqnarray}
  \widetilde{w}_{\bar{f}f}^{(A)}&=&
   \left | \frac{C_{P}^{ffA}\: C_{P}^{\chi\chi A}}
    {s-\mA^{2}+i\, \Gamma_{A}\, \mA} \right |^{2} \,s^{2}\, ;
\end{eqnarray}
\vspace{0.25cm}
$\bullet$
 \underline{$Z$--boson exchange:}
\begin{eqnarray}
   \widetilde{w}_{\bar{f}f}^{(Z)}& =&
     \frac{4}{3} \left | \frac{C_{A}^{\chi\chi Z}}
     { s-\mZ^{2}+i\, \Gamma_{Z}\, \mZ} \right |^{2} \nonumber \\
  & & \times \bigg\{12\,|C_{A}^{ffZ}|^{2}
    \frac{\mchi^{2}\,m_{f}^{2}}{\mZ^{4}}(s-\mZ^{2})^{2} \nonumber \\
  & & \hspace{0.2in}+\Big[|C_{V}^{ffZ}|^{2}\,(s+2\,m_{f}^{2})
    +|C_{A}^{ffZ}|^{2}\,(s-4\,m_{f}^{2})\Big](s-4\,m_{\chi}^{2})\bigg\};
\end{eqnarray}
\vspace{0.25cm}
$\bullet$
 \underline{sfermion\ ($\widetilde{f}_{a}$) exchange:}
\begin{eqnarray}
   \widetilde{w}_{\bar{f}f}^{(\widetilde{f})}& =&
     \frac{1}{4}\sum_{a,b} \bigg[ (C_{+}^{a}C_{+}^{b}
       +D_{+}^{a}D_{+}^{b})
          \Big[\T_{2}-2\,(\mchi^{2}+m_{f}^{2})\,\T_{1} \nonumber \\
  & & \hspace{2.0in} +\Big\{(\mchi^{2}+m_{f}^{2})^{2}
      +4\,\mchi^{2}\,m_{f}^{2}\Big\}\T_{0}\Big] \nonumber \\
  & & + (C_{-}^{a}C_{-}^{b}-D_{-}^{a}D_{-}^{b})
          \Big[\T_{2}-2\,(\mchi^{2}+m_{f}^{2})\,\T_{1}
                +(\mchi^{2}-m_{f}^{2})^{2}\,\T_{0}\Big] \nonumber \\
  & & -(C_{+}^{a}C_{+}^{b}-D_{-}^{a}D_{-}^{b})\,
         4\,\mchi^{2}\,m_{f}^{2}\,\Y_{0} \nonumber \\
  & & + (C_{+}^{a}C_{+}^{b}+C_{-}^{a}C_{-}^{b})\,
      s\,\mchi^{2}\,\Y_{0}+(C_{+}^{a}C_{+}^{b}
       -C_{-}^{a}C_{-}^{b})\,s\,m_{f}^{2}\,\Y_{0} \nonumber \\
  & & + (D_{+}^{a}D_{+}^{b}-D_{-}^{a}D_{-}^{b})
      \Big[\Y_{2}+(\mchi^{2}+m_{f}^{2})^{2}\Y_{0}\Big] \nonumber \\
  & & +(C_{+}^{a}D_{+}^{b}+D_{+}^{a}C_{+}^{b}) \mchi m_f
      \Big[4\,(\mchi^{2}+m_{f}^{2})\T_{0}
            -4\,\T_{1}+s\,\Y_{0}\Big] \nonumber \\
  & & -(C_{+}^{a}D_{+}^{b}-D_{+}^{a}C_{+}^{b})
         \mchi\,m_{f}\,\Y_{1} \bigg]
        (s,\mchi^{2},m_{f}^{2},m_{f}^{2},\msfa^{2},\msfb^{2}),
\label{generation:eq}
\end{eqnarray}
where
\begin{eqnarray}
\label{ff-coupl1:eq} 
C_{\pm}^{a} &=& |\Lambda_{fL}^{a}|^{2}\pm|\Lambda_{fR}^{a}|^{2}, \\
\label{ff-coupl2:eq}
D_{\pm}^{a} &=& \Lambda_{fL}^{a}(\Lambda_{fR}^{a})^{*}
                \pm(\Lambda_{fL}^{a})^{*}\Lambda_{fR}^{a}.
\end{eqnarray}
In the above, $a$ is the index for sfermion mass eigenstates so that
$a$ $=$ 1, $\cdots$, 6 for squarks and charged sleptons, and $a$ $=$
1, 2, 3 for sneutrinos.

The symbol $f$ represents each fermion: $f$ $=$ $u$, $c$, $t$,
$\cdots$. The symbol $\widetilde{f}_{a}$ should be understood as
follows. For up--type quarks, down--type quarks and charged leptons,
the corresponding symbol $\widetilde{f}_{a}$ represents
$\widetilde{u}_a$, $\widetilde{d}_a$ and $\widetilde{e}_a$ ($a$ $=$ 1,
$\cdots$, 6), respectively.  For neutrinos, $\widetilde{f}_{a}$
represents $\widetilde{\nu}_a$ ($a$ $=$ 1, 2, 3). For eg. for
$\widetilde{w}_{\bar{c}c}^{(\widetilde u)}$, the last argument at the
end of eq.~(\ref{generation:eq}) should read
$(s,\mchi^{2},m_{c}^{2},m_{c}^{2},m_{\widetilde u_a}^{2},
m_{\widetilde u_b}^{2})$ when $a$ $=$ 1, $\cdots$, 6.

The coupling $\Lambda_{fL}^{a}$ for each fermion is defined by
$\Lambda_{nai}^{(f)L}$ in Appendix~A as $\Lambda_{uL}^{a}$ $=$
$\Lambda_{1a1}^{(u)L}$, $\Lambda_{cL}^{a}$ $=$ $\Lambda_{2a1}^{(u)L}$,
$\Lambda_{tL}^{a}$ $=$ $\Lambda_{3a1}^{(u)L}$, $\Lambda_{dL}^{a}$ $=$
$\Lambda_{1a1}^{(d)L}$, $\Lambda_{sL}^{a}$ $=$ $\Lambda_{2a1}^{(d)L}$,
$\Lambda_{bL}^{a}$ $=$ $\Lambda_{3a1}^{(d)L}$, etc. Similarly, the coupling
$\Lambda_{fR}^{a}$ is defined by
$\Lambda_{nai}^{(f)R}$. \\
\vspace{0.35cm}
%
$\bullet$
 \underline{Higgs $(h,H)$--sfermion\ ($\widetilde{f}_{a}$) interference term:}
 \begin{eqnarray}
    \widetilde{w}_{\bar{f}f}^{(h,H-\widetilde{f})} & =&
     \sum_{a} Re \left[\sum_{r=h,H} \frac{C_{S}^{ffr}\:
      C_{S}^{\chi\chi r}}{ s-m_{r}^{2}+i\, \Gamma_{r}\, m_{r}} \right ]
        \nonumber \\
    & & \times \bigg[C_{+}^{a} 2\,m_{\chi}\,m_{f}
    \Big\{2+[s-2\,(m_{\chi}^{2}
      +m_{f}^{2}-m_{\widetilde{f}_{a}}^{2})] \nonumber \\
    & & \hspace{1.3in}\times
       \F(s,m_{\chi}^{2},m_{f}^{2},m_{f}^{2},m_{\widetilde{f}_{a}}^{2})
             \Big\} \nonumber \\
    & & \hspace{0.2in} +D_{+}^{a}
    \Big\{s+[s\,(m_{\chi}^{2}+m_{f}^{2}+m_{\widetilde{f}_{a}}^{2})
          -8\,m_{\chi}^{2}\,m_{f}^{2}]  \nonumber \\
    & & \hspace{1.3in} \times
       \F(s,m_{\chi}^{2},m_{f}^{2},m_{f}^{2},m_{\widetilde{f}_{a}}^{2})
             \Big\}\bigg];
\end{eqnarray}
\vspace{0.25cm}
$\bullet$
 \underline{Higgs ($A$)--$Z$ interference term:}
\begin{eqnarray}
  \widetilde{w}_{\bar{f}f}^{(A-Z)}& =&
   Re \left [\left(\frac{C_{P}^{ffA}\:C_{P}^{\chi\chi A}}
    {s-\mA^{2}+i\, \Gamma_{A}\, \mA} \right)^{*}
    \left(\frac{C_{A}^{ffZ}\:C_{A}^{\chi\chi Z}}
    { s-\mZ^{2}+i\, \Gamma_{Z}\, \mZ} \right)\right] \nonumber \\
 & & \hspace{0.2in} \times \frac{8\,\mchi\,m_{f}}{\mZ^{2}}\,s\,(\mZ^{2}-s);
\end{eqnarray}
\vspace{0.25cm}
$\bullet$
 \underline{Higgs ($A$)--sfermion\ ($\widetilde{f}_{a}$) interference term:}
\begin{eqnarray}
  \widetilde{w}_{\bar{f}f}^{(A-\widetilde{f})}& =&
   - \sum_{a} Re \left[ \frac{C_{P}^{ffA}\:
     C_{P}^{\chi\chi A}}{ s-\mA^{2}+i\, \Gamma_{A}\, \mA} \right ]
       \nonumber \\
    & & \times \bigg[C_{+}^{a} 2\,m_{\chi}\,m_{f}
  s \F (s,m_{\chi}^{2},m_{f}^{2},m_{f}^{2},m_{\widetilde{f}_{a}}^{2}) 
         \nonumber \\
    & & \hspace{0.2in} +D_{+}^{a}
    \Big\{-s+s\,(m_{\chi}^{2}+m_{f}^{2}-m_{\widetilde{f}_{a}}^{2})
         \nonumber \\
    & & \hspace{1.3in} \times
    \F (s,m_{\chi}^{2},m_{f}^{2},m_{f}^{2},m_{\widetilde{f}_{a}}^{2})
       \Big\}\bigg];
\end{eqnarray}
\vspace{0.25cm}
$\bullet$
 \underline{$Z$--sfermion\ ($\widetilde{f}_{a}$) interference term:}
\begin{eqnarray}
  \widetilde{w}_{\bar{f}f}^{(Z-\widetilde{f})} & = &
    \sum_{a} \Bigg\{ Re \left[ \frac{C_{A}^{ffZ}\:
         C_{A}^{\chi\chi Z}}{ s-\mZ^{2}+i\, \Gamma_{Z}\, \mZ} \right ]
     Z^{(+)\bar{f}f}_{a}+Re \left[ \frac{C_{V}^{ffZ}\:
     C_{A}^{\chi\chi Z}}{ s-\mZ^{2}+i\, \Gamma_{Z}\, \mZ} \right ]
     Z^{(-)\bar{f}f}_{a} \Bigg\}, \nonumber \\
   & &
\end{eqnarray}
where
\begin{eqnarray}
Z^{(+)\bar{f}f}_{a} &=&
  C_{+}^{a}\Big\{s+2\,(\mchi^{2}+m_{f}^{2}-\msfa^{2})
    - \Big[-2\,s\,(\mchi^{2}+m_{f}^{2})
    + 2\,(\mchi^{2}+m_{f}^{2}-\msfa^{2})^{2} \nonumber \\
 & & \hspace{0.4in} +8\,\mchi^{2}\,m_{f}^{2}\,
    \Big(2-\frac{s}{\mZ^{2}}\Big)\Big]\,
     \F(s,\mchi^{2},m_{f}^{2},m_{f}^{2},\msfa^{2})\Big\} \nonumber \\
 & & - D_{+}^{a} 4\,\mchi\,m_{f} \Big\{\frac{s}{\mZ^{2}}-3 \nonumber \\
 & & \hspace{0.5in} -\Big[s+\Big(\frac{s}{\mZ^{2}}-3\Big)\,
     (\mchi^{2}+m_{f}^{2}-\msfa^{2})\Big]
   \F (s,\mchi^{2},m_{f}^{2},m_{f}^{2},\msfa^{2}) \Big\}, \nonumber \\
 & & \\
Z^{(-)\bar{f}f}_{a} &=&
 C_{-}^{a}\Big\{s+2\,(\mchi^{2}+m_{f}^{2}-\msfa^{2}) \nonumber \\
 & & \hspace{0.2in} + [2\,s\,(\mchi^{2}-m_{f}^{2})
  -2\,(\mchi^{2}+m_{f}^{2}-\msfa^{2})^2+8\,\mchi^{2}\,m_{f}^{2}] \nonumber \\
 & & \hspace{1in} \times \F (s,\mchi^{2},m_{f}^{2},m_{f}^{2},\msfa^{2})\Big\}.
\end{eqnarray}

This completes the list of all the tree-level two-body neutralino
pair-annihilation channels in the MSSM.

%
\section{Partial Wave Expansion}\label{expansion:sec}
In the literature one still often uses the usual approximation
in terms of the expansion in powers of $x$ (or, equivalently, WIMP
velocity-square), $\langle\sigma v_{\rm M\o l}\rangle \simeq a +
bx$. This is because in the early days it was often much easier to
compute the coefficients $a$ and $b$, rather than the cross section
itself~\cite{simple:ref}. 
Furthermore, the partial wave expansion gives a rather good
approximation to the exact result but only far enough from $s$--channel
resonances and thresholds for new final states. (For a recent detailed
study, see Ref.~\cite{nrr1}.)

The expansion of $\langle\sigma v_{\rm M\o l}\rangle$, as defined
in eq.~(\ref{sigmavdef:eq}), is given by~\cite{swo88,gg91}
\begin{equation}
\langle\sigma v_{\rm M\o l}\rangle=\frac{1}{m_\chi^2}
\left[w-\frac{3}{2}\left(2w-w'\right) x +
{\mathcal O} (x^2)\right]_{s=4m_\chi^2} \equiv a + bx + {\mathcal
O}(x^2)
\label{expansion:eq}
\end{equation}
where $w^{\prime}(s)$ denotes $d\,w(s)/d\,(s/4\mchi^2)$.

Analogously to the function $w(s)$~(eq.~(\ref{wtowtilde:eq})), the
coefficients $a$ and $b$ need to be summed over all possible final
states $f_1 f_2$. One can write them as~\cite{swo88} 
\begin{eqnarray}
a  &=& 
\sum_{f_1 f_2}c \, \theta\left(4\mchi^2-(m_{f_1}+m_{f_2})^2 \right)\, 
v_{f_1 f_2}\,\widetilde{a}_{f_1 f_2},  \\
b  &=& 
\sum_{f_1 f_2}c \, \theta\left(4\mchi^2-(m_{f_1}+m_{f_2})^2 \right)\, 
v_{f_1 f_2} \Bigg\{ \widetilde{b}_{f_1 f_2}+ \\
  && \hspace{4cm}    \widetilde{a}_{f_1 f_2}
\left[ -3+\frac{3}{4}v_{f_1 f_2}^{-2} \left(
      \frac{m_{f_1}^{2}+m_{f_2}^{2}}{2 \mchi^{2}}
   +\frac{(m_{f_1}^{2}-m_{f_2}^{2})^{2}}{8 \mchi^{4}}\right)\right]
                     \Bigg\}, \nonumber 
\end{eqnarray}
where the summation extends over all possible two-body final states
$f_1f_2$, the coefficient $c$ is defined in eq.~(\ref{color:eq}), and
\begin{eqnarray} 
v_{f_1 f_2} &=& \beta_f(4\mchi^2,m_{f_1},m_{f_2}), 
\end{eqnarray} 
and the velocity
 $\beta_f(s,m_{f_1},m_{f_2})$ was defined in eq.~(\ref{kdef:eq}).  
The ``reduced'' coefficients $\widetilde{a}_{f_1 f_2}$ and
$\widetilde{b}_{f_1 f_2}$ are given by
\begin{eqnarray}
\widetilde{a}_{f_1 f_2} & = & \frac{1}{32 \pi \mchi^2}
    \widetilde{w}_{f_1 f_2} (4 \mchi^2), \\
\widetilde{b}_{f_1 f_2} & = &  \frac{3}{64 \pi \mchi^2}
    \widetilde{w}^{\prime}_{f_1 f_2}(4 \mchi^2), 
\end{eqnarray}
where $\widetilde{w}(s)$ was defined in eq.~(\ref{wtildedef:eq}) and
$\widetilde{w}^{\prime}(s)\equiv d\,\widetilde{w}(s)/d\,(s/4\mchi^2)$.

In this Section, we provide a set of expressions for the coefficients
$a$ and $b$ in the case of equal-mass final states. Using
eq.~(\ref{expansion:eq}), we have derived the coefficients for all the
final states by using the analytic expressions for $w(s)$ presented in
the previous Section. In the case of unequal masses of the final
state particles the resulting formulae are exceedingly lengthly and we
will not include them here.

In the literature one can find several analytic formulae for the
expansion coefficients, including~\cite{erl90,os91,dn93,jkg96}, but,
due to different conventions and complexity of expressions, comparison
is not always doable. We have checked our results for the
$a$--coefficients in appropriate limits against published results and
agreed in all cases.

All the couplings and auxiliary functions appearing below are listed
in the Appendices.
\vspace{2.0cm} \\
\begin{center}
\fbox{\boldmath ${1.\:\chi\chi\rightarrow hh}$}
\end{center}
This process involves the $s$-channel CP--even Higgs boson ($h$
and $H$) exchange and the $t$- and $u$-channel neutralino
($\chi_{i}^{0}$,$i=1,\ldots,4$) exchange
\begin{eqnarray}
 \widetilde{a}_{hh} &=&
 \widetilde{a}_{hh}^{(h,H)}
+\widetilde{a}_{hh}^{(\chi^{0})}
+\widetilde{a}_{hh}^{(h,H-\chi^{0})}, \\
 \widetilde{b}_{hh} &=&
 \widetilde{b}_{hh}^{(h,H)}
+\widetilde{b}_{hh}^{(\chi^{0})}
+\widetilde{b}_{hh}^{(h,H-\chi^{0})}:
\end{eqnarray}
\vspace{0.25cm}
$\bullet$
\underline{CP-even Higgs-boson ($h,H$) exchange:}
   \begin{eqnarray}
   \widetilde{a}_{hh}^{(h,H)}& = & 0, 
\\
   \widetilde{b}_{hh}^{(h,H)}& = & \frac{3}{64\,\pi} \left |
     \sum_{r=h,H} \frac{C^{hhr}\: C_{S}^{\chi\chi r}}
       {4\, \mchi^{2}-m_{r}^{2}+i\, \Gamma_{r}\, m_{r}}
                 \right |^{2} ;
    \end{eqnarray}
\vspace{0.25cm}
$\bullet$
\underline{neutralino\ ($\chi_{i}^{0}$) exchange:}
    \begin{eqnarray}
    \widetilde{a}_{hh}^{(\chi^0)}& = & 0, 
\\
    \widetilde{b}_{hh}^{(\chi^0)}& = &  \frac{1}{16\,\pi}
      \sum_{i,j=1}^{4}  (C_{S}^{\chi_{i}^0 \chi\, h })^2
        (C_{S}^{\chi_{j}^0 \chi\, h * })^2
         \frac{1}{\Delta_{hi}^{2}\,\Delta_{hj}^{2} } \nonumber 
\\
       & & \times \Big[4 \,\mchi^{2}\,(\mchi^{2}-\mh^{2})^{2}
           +4\,\mchi\,(\mchi^{2}-\mh^{2})\,
           (\mchi+\mchii)\,\Delta_{hi} \nonumber \\
       & &\hspace{0.3in} +3\,(\mchi+\mchii)\,(\mchi+\mchij)\,
             \Delta_{hi}\,\Delta_{hj}\Big],
     \end{eqnarray}
\vspace{0.2cm}
where $\Delta_{hi}\equiv\,\mh^{2}-\mchi^{2}-m_{\chi^0_{i}}^{2}$. \\
$\bullet$
\underline{Higgs ($h,H$)--neutralino\ ($\chi_{i}^{0}$) interference term:}
   \begin{eqnarray}
    \widetilde{a}_{hh}^{(h,H-\chi^0)}& = & 0, 
\\
    \widetilde{b}_{hh}^{(h,H-\chi^0)}& = &  \frac{1}{16\,\pi}
      \sum_{i=1}^{4} Re \left[\sum_{r=h,H} \left(\frac{C^{hhr}\:
       C_{S}^{\chi\chi r}}{4\, \mchi^{2}-m_{r}^{2}
            +i\, \Gamma_{r}\, m_{r}}\right)^{*}\, C_{S}^{\chi^0_{i} \chi\, h}
       C_{S}^{\chi^0_{i} \chi\, h} \right] \nonumber \\
       & & \times \frac{[2\,\mchi\,(\mchi^{2}-\mh^{2})
         +3\,(\mchi+\mchii)\,\Delta_{hi}]}{\Delta_{hi}^{2}}.
     \end{eqnarray}

The expressions for $HH$ final state are obtained from the above by
replacing $m_h$, $C^{hhr}$, $C_{S}^{\chi_{i}^{0} \chi h }$,
$C_{S}^{\chi_{j}^{0} \chi h }$ with $m_H$, $C^{HHr}$,
$C_{S}^{\chi_{i}^{0} \chi H }$, $C_{S}^{\chi_{j}^{0} \chi H }$,
respectively. 

\vspace{0.3cm}
\begin{center}
\fbox{\boldmath ${2.\:\chi\chi\rightarrow AA}$}
\end{center}
Similarly to the final state $hh$,
this process proceeds via the $s$-channel CP--even Higgs boson ($h$
and $H$) exchange and the $t$- and $u$-channel neutralino
($\chi_{i}^{0}$,$i=1,\ldots,4$) exchange
\begin{eqnarray}
 \widetilde{a}_{AA} &=&
 \widetilde{a}_{AA}^{(h,H)}
+\widetilde{a}_{AA}^{(\chi^{0})}
+\widetilde{a}_{AA}^{(h,H-\chi^{0})}, \\
 \widetilde{b}_{AA} &=&
 \widetilde{b}_{AA}^{(h,H)}
+\widetilde{b}_{AA}^{(\chi^{0})}
+\widetilde{b}_{AA}^{(h,H-\chi^{0})}:
\end{eqnarray}
\vspace{0.25cm}
$\bullet$
\underline{CP--even Higgs--boson ($h,H$) exchange:}
   \begin{eqnarray}
   \widetilde{a}_{AA}^{(h,H)}& = & 0, 
\\
   \widetilde{b}_{AA}^{(h,H)}& = & \frac{3}{64\,\pi} \left |
      \sum_{r=h,H} \frac{C^{AAr}\: C_{S}^{\chi\chi r}}
     {4\, \mchi^{2}-m_{r}^{2}+i\, \Gamma_{r}\, m_{r}} \right |^{2} ;
    \end{eqnarray}
\vspace{0.25cm}
$\bullet$
\underline{neutralino\ ($\chi_{i}^{0}$) exchange:}
    \begin{eqnarray}
    \widetilde{a}_{AA}^{(\chi^{0})}& = & 0, 
\\
    \widetilde{b}_{AA}^{(\chi^{0})}& = &  \frac{1}{16\,\pi}
      \sum_{i,j=1}^{4}  (C_{P}^{\chi^0_{i} \chi\, A})^{2}
        (C_{P}^{\chi^0_{j} \chi\, A *})^{2}
         \frac{1}{\Delta_{Ai}^{2}\,\Delta_{Aj}^{2} } \nonumber \\
     & & \times \Big[4 \,\mchi^{2}\,(\mchi^{2}-\mA^{2})^{2}
            +4\,\mchi\,(\mchi^{2}-\mA^{2})\,
                    (\mchi-\mchii)\,\Delta_{Ai} \nonumber \\
     & & \hspace{0.3in} +3\,(\mchi-\mchii)\,(\mchi-\mchij)\,
             \Delta_{Ai}\,\Delta_{Aj}\Big],
     \end{eqnarray}
\vspace{0.2cm}
where $\Delta_{Ai}\equiv\,\mA^{2}-\mchi^{2}-m_{\chi^0_{i}}^{2}$. \\
$\bullet$
\underline{Higgs ($h,H$)--neutralino\ ($\chi_{i}^{0}$) interference term:}
   \begin{eqnarray}
    \widetilde{a}_{AA}^{(h,H-\chi^0)}& = & 0, 
\\
    \widetilde{b}_{AA}^{(h,H-\chi^0)}& = &  \frac{1}{16\,\pi}
      \sum_{i=1}^{4} Re \left[\sum_{r=h,H} \left(\frac{C^{AAr}\:
       C_{S}^{\chi\chi r}}{4\, \mchi^{2}-m_{r}^{2}
         +i\, \Gamma_{r}\, m_{r}}\right)^{*}\,
            (C_{P}^{\chi^0_{i} \chi\, A})^2 \right] \nonumber \\
      & & \times \frac{[-2\,\mchi\,(\mchi^{2}-\mA^{2})
          -3\,(\mchi-\mchii)\,\Delta_{Ai}]}{\Delta_{Ai}^{2}}.
     \end{eqnarray}

\vspace{0.3cm}
\begin{center}
\fbox{\boldmath ${3.\:\chi\chi\rightarrow H^{+}H^{-}}$}
\end{center}
This process proceeds via the $s$-channel $Z$ and CP--even
Higgs boson ($h,H$) exchange as well as the $t$- and $u$-channel
chargino ($\chi_{k}^{\pm}$,
$k=1,2$) exchange
\begin{eqnarray}
 \widetilde{a}_{H^+H^-} &=&
 \widetilde{a}_{H^+H^-}^{(h,H)}
+\widetilde{a}_{H^+H^-}^{(Z)}
+\widetilde{a}_{H^+H^-}^{(\chi^{\pm})}
+\widetilde{a}_{H^+H^-}^{(h,H-\chi^{\pm})}
+\widetilde{a}_{H^+H^-}^{(Z-\chi^{\pm})}, \\
 \widetilde{b}_{H^+H^-} &=&
 \widetilde{b}_{H^+H^-}^{(h,H)}
+\widetilde{b}_{H^+H^-}^{(Z)}
+\widetilde{b}_{H^+H^-}^{(\chi^{\pm})}
+\widetilde{b}_{H^+H^-}^{(h,H-\chi^{\pm})}
+\widetilde{b}_{H^+H^-}^{(Z-\chi^{\pm})}:
\end{eqnarray}
\vspace{0.25cm}
$\bullet$
\underline{CP--even Higgs--boson ($h,H$) exchange:}
   \begin{eqnarray}
   \widetilde{a}_{H^{+}H^{-}}^{(h,H)}& = & 0, 
\\
   \widetilde{b}_{H^{+}H^{-}}^{(h,H)}& = &
      \frac{3}{32\,\pi} \left | \sum_{r=h,H} \frac{C^{H^{+}H^{-} r}\:
        C_{S}^{\chi\chi r}}
   {4\, \mchi^{2}-m_{r}^{2}+i\, \Gamma_{r}\, m_{r}} \right |^{2} ;
    \end{eqnarray}
\vspace{0.25cm}
$\bullet$
\underline{$Z$--boson exchange:}
   \begin{eqnarray}
   \widetilde{a}_{H^{+}H^{-}}^{(Z)}& = & 0, 
\\
   \widetilde{b}_{H^{+}H^{-}}^{(Z)}& = &
       \frac{1}{4\,\pi} \left | \frac{C^{H^{+}H^{-} Z}\:
         C_{A}^{\chi\chi Z}}
         {4\, \mchi^{2}-\mZ^{2}+i\, \Gamma_{Z}\, \mZ} \right |^{2}
           (\mchi^{2}-\mcH^{2});
    \end{eqnarray}
\vspace{0.25cm}
$\bullet$
\underline{chargino ($\chi_{k}^{\pm})$ exchange:}
    \begin{eqnarray}
    \widetilde{a}_{H^{+}H^{-}}^{(\chi^{\pm})}& = & 0, 
\\
     \widetilde{b}_{H^{+}H^{-}}^{(\chi^{\pm})}& = &  \frac{1}{8\,\pi}
       \sum_{k,l=1}^{2} \frac{1}
         {\Delta_{H^{\pm}k}^{2}\,\Delta_{H^{\pm}l}^{2} }
       \bigg\{C_{+k}^{H}\,C_{+l}^{H}\mchi^{2}\Big[4\,(\mchi^{2}-\mcH^{2})^2
           \nonumber \\
       & & \hspace{0.4in} +4\,(\mchi^{2}-\mcH^{2})\,\Delta_{H^{\pm}k}
             +3\,\Delta_{H^{\pm}k}\,\Delta_{H^{\pm}l}\Big] \nonumber \\
       & & +D_{-k}^{H}\,D_{-l}^{H}
 \Big[4\,\mchi^{2}\,\mchargk\,\mchargl\,(\mchi^{2}-\mcH^{2})\Big]
           \nonumber \\
       & & +C_{+k}^{H}\,C_{-l}^{H}\,
          2\,\mchi\,\mchargk\,\Big[2\,(\mchi^{2}-\mcH^{2})\,
          \Delta_{H^{\pm}k}+3\,\Delta_{H^{\pm}k}^{2}\Big] \nonumber \\
       & & +C_{-k}^{H}\,C_{-l}^{H}
          \Big[\,3\,\mchargk\,\mchargl\,\Delta_{H^{\pm}k}\,
                \Delta_{H^{\pm}l}\Big] \nonumber \\
       & & +D_{+k}^{H}\,D_{+l}^{H}\,\Big[2\,(\mchi^{2}-\mcH^{2})\,
            \Delta_{H^{\pm}k}\,\Delta_{H^{\pm}l}\Big]\bigg\},
\end{eqnarray}
where $\Delta_{H^{\pm}k}\equiv\,\mcH^{2}-\mchi^{2}-\mchargk^{2}$.
The couplings $C_{\pm k}^{H}$ and $D_{\pm k}^{H}$ are given
in eqs.~(\ref{chch-coupl1:eq}) and~(\ref{chch-coupl2:eq}). \vspace{0.2cm} \\
$\bullet$
\underline{Higgs ($h,H$)--chargino\ ($\chi_{k}^{\pm}$) interference term:}
   \begin{eqnarray}
    \widetilde{a}_{H^{+}H^{-}}^{(h,H-\chi^{\pm})}& = & 0, 
\\
    \widetilde{b}_{H^{+}H^{-}}^{(h,H-\chi^{\pm})}& = &  \frac{1}{8\,\pi}
      \sum_{k=1}^{2} Re \left[\sum_{r=h,H}
        \left(\frac{C^{H^{+}H^{-}r}\: C_{S}^{\chi\chi r}}
           {4\, \mchi^{2}-m_{r}^{2}+i\, \Gamma_{r}\, m_{r}}\right)^{*}\right]
             \frac{1}{\Delta_{H^{\pm}}^{2}} \nonumber \\
      & & \times \bigg\{C^{H}_{+k} \mchi\,\Big[2\,(\mchi^{2}-\mcH^{2})
               +3\,\Delta_{H^{\pm}_k}\Big]+C^{H}_{-k}
                    3\,\mchargk \Delta_{H^{\pm}_k} \bigg\};
     \end{eqnarray}
\vspace{0.25cm}
$\bullet$
\underline{$Z$--chargino\ ($\chi_{k}^{\pm}$) interference term:}
   \begin{eqnarray}
    \widetilde{a}_{H^{+}H^{-}}^{(Z-\chi^{\pm})}& = & 0, 
\\
    \widetilde{b}_{H^{+}H^{-}}^{(Z-\chi^{\pm})}& = & - \frac{1}{2\,\pi}
      \sum_{k=1}^{2} Re \left[\left(\frac{C^{H^{+}H^{-} Z}\:
        C_{A}^{\chi\chi Z}}{4\, \mchi^{2}-\mZ^{2}
            +i\, \Gamma_{Z}\, \mZ}\right)^{*} D^{H}_{+k} \right]
                 \frac{(\mchi^{2}-\mcH^{2})}{\Delta_{H^{\pm}k}}.
     \end{eqnarray}
%
\vspace{2.0cm} \\
\begin{center}
\fbox{\boldmath ${4.\:\:\chi\chi\rightarrow W^{+}W^{-}}$}
\end{center}
This process involves
the $s$-channel CP--even Higgs boson ($h$ and $H$) and $Z$ exchange,
and the $t$- and $u$-channel chargino ($\chi_{k}^{\pm}$, $k=1,2$) exchange
\begin{eqnarray}
 \widetilde{a}_{WW} &=&
 \widetilde{a}_{WW}^{(h,H)}
+\widetilde{a}_{WW}^{(Z)}
+\widetilde{a}_{WW}^{(\chi^{\pm})}
+\widetilde{a}_{WW}^{(h,H-\chi^{\pm})}
+\widetilde{a}_{WW}^{(Z-\chi^{\pm})}, \\
 \widetilde{b}_{WW} &=&
 \widetilde{b}_{WW}^{(h,H)}
+\widetilde{b}_{WW}^{(Z)}
+\widetilde{b}_{WW}^{(\chi^{\pm})}
+\widetilde{b}_{WW}^{(h,H-\chi^{\pm})}
+\widetilde{b}_{WW}^{(Z-\chi^{\pm})};
\end{eqnarray}
\vspace{0.25cm}
$\bullet$
\underline{CP--even Higgs--boson ($h,H$) exchange:}
   \begin{eqnarray}
    \widetilde{a}_{WW}^{(h,H)} & = & 0, 
\\
    \widetilde{b}_{WW}^{(h,H)} & = &
      \frac{3}{32\,\pi} \left |\sum_{r=h,H} \frac{C^{WWr}\:
      C_{S}^{\chi\chi r}}{ s-m_{r}^{2}+i\, \Gamma_{r}\, m_{r}} \right |^{2}
              \frac{3\,\mW^4-4\,\mW^2\,\mchi^2+4\,\mchi^4}{\mW^4};
   \end{eqnarray}
\vspace{0.25cm}
$\bullet$
\underline{$Z$-boson exchange:}
   \begin{eqnarray}
   \hspace{-0.2in} \widetilde{a}_{WW}^{(Z)} & = & 0, 
\\
   \hspace{-0.2in} \widetilde{b}_{WW}^{(Z)}& = & \frac{1}{4\,\pi}
      \left |\frac{C^{WWZ}\: C_{A}^{\chi\chi Z}}
       { s-\mZ^{2}+i\, \Gamma_{Z}\, \mZ} \right |^{2}
             \frac{(3\,\mW^4+20\,\mW^2\,\mchi^2+4\,\mchi^4)
           (\mchi^2-\mW^2)}{\mW^4};
   \end{eqnarray}
\vspace{0.25cm}
$\bullet$
\underline{chargino\ ($\chi_{k}^{\pm}$) exchange:}
   \begin{eqnarray}
    \widetilde{a}_{WW}^{(\chi^{\pm})} & = & \frac{1}{4\,\pi}
     \sum_{k,l=1}^{2} \frac{1}{\mW^4 \Delta_{Wk}\Delta_{Wl}}
      \bigg\{C_{+k}^{W\!*}C_{+l}^{W}\Big[ 2\,\mW^4\,(\mchi^2-\mW^2)
       \Big]    
\\
     & & \hspace{0.2in} +D_{+k}^{W\!*}D_{+l}^{W}
     \Big[\mchargk\,\mchargl\,(4 \, \mchi^4
                 +3\,\mW^4-4\,\mW^2\,\mchi^2)\Big]\bigg\}, \nonumber \\
    \widetilde{b}_{WW}^{(\chi^{\pm})} & = & \frac{1}{8\,\pi}
    \sum_{k,l=1}^{2} \frac{1}{\mW^4\Delta_{Wk}^3 \Delta_{Wl}^3}
      \bigg\{C_{+k}^{W\!*} C_{+l}^{W}
     \Big[16\,\mchi^2\,\mW^4\,(\mchi^2-\mW^2)^2 \Delta_{Wk}^2 \nonumber \\
    & & \hspace{0.2in}+4\,\mchi^2\,\mW^2\,(3\,\mW^2+4\,\mchi^2)\,
      (\mchi^2-\mW^2)^2 \,\Delta_{Wk}\,\Delta_{Wl} \nonumber \\
    & & \hspace{0.2in} -4\,\mchi^2\,\mW^2\,(7\,\mW^4+\mchi^2\,\mW^2
      -8\,\mchi^4)\,\Delta_{Wk}^2\, \Delta_{Wl} \nonumber \\
    & & \hspace{0.2in}+(8\,\mW^6+3\,\mW^4\,\mchi^2+8\,\mW^2\,\mchi^4
      +8\,\mchi^6)\,\Delta_{Wk}^2\,\Delta_{Wl}^2 \Big] \nonumber \\
    & & +C_{+k}^{W\!*} C_{-l}^{W}\Big[48\,\mW^2\,\mchi^3\,\mchargk\,
      (\mchi^2-\mW^2)^2\,\Delta_{Wk}\,\Delta_{Wl} \nonumber \\
    & & \hspace{0.2in}+4\,\mW^2\,\mchi\,\mchargk\,(3\,\mW^4
       -5\,\mW^2\,\mchi^2+2\,\mchi^4)\, \Delta_{Wk}^2\,
          \Delta_{Wl} \nonumber \\
    & & \hspace{0.2in}+8\,\mW^2\,\mchi\,\mchargk\,(3\,\mW^4-8\,\mW^2\,\mchi^2
       +5\,\mchi^4)\, \Delta_{Wk}\, \Delta_{Wl}^2 \nonumber \\
    & & \hspace{0.2in} +2\,\mchi\,\mchargk\,(-5\,\mW^4+10\,\mW^2\,\mchi^2
       -8\,\mchi^4)\, \Delta_{Wk}^2\, \Delta_{Wl}^2 \Big] \nonumber \\
    & & +C_{-k}^{W\!*} C_{-l}^{W}\Big[16\,\mchi^2\,\mchargk\,\mchargl\,
       (\mchi^2-\mW^2)^2\, (\mW^2+\mchi^2)\, \Delta_{Wk}\,
          \Delta_{Wl} \nonumber \\
    & & \hspace{0.2in} +8\,\mchi^2\,\mchargk\,\mchargl\,
       (\mW^4-3\,\mW^2\,\mchi^2+2\,\mchi^4)\, \Delta_{Wk}^2\,
          \Delta_{Wl} \nonumber \\
    & & \hspace{0.2in} +3\,\mchargk\,\mchargl\,(3\,\mW^4-4\,\mW^2\,\mchi^2
       +4\,\mchi^4)\,\Delta_{Wk}^2\,\Delta_{Wl}^2 \Big] \nonumber \\
    & & +D_{+k}^{W\!*} D_{+l}^{W}\Big[16\,\mchi^2\,\mW^2\,(\mchi^2-\mW^2)^3\,
       \Delta_{Wk} \,\Delta_{Wl} \nonumber \\
    & & \hspace{0.2in} +32\,\,\mchi^2\,\mW^2\,(\mchi^2-\mW^2)^2\,
       \Delta_{Wk}^2\, \Delta_{Wl} \nonumber \\
    & &\hspace{0.2in} +2\,(-3\,\mW^6-17\,\mW^4\,\mchi^2+16\,\mW^2\,\mchi^4
       +4\,\mchi^6)\, \Delta_{Wk}^2\, \Delta_{Wl}^2 \Big] \nonumber \\
    & &  +D_{-k}^{W\!*} D_{-l}^{W}\Big[8\,\mchi^2\,\mchargk\,\mchargl
      (-3\,\mW^6+7\,\mW^4\,\mchi^2-8\,\mW^2\,\mchi^4+4\,\mchi^6)\,
         \Delta_{Wk}^2 \nonumber \\
    & & \hspace{0.2in} -8\,\mW^2\,\mchi^2\,\mchargk\,\mchargl\,
       (\mW^4+\mW^2\,\mchi^2-2\,\mchi^4)\,\Delta_{Wk}\,
         \Delta_{Wl} \nonumber \\
    & & \hspace{0.2in}  +4\,\mchi^2\,\mchargk\,\mchargl\,(11\,\mW^4
        -18\,\mW^2\,\mchi^2+16\,\mchi^4)\,\Delta_{Wk}^2\,
           \Delta_{Wl} \nonumber \\
    & & \hspace{0.2in}+3\,\mchargk\,\mchargl\,(3\,\mW^4-8\,\mW^2\,\mchi^2
        +12\,\mchi^4)\,\Delta_{Wk}^2\, \Delta_{Wl}^2 \Big] \bigg\},
\end{eqnarray}
where $\Delta_{Wk}\equiv\,\mW^{2}-\mchi^{2}-\mchargk^{2}$.
The couplings  $C_{\pm k}^{W}$ and  $D_{\pm k}^{W}$ are given
in eqs.~(\ref{WW-coupl1:eq}) and~(\ref{WW-coupl2:eq}). \vspace{0.2cm} \\
$\bullet$
\underline{Higgs ($h,H$)--chargino\ ($\chi_{k}^{\pm}$) interference term:}
   \begin{eqnarray}
    \widetilde{a}_{WW}^{(h,H-\chi^{\pm})}& = & 0, 
\\
    \widetilde{b}_{WW}^{(h,H-\chi^{\pm})}& = & \frac{1}{8\,\pi}
       \sum_{k=1}^{2} Re \left[\left(\sum_{r=h,H} \frac{C^{WWr}\:
       C_{S}^{\chi\chi r}}{ s-m_{r}^{2}
        +i\, \Gamma_{r}\, m_{r}} \right )^{*}\right]
           \frac{1}{\mW^4\,\Delta_{Wk}^2} \nonumber \\
       & & \times  \bigg\{C_{+k}^{W}\,
   \Big[ 2\,\mchi\,\mW^2\,(3\,\mW^4-5\,\mW^2\,\mchi^2+2\,\mchi^4) \nonumber \\
    & & \hspace{0.6in} +\mchi\,(-5\,\mW^4+10\,\mW^2\,\mchi^2
       -8\,\mchi^4)\,\Delta_{Wk}\Big] \nonumber \\
    & &\hspace{0.2in} +C_{-k}^{W}\Big[4\,\mchi^2 \,\mchargk \,
      (\mW^4-3\,\mW^2\,\mchi^2+2\,\mchi^4) \nonumber \\
    & & \hspace{0.6in} +3\,\mchargk\,(3\,\mW^4-4\,\mW^2\,\mchi^2
     +4\,\mchi^4)\,\Delta_{Wk}\Big]\bigg\};
   \end{eqnarray}
$\bullet$
\underline{$Z$--chargino\ ($\chi_{k}^{\pm}$) interference term:}
   \begin{eqnarray}
    \widetilde{a}_{WW}^{(Z-\chi^{\pm})} & = & 0, 
\\
    \widetilde{b}_{WW}^{(Z-\chi^{\pm})} & = & \frac{1}{2\,\pi}
     \sum_{k=1}^{2} Re \left[\left(\frac{C^{WWZ}\:
      C_{A}^{\chi\chi Z}}{ s-\mZ^{2}+i\, \Gamma_{Z}\, \mZ} \right )^{*}\right]
        \frac{1}{\mW^4\,\Delta_{Wk}^2} D_{+k}^{W} \nonumber \\
 & & \times \Big[ -8\,\mW^2\,\mchi^2\,(\mW^2-\mchi^2)^2 \nonumber \\
 & & \hspace{0.2in} +(\mW^2-\mchi^2)\,(3\,\mW^4+20\,\mW^2\,\mchi^2+4\,\mchi^4)
                                  \Delta_{Wk} \Big].
   \end{eqnarray}

\vspace{0.3cm}
\begin{center}
\fbox{\boldmath ${5.\:\chi\chi\rightarrow ZZ}$}
\end{center}
This process involves the $s$-channel CP--even Higgs boson ($h$
and $H$) exchange and the $t$- and $u$-channel neutralino
($\chi_{i}^{0}$,$i=1,\ldots,4$) exchange
\begin{eqnarray}
 \widetilde{a}_{ZZ} &=&
 \widetilde{a}_{ZZ}^{(h,H)}
+\widetilde{a}_{ZZ}^{(\chi^0)}
+\widetilde{a}_{ZZ}^{(h,H-\chi^0)}, \\
 \widetilde{b}_{ZZ} &=&
 \widetilde{b}_{ZZ}^{(h,H)}
+\widetilde{b}_{ZZ}^{(\chi^0)}
+\widetilde{b}_{ZZ}^{(h,H-\chi^0)};
\end{eqnarray}
\vspace{0.25cm}
$\bullet$
\underline{CP--even Higgs--boson $(h,H)$ exchange:}
   \begin{eqnarray}
    \widetilde{a}_{ZZ}^{(h,H)}& = & 0, 
\\
    \widetilde{b}_{ZZ}^{(h,H)}& = & \frac{3}{64\,\pi}
      \left |\sum_{r=h,H} \frac{C^{ZZr}\:
       C_{S}^{\chi\chi r}}{ s-m_{r}^{2}+i\, \Gamma_{r}\, m_{r}} \right |^{2}
               \frac{3\,\mZ^4-4\,\mZ^2\,\mchi^2+4\,\mchi^4}{\mZ^4};
   \end{eqnarray}
\vspace{0.25cm}
$\bullet$
\underline{neutralino\ ($\chi_{i}^{0}$) exchange:}
   \begin{eqnarray}
    \widetilde{a}_{ZZ}^{(\chi^0)} & = & \frac{1}{4\,\pi}\sum_{i,j=1}^{4}
    C_{+i}^{Z*}C_{+j}^{Z}
   \frac{(\mchi^2-\mZ^2)}{\Delta_{Zi}\Delta_{Zj}}, 
\\
    \widetilde{b}_{ZZ}^{(\chi^0)} & = & \frac{1}{16\,\pi}
       \sum_{i,j=1}^{4} \frac{1}{\mZ^4\Delta_{Zi}^3 \Delta_{Zj}^3} 
\\
      & & \times  C_{+i}^{Z*}C_{+j}^{Z}\Big[D_{ij}^{(1)}\,\Delta_{Zi}^2
       +D_{ij}^{(2)}\,\Delta_{Zi} \Delta_{Zj}
       +D_{ij}^{(3)}\,\Delta_{Zi}^2 \Delta_{Zj}
       +D_{ij}^{(4)}\,\Delta_{Zi}^2 \Delta_{Zj}^2\Big], \nonumber
   \end{eqnarray}
where
   \begin{eqnarray}
     D_{ij}^{(1)} \! &=& \!
16\,\mchi^2\,\mZ^4\,(\mZ^2-\mchi^2)^2,  \\
     D_{ij}^{(2)} \! &=& \!
4\,\mchi^2\,(\mZ^2-\mchi^2)^2 \{3\,\mZ^4
                              +4\,\mchi^2\,\mchii\,\mchij  \nn \\
& & \hspace{0.2in}
 +\mZ^2\,[4\,\mchi^2+4\,\mchii\,\mchij-6\,\mchi\,(\mchii+\mchij)]\}, \\
     D_{ij}^{(3)} \! &=& \!
   -4\,\mchi\,(\mZ^2-\mchi^2)
   \{4\,\mchi^3\,\mchii\,\mchij+\mZ^4\,(7\,\mchi+3\,\mchii
                                                 +6\,\mchij) \nn \\
& & \hspace{0.2in}
+2\,\mZ^2\,\mchi\,[4\,\mchi^2-\mchii\,\mchij-\mchi\,(\mchii+5\,\mchij)]\}, \\
     D_{ij}^{(4)} \! &=& \!
    8\,\mZ^6+2\,\mZ^2\,\mchi^2
    [4\,\mchi^2-6\,\mchii\,\mchij-5\,\mchi\,(\mchii+\mchij)] \nn \\
& & +4\, \mchi^4\,[2\,\mchi^2+3\,\mchii\,\mchij+
                                   2\,\mchi\,(\mchii+\mchij)] \nn \\
& & +\mZ^4\,[3\,\mchi^2+9\,\mchii\,\mchij
                                 +5\,\mchi\,(\mchii+\mchij)],
   \end{eqnarray}
where $\Delta_{Zi}\equiv\,\mZ^{2}-\mchi^{2}-\mchii^{2}$.
The couplings $C_{\pm i}^{Z}$ and $D_{\pm i}^{Z}$ are given in
eqs.~(\ref{ZZ-coupl1:eq}) and~(\ref{ZZ-coupl2:eq}).
\vspace{0.2cm} \\
$\bullet$
\underline{Higgs $(h,H)$--neutralino\ ($\chi_{i}^{0}$) interference term:}
   \begin{eqnarray}
    \widetilde{a}_{ZZ}^{(h,H-\chi^{0})}& = & 0, 
\\
   \widetilde{b}_{ZZ}^{(h,H-\chi^{0})}& = & \frac{1}{16\,\pi}
      \sum_{i=1}^{4} Re \left[\left(\sum_{r=h,H} \frac{C^{ZZr}\:
       C_{S}^{\chi\chi r}}
         { s-m_{r}^{2}+i\, \Gamma_{r}\, m_{r}} \right )^{*}\right]
       \frac{1}{\mZ^4\,\Delta_{Zi}^2}
     C_{S}^{\chi\chi r} \nonumber \\
 & & \times \bigg\{
      2\,\mchi\,(\mchi^2-\mZ^2)\,[-3\,\mZ^4
       -4\,\mchi^3\,\mchii+2\,\mZ^2\,\mchi\,(\mchi+\mchii)]
            . \nonumber \\
 & & \hspace{1cm} + \Delta_{Zi}
      \left[ -4\,\mchi^4\,(2\,\mchi+3\,\mchi)
         +2\,\mZ^2\,\mchi^2\,(5\,\mchi+6\,\mchii) \right. \nonumber \\
 & &  \left. \hspace{5cm}
         -\mZ^4\,(5\,\mchi+9\,\mchii) \right] \bigg\}.
    \end{eqnarray}

\vspace{0.3cm}
\begin{center}
\fbox{\boldmath ${6.\:\:\chi\chi\rightarrow \bar{f}f}$}
\end{center}
This process involves the $s$-channel Higgs boson ($h$, $H$ and $A$)
and $Z$ boson exchange and the $t$- and $u$-channel sfermion
($\widetilde{f}_{a}$) exchange
\begin{eqnarray}
 \widetilde{a}_{\bar{f}f} &=&
 \widetilde{a}_{\bar{f}f}^{(h,H)}
+\widetilde{a}_{\bar{f}f}^{(A)}
+\widetilde{a}_{\bar{f}f}^{(Z)}
+\widetilde{a}_{\bar{f}f}^{(\widetilde{f})}
+\widetilde{a}_{\bar{f}f}^{(h,H-\widetilde{f})}
+\widetilde{a}_{\bar{f}f}^{(A-Z)}
+\widetilde{a}_{\bar{f}f}^{(A-\widetilde{f})}
+\widetilde{a}_{\bar{f}f}^{(Z-\widetilde{f})}, \\
 \widetilde{b}_{\bar{f}f} &=&
 \widetilde{b}_{\bar{f}f}^{(h,H)}
+\widetilde{b}_{\bar{f}f}^{(A)}
+\widetilde{b}_{\bar{f}f}^{(Z)}
+\widetilde{b}_{\bar{f}f}^{(\widetilde{f})}
+\widetilde{b}_{\bar{f}f}^{(h,H-\widetilde{f})}
+\widetilde{b}_{\bar{f}f}^{(A-Z)}
+\widetilde{b}_{\bar{f}f}^{(A-\widetilde{f})}
+\widetilde{b}_{\bar{f}f}^{(Z-\widetilde{f})}:
\end{eqnarray}
\vspace{0.25cm}
$\bullet$
\underline{CP--even Higgs--boson $(h,H)$ exchange:}
   \begin{eqnarray}
   \widetilde{a}_{\bar{f}f}^{(h,H)}& = & 0, 
\\
   \widetilde{b}_{\bar{f}f}^{(h,H)}& = & \frac{3}{4\, \pi}
     \left | \sum_{r=h,H} \frac{C_{S}^{ffr}\:
      C_{S}^{\chi\chi r}}{4\, \mchi^{2}-m_{r}^{2}
        +i\, \Gamma_{r}\, m_{r}} \right |^{2} (\mchi^{2}-m_{f}^{2});
    \end{eqnarray}
\vspace{0.25cm}
$\bullet$
\underline{CP--odd Higgs--boson ($A$) exchange:}
  \begin{eqnarray}
   \widetilde{a}_{\bar{f}f}^{(A)} &=& \frac{1}{2\, \pi}
     \left |  \frac{C_{P}^{ffA}\:
      C_{P}^{\chi\chi A}}{4\, \mchi^{2}-\mA^{2}
       +i\, \Gamma_{A}\, \mA} \right |^{2} \mchi^{2}, 
\\
   \widetilde{b}_{\bar{f}f}^{(A)} &=& \frac{3}{2 \,\pi}
      \left |  \frac{C_{P}^{ffA}\:
       C_{P}^{\chi\chi A}}{4\, \mchi^{2}-\mA^{2}
        +i\, \Gamma_{A}\, \mA} \right |^{2}
         \frac{\mchi^{2}\, \mA^{2}\,(\mA^{2}-4\, \mchi^{2}
           +\Gamma_{A}^{2})}{(4\, \mchi^{2}-\mA^{2})^{2}
              +(\Gamma_{A}\, \mA)^{2}};
  \end{eqnarray}
$\bullet$
\underline{$Z$--boson exchange:}
  \begin{eqnarray}
      \widetilde{a}_{\bar{f}f}^{(Z)} &=& \frac{1}{2\, \pi}
       \left |  \frac{C_{A}^{ffZ}\: C_{A}^{\chi\chi Z}}
         {4 \mchi^{2}-\mZ^{2}+i\, \Gamma_{Z}\, \mZ} \right |^{2}
       \frac{m_{f}^{2}\,(\mZ^{2}-4\, \mchi^{2})^{2}}{\mZ^{4}}, 
\\
     \widetilde{b}_{\bar{f}f}^{(Z)} &=& \frac{1}{2\, \pi}
      \left |  \frac{C_{A}^{ffZ}\: C_{A}^{\chi\chi Z}}
       {4\, \mchi^{2}-\mZ^{2}+i\, \Gamma_{Z}\, \mZ} \right |^{2}
        \frac{1}{\mZ^{2}\,((4\, \mchi^{2}-\mZ^{2})^{2}
          +(\Gamma_{Z}\, \mZ)^{2})} \nonumber \\
      & &  \times \bigg[2 \,|C_{A}^{ffZ}|^{2}\, \Big\{\mZ^2\,
             (\mchi^{2}-m_{f}^{2})\,(\mZ^{2}-4 \mchi^{2})^{2}  \nonumber \\
      & & \hspace{1.1in} +\Gamma_{Z}^{2} [\mchi^{2}\, \mZ^{4} + m_{f}^{2}\,
           (24\, \mchi^{4}-6\, \mZ^{2}\, \mchi^{2}-\mZ^{4})] \Big\}
             \nonumber \\
         & & \hspace{0.2in} +\mZ^{2}\, |C_{V}^{ffZ}|^{2}\,
        \Big\{(2\, \mchi^{2}+m_{f}^{2})\,[(4\, \mchi^{2}-\mZ^{2})^{2}
             + \mZ^{2}\,\Gamma_{Z}^{2}]\Big\} \bigg];
  \end{eqnarray}
$\bullet$
\underline{sfermion\ ($\widetilde{f}_{a}$) exchange:}
   \begin{eqnarray}
    \widetilde{a}_{\bar{f}f}^{(\widetilde{f})} &=& \frac{1}{32\, \pi}
      \sum_{a,b}\frac{(m_{f}\,C_{+}^{a}
           +\mchi\,D_{+}^{a})\,(m_{f}\,C_{+}^{b}
           +\mchi\,D_{+}^{b})}
         {\Delta_{{\widetilde f}_{a}}\, \Delta_{{\widetilde f}_{b}}},
\\
     \widetilde{b}_{\bar{f}f}^{(\widetilde{f})} &=& \frac{1}{64\, \pi}
      \sum_{a,b} \frac{1}{\Delta_{{\widetilde f}_{a}}^{3}\,
         \Delta_{{\widetilde f}_{b}}^{3}}
     \bigg\{C_{+}^{a}\,C_{+}^{b}\Big[8\, m_{f}^{2}\, \mchi^{2}\,
       (\mchi^{2}-m_{f}^{2})\, \Delta_{{\widetilde f}_{a}}^{2} \nonumber \\
      & & \hspace{0.4in}- 4\, \mchi^{2}\, (m_{f}^{4}+\mchi^{2}\, m_{f}^{2}
       -2\, \mchi^{4})\, \Delta_{{\widetilde f}_{a}}\,
         \Delta_{{\widetilde f}_{b}} \nonumber \\
     & & \hspace{0.4in}+4\,  \mchi^{2}\, ( m_{f}^{2}+2\, \mchi^{2})
        \Delta_{{\widetilde f}_{a}}^{2}\, \Delta_{{\widetilde f}_{b}}
         +4\, ( \mchi^{2}-m_{f}^{2})\,\Delta_{{\widetilde f}_{a}}^{2}\,
          \Delta_{{\widetilde f}_{b}}^{2} \Big] \nonumber \\
     & & +D_{+}^{a}\, D_{+}^{b} \Big[ 8\, \mchi^{4}\,
          ( \mchi^{2}-m_{f}^{2})\,  \Delta_{{\widetilde f}_{a}}^{2}
        +4\, \mchi^{2}\, (\mchi^{4}+\mchi^{2}\, m_{f}^{2}
        -2\, m_{f}^{4})\Delta_{{\widetilde f}_{a}}\,
             \Delta_{{\widetilde f}_{b}} \nonumber \\
  & &  \hspace{0.4in} +4\, \mchi^{2} (5\, \mchi^{2}-2\, m_{f}^{2})\,
      \Delta_{{\widetilde f}_{a}}^{2}\, \Delta_{{\widetilde f}_{b}}
        -3\, (m_{f}^{2}-3\,  \mchi^{2})\, \Delta_{{\widetilde f}_{a}}^{2}\,
          \Delta_{{\widetilde f}_{b}}^{2} \Big] \nonumber \\
  & & +C_{-}^{a}\,C_{-}^{b} \Big[8\, \mchi^{2}\, ( \mchi^{2}
     -m_{f}^{2})^{2}\,\Delta_{{\widetilde f}_{a}}\,\Delta_{{\widetilde f}_{b}}
     +8\, \mchi^{2}\, (\mchi^2-m_{f}^{2})\Delta_{{\widetilde f}_{a}}^{2}\,
      \Delta_{{\widetilde f}_{b}} \nonumber \\
    & & \hspace{0.4in}+2\, (m_{f}^{2}+2\, \mchi^{2})\,
       \Delta_{{\widetilde f}_{a}}^{2}\,\Delta_{{\widetilde f}_{b}}^{2} \Big]
       \nonumber \\
     & & + C_{+}^{a}\,D_{+}^{b}\,2\,m_{f}\, \mchi\,
      \Big[ 8 \,  \mchi^{2} (\mchi^{2}-m_{f}^{2})\,
      \Delta_{{\widetilde f}_{a}}^{2} +12 \,\mchi^{2} (\mchi^{2}-m_{f}^{2})\,
       \Delta_{{\widetilde f}_{a}}\,\Delta_{{\widetilde f}_{b}}
         \nonumber \\
   & & \hspace{0.4in} +3 \,
        \Delta_{{\widetilde f}_{a}}^{2}\,\Delta_{{\widetilde f}_{b}}^{2}
           -2\,(m_{f}^{2}-4\, \mchi^{2})\,
        \Delta_{{\widetilde f}_{a}}\,\Delta_{{\widetilde f}_{b}}^{2}
          -6\,(m_{f}^{2}-2\, \mchi^{2})\,\Delta_{{\widetilde f}_{a}}^{2}\,
             \Delta_{{\widetilde f}_{b}} \Big] \bigg\}, \nonumber \\
& &
   \end{eqnarray}
where $\Delta_{{\widetilde f}_{a}}\equiv
m_{f}^{2}-\mchi^{2}-m_{\widetilde{f}_{a}}^{2}$. The index
$a$ counts sfermions so that $a$ $=$ 1, $\ldots$, 6 for squarks and
charged sleptons, and $a$ $=$ 1, 2, 3 for sneutrinos. 
The couplings $C_\pm^a$ and
$D_\pm^a$ are given in eqs.~(\ref{ff-coupl1:eq}) and~(\ref{ff-coupl2:eq}).
\vspace{0.2cm} \\
$\bullet$
\underline{Higgs $(h,H)$--sfermion\ ($\widetilde{f}_{a}$) interference term:}
   \begin{eqnarray}
     \widetilde{a}_{\bar{f}f}^{(h,H-\widetilde{f})}& = & 0, 
\\
     \widetilde{b}_{\bar{f}f}^{(h,H-\widetilde{f})}& = &-\frac{1}{8\,\pi}\,
       \sum_{a} Re\,\left[ \sum_{r=h,H} \frac{C_{S}^{ffr}\:
          C_{S}^{\chi\chi r}}{4\, \mchi^{2}-m_{r}^{2}
          +i\, \Gamma_{r}\, m_{r}} \right]
          \frac{(\mchi^2-m_{f}^{2})}{\Delta_{{\widetilde f}_{a}}^2}
            \nonumber \\
      & & \times \left[C_{+}^{a}\,2\,m_{f}\,\mchi
            + D_{+}^{a}\,(2\,\mchi^{2}+3\,\Delta_{{\widetilde f}_{a}})
                  \right];
     \end{eqnarray}
$\bullet$
\underline{Higgs ($A$)--$Z$ interference term:}
   \begin{eqnarray}
   \widetilde{a}_{\bar{f}f}^{(A-Z)} &=& \frac{1}{\pi}
     Re \left [ \left( \frac{C_{P}^{ffA}\:
      C_{P}^{\chi\chi A}}{4\, \mchi^{2}-\mA^{2}
          +i\, \Gamma_{A}\, \mA} \right)^{*}
      \left(\frac{C_{A}^{ffZ}\: C_{A}^{\chi\chi Z}}
     {4\, \mchi^{2}-\mZ^{2}+i\, \Gamma_{Z}\, \mZ} \right )\right] \nonumber \\
     & & \times\,\frac{\mchi\, m_{f}\,
         (\mZ^{2}-4\,\mchi^{2})}{\mZ^{2}}, 
\\
   \widetilde{b}_{\bar{f}f}^{(A-Z)} &=& \frac{3}{2\, \pi}
     Re \left [ \left( \frac{C_{P}^{ffA}\:
       C_{P}^{\chi\chi A}}{4\, \mchi^{2}-\mA^{2}
           +i\, \Gamma_{A}\, \mA} \right)^{*}
       \left(\frac{C_{A}^{ffZ}\:
         C_{A}^{\chi\chi Z}}{4\, \mchi^{2}-\mZ^{2}
           +i\, \Gamma_{Z}\, \mZ} \right )\right. \nonumber \\
      & & \times \,\frac{\mchi m_{f}} {\mZ^{2}} \bigg[ \mA\,
         (\mZ^{2}-4\, \mchi^{2})^{2} (\mA+ i\, \Gamma_{A}) \nonumber \\
      & & \hspace{0.2in}+\mZ\, \Gamma_{Z} \Big\{\mA\, \Gamma_{A}\,
          (\mZ^{2}-8\, \mchi^{2})-i\,[16\, \mchi^{4}+\mA^{2}\,
            (\mZ^{2}-8\, \mchi^{2})]\Big\} \bigg] \Bigg ]; \nn \\
   \end{eqnarray}
$\bullet$
\underline{Higgs ($A$)--sfermion\ ($\widetilde{f}_{a}$) interference term:}
    \begin{eqnarray}
     \widetilde{a}_{\bar{f}f}^{(A-\widetilde{f})}& =& -\frac{1}{4\,\pi}
      \sum_{a} Re \left[\frac{C_{P}^{ffA}\:
       C_{P}^{\chi\chi A}}{4\, \mchi^{2}-\mA^{2}
          +i\, \Gamma_{A}\, \mA} \right]
            \frac{(m_{f}\,C_{+}^{a}+\mchi\,D_{+}^{a})}
            {\Delta_{{\widetilde f}_{a}}}, 
\\
     \widetilde{b}_{\bar{f}f}^{(A-\widetilde{f})} &=& -\frac{1}{8\,\pi}
        \sum_{a} Re\left[ \left(\frac{C_{P}^{ffA}\:
         C_{P}^{\chi\chi A}}{(4\, \mchi^{2}-\mA^{2}
           +i\, \Gamma_{A}\, \mA)^{2}} \right) \,
          \frac{\mchi}{\Delta_{{\widetilde f}_{a}}^{3}}\right. \nonumber \\
        & & \times \bigg\{ C_{+}^{a} \Big[ 4 \,m_{f}\,\mchi^{2}\,
           (\mchi^2-m_{f}^{2})\,P_A
              +6\,m_{f}\,\mchi^{2}\,P_A\,
                     \Delta_{{\widetilde f}_{a}}  \nonumber \\
        & & \hspace{0.5in}-3\,m_{f}\,\mA\,(\mA-i\,\Gamma_{A})\,
              \Delta_{{\widetilde f}_{a}}^{2} \Big]\nonumber \\
        & & \hspace{0.1in} +\,D_{+}^{a} \Big[ 4 \,\mchi^{3}\,(\mchi^2
             -m_{f}^{2})\,P_A-2\,\mchi\,(m_{f}^{2}
             -4\,\mchi^{2})\,P_A\,\Delta_{{\widetilde f}_{a}}  \nonumber \\
        & & \hspace{0.5in}+6\,\mchi\,(2\,\mchi^{2}-\mA^{2}
           +i\, \Gamma_{A}\, \mA)\,\Delta_{{\widetilde f}_{a}}^{2}
                  \Big]\bigg\}\Bigg];
    \end{eqnarray}
\vspace{0.25cm}
where $P_A\equiv 4\,\mchi^{2}-\mA^{2}+i\, \Gamma_{A}\, \mA\, .$\\
$\bullet$
\underline{$Z$--sfermion\ ($\widetilde{f}_{a}$) interference term:}
    \begin{eqnarray}
      \widetilde{a}_{\bar{f}f}^{(Z-\widetilde{f})}&=&-\frac{1}{4\, \pi}
        \sum_{a} Re \left [  \frac{C_{A}^{ffZ}\:
           C_{A}^{\chi\chi Z}}{4\, \mchi^{2}-\mZ^{2}
             +i\, \Gamma_{Z}\, \mZ} \right] \frac{m_{f}\,
                 (\mZ^{2}-4\,\mchi^{2})}{\mZ^{2}} \nonumber \\
        & & \times \frac{(m_{f}\,C_{+}^{a}
           +\mchi\,D_{+}^{a})}{\Delta_{{\widetilde f}_{a}}}, 
\\
      \widetilde{b}_{\bar{f}f}^{(Z-\widetilde{f})} &=&-\frac{1}{8\, \pi}
        \sum_{a} Re \left [ \left( \frac{ C_{A}^{\chi\chi Z}}
          {(4\, \mchi^{2}-\mZ^{2}+i\, \Gamma_{Z}\, \mZ)^{2}} \right) \,
           \frac{1}{\mZ^{2}\, \Delta_{{\widetilde f}_{a}}^{3}}\right.
             \nonumber \\
      & &  \times \bigg [C_{V}^{ffZ}\,C_{-}^{a}\,
           \Big\{2\,\mZ^{2}\,P_Z\,\Delta_{{\widetilde f}_{a}}
            [2\,\mchi^{2}\,(\mchi^{2}+\Delta_{{\widetilde f}_{a}})
             +m_{f}^{2}\,(-2\,\mchi^{2}+\Delta_{{\widetilde f}_{a}})]\,\Big\}
          \nonumber \\
      & & \hspace{0.3in}+\,C_{A}^{ffZ}\Big\{C_{+}^{a}
         \Big[2\,m_{f}^{2}\,\mchi^{2}\,(\mchi^2-m_{f}^{2})\,
            (\mZ^{2}-4\,\mchi^{2})\,P_Z \nonumber  \\
      & & \hspace{1.in}+\, \mchi^{2} [m_{f}^{2} \mZ^{2}
         +\,2\, \mchi^{2}\,(\mZ^{2}-6\,m_{f}^{2})]\,P_Z\,
         \Delta_{{\widetilde f}_{a}} \nonumber \\
      & & \hspace{1.1in}  + \,2\,\mZ\,\{-\mZ\,(\mchi^2-m_{f}^{2})\,
         (\mZ^{2}-4\,\mchi^{2})  \nonumber \\
      & & \hspace{1.in}+\,i\,\Gamma_{Z}\,[\mZ^{2}\,\mchi^{2}
         -m_{f}^{2}(\mZ^{2}+3\,\mchi^{2})]\}\,
            \Delta_{{\widetilde f}_{a}}^{2}\Big] \nonumber \\
      & &\hspace{0.9in} +m_{f}\, \mchi\,D_{+}^{a}\Big[4\,\mchi^{2}\,
         (\mchi^2-m_{f}^{2})\,(\mZ^{2}-4\,\mchi^{2})\,P_Z \nonumber \\
      & &\hspace{1.in} +2\,[6\,\mZ^{2}\,\mchi^{2}
        -16\ \mchi^{4}-m_{f}^{2}\,(3\,\mZ^{2}
        -4\,\mchi^{2})]\,P_Z\,\Delta_{{\widetilde f}_{a}} \nonumber \\
      & &\hspace{1.in}  -3\,[(\mZ^{2}-4\,\mchi^{2})^{2}
        -i\, \mZ\,\Gamma_{Z}\,(\mZ^{2}-8\,\mchi^{2})]\,
         \Delta_{{\widetilde f}_{a}}^{2}\Big]\bigg\}\Bigg] \Bigg]; \nn \\
\end{eqnarray}
where $P_Z\equiv 4\,\mchi^{2}-\mZ^{2}+i\, \Gamma_{Z}\, \mZ$.

%
\section{Summary}\label{summary:sec}
The neutralino is undoubtedly the most popular
candidate for a WIMP dark matter in the Universe. Supersymmetry
remains arguably the most promising extension of the Standard
Model. The next several years will witness extensive searches for
supersymmetry in colliders as well as for WIMPs in underground
detectors. Measurements of the cosmological parameters, and in
particular of the relic abundance of the dark matter, have already
reached the accuracy of a few per cent and more progress is expected.

In light of this, theoretical computations of the neutralino relic
abundance need to be now performed with at least the same, if not
better, level of precision, if one wants to reliably compare
theoretical predictions with observations.  Motivated by this goal, we
have derived a full set of exact, analytic expressions for the
neutralino pair-annihilation cross sections into all tree-level
two-body final states in the framework of the MSSM.

%
\bigskip

\acknowledgments
T.N. is grateful to Lancaster University for kind hospitality
extended during his visit. The authors appreciate Dr. A. Birkedal-Hansen 
to point out several minor typographical errors in the first version 
of this paper.

%
\newpage

\appendix
\section{The MSSM Lagrangian}\label{appxa}


In this Appendix, the MSSM Lagrangian is given explicitly in the mass
eigenstates, which makes it easier to read out relevant Feynman
rules. 

\vspace*{0.2cm}\noindent
\underline{{\bf Z-boson-Fermion-Fermion}}\par
\begin{eqnarray}
  {\cal L} &=& \sum_f \bar f \,\gamma^\mu
    \left[ \,C_{V}^{ffZ}-C_{A}^{ffZ}\gamma_5 \right]f Z_\mu,
\end{eqnarray}
where
\begin{eqnarray}
C_{V}^{ffZ} & = & -\frac{g}{2 \cw} (T_{3f}-2\swsq \,Q_f),
\\
C_{A}^{ffZ} & = & -\frac{g}{2 \cw} T_{3f}, 
\end{eqnarray}
and $g$ is the usual gauge coupling of $SU(2)_L$ whereas $Q_f$ and 
$T_{3f}$ were defined below eq.~(\ref{eqn:squarkmass}).

\vspace*{0.2cm}\noindent
\underline{{\bf W-Chargino-Neutralino}}\par
\begin{eqnarray}
     {\cal L} &=& \sum_{k=1}^{2} \sum_{i=1}^{4}
    \overline{\chi^-_k} \gamma^\mu
    \left[ C_{V}^{\chi_{k}^{+}\chi_{i}^0 W^-}
          -C_{A}^{\chi_{k}^{+}\chi_{i}^0 W^-}\gamma_5 \right]
     \chi_i^0 W_\mu^- + h.c.,
\end{eqnarray}
where
\begin{eqnarray}
C_{V}^{\chi_{k}^{+}\chi_{i}^0 W^-}
  & = & - \textstyle{\frac{1}{2}} g (O^L_{ik}+O^R_{ik}),  \\
C_{A}^{\chi_{k}^{+}\chi_{i}^0 W^-}
  & = & \textstyle{\frac{1}{2}} g (O^L_{ik}-O^R_{ik}),
\end{eqnarray}
and
\begin{eqnarray}
O^L_{ik} &=& -\sqrt{\textstyle{\frac{1}{2}}}
                    N_{i4} V_{k2}^* + N_{i2} V_{k1}^* ,  \\
O^R_{ik} &=& \sqrt{\textstyle{\frac{1}{2}}}
                    N_{i3}^* U_{k2} + N_{i2}^* U_{k1}^*, 
\end{eqnarray}
where the matrices $N, U, V$ were defined via eqs.~(\ref{neutdiag})
and~(\ref{chardiag}).

\vspace*{0.2cm}\noindent
\underline{{\bf Higgs-Chargino-Neutralino}}\par
\begin{eqnarray}
  {\cal L} &=& \sum_{k=1}^{2} \sum_{i=1}^{4}
    \overline{\chi_k^-}
    \left[ C_{S}^{\chi_{k}^{+}\chi_{i}^0 H^-}
          -C_{P}^{\chi_{k}^{+}\chi_{i}^0 H^-}\gamma_5 \right]
     \chi_i^0 H^- + h.c.,
\end{eqnarray}
where
\begin{eqnarray}
C_{S}^{\chi_{k}^{+}\chi_{i}^0 H^-}
 & = & -\textstyle{\frac{1}{2}} (Q_{ik}^{'L}+Q_{ik}^{'R}),  \\
C_{P}^{\chi_{k}^{+}\chi_{i}^0 H^-}
 & = & -\textstyle{\frac{1}{2}} (Q_{ik}^{'L}-Q_{ik}^{'R}),
\end{eqnarray}
and
\begin{eqnarray}
Q_{ik}^{'L} &=&~ g \cos\beta \left[ N_{i4}^* V_{k1}^*
                   +\sqrt{\textstyle{\frac{1}{2}}}
                    (N_{i2}^* + N_{i1}^* \tw) V_{k2}^* \right],  \\
Q_{ik}^{'R} &=&~ g \sin\beta \left[ N_{i3} U_{k1}
                   -\sqrt{\textstyle{\frac{1}{2}}}
                    (N_{i2} + N_{i1} \tw) U_{k2} \right].
\end{eqnarray}

\vspace*{0.2cm}\noindent
\underline{{\bf Z-Neutralino-Neutralino}}\par
\begin{eqnarray}
   {\cal L}& =& {1\over 2} \sum_{i,j=1}^{4}
        \overline{\chi_i^0}\, \gamma^\mu\,
            \left[ C_{V}^{\chi_{i}^0\chi_{j}^0 Z}
                 - C_{A}^{\chi_{i}^0\chi_{j}^0 Z}\gamma_5 \right]
                                 \,\chi_j^0 \,Z_\mu \, ,
\end{eqnarray}
where
\begin{eqnarray}
         C_{V}^{\chi_{i}^0\chi_{j}^0 Z} &=&
 \frac{g}{2\cw} ( O_{ij}^{''L}-O_{ij}^{''L*} ),  \\
         C_{A}^{\chi_{i}^0\chi_{j}^0 Z} &=&
 \frac{g}{2\cw} ( O_{ij}^{''L}+O_{ij}^{''L*} ),
\end{eqnarray}
and
\begin{eqnarray}
O_{ij}^{''L} &=& \textstyle{\frac{1}{2}}
              ( - N_{i3}\,N_{j3}^* + N_{i4}\,N_{j4}^* ).
\end{eqnarray}
We neglect CP violation hence $C_V^{\chi^0_i\chi^0_j Z}$ $=$ 0.

\vspace*{0.2cm}\noindent
\underline{{\bf Gauge-Higgs-Higgs}}\par
\begin{eqnarray}
   {\cal L} &=&
        i\left[ C^{hAZ}(A \delmux h)
               +C^{HAZ}(A \delmux H)
               +C^{H^{+}H^{-}Z}(H^+ \delmux H^-) \right] Z_\mu \nn \\
 & &   + \left\{ i W_\mu^- \left[C^{W^- H^+ h}(h \delmux H^+)
        + C^{W^- H^+ H}(H \delmux H^+) \right. \right. \nn \\
 & &  \hspace{2.1in} \left. \left.
     + C^{W^- H^+ A}(A \delmux H^+) \right] + h.c. \right\},
\end{eqnarray}
where
\begin{eqnarray}
  C^{hAZ} & = & -\frac{ig}{2 \cw}\cos(\alpha-\beta)\, ,   \\
  C^{HAZ} & = & -\frac{ig}{2 \cw} \sin(\alpha-\beta)\, ,  \\
  C^{H^{+}H^{-}Z} & = & \frac{g}{2\cw}(1-2\swsq)\, ,      \\
  C^{W^-H^+h} & = & -\frac{g}{2} \cos (\alpha-\beta) ,    \\
  C^{W^-H^+H} & = & -\frac{g}{2} \sin (\alpha-\beta) ,    \\
  C^{W^-H^+A} & = &  \frac{ig}{2}. 
\end{eqnarray}

\vspace*{0.2cm}\noindent
\underline{{\bf Higgs-Fermion-Fermion}}\par
\begin{eqnarray}
    {\cal L} &=& \sum_{f}\,
           \left[ C_{S}^{ffh} \bar f f h + C_{S}^{ffH} \bar f f H
                 +C_{P}^{ffA} \bar f\gamma_{5} f A \right]  \\
   & \equiv &
\sum_{n=1}^{3} \left[
\ {{g\,m_{e_n}\sin\alpha}\over{2\,m_W\cos\beta}} \bar e_n\, e_n\, h
     -{{g\,m_{e_n}\cos\alpha}\over{2\,m_W\cos\beta}} \bar e_n\, e_n\, H
     +i{{g\,m_{e_n}}\over{2\,m_W}}\tan\beta\,\bar e_n\, \gamma_5\, e_n\, A
       \right. \nonumber \\
   & & \hspace{8mm}
     -\,{{g\,m_{u_n}\cos\alpha}\over{2\,m_W\sin\beta}} \bar u_n\, u_n\, h
     -{{g\,m_{u_n}\sin\alpha}\over{2\,m_W\sin\beta}} \bar u_n\, u_n\, H
     +i{{g\,m_{u_n}}\over{2\,m_W}}\cot\beta\, \bar u_n \,\gamma_5 u_n\, A
       \nonumber \\
   & & \left. \hspace{8mm}
     +\,{{g\,m_{d_n}\sin\alpha}\over{2\,m_W\cos\beta}} \bar d_n\, d_n\, h
     -{{g\,m_{d_n}\cos\alpha}\over{2\,m_W\cos\beta}} \bar d_n\, d_n\, H
     +i{{g\,m_{d_n}}\over{2\,m_W}}\tan\beta \,\bar d_n \,\gamma_5\, d_n \,A
           \right],  \nonumber \\
   \end{eqnarray}
where $n$ $=$ $1,2,3$ is an index for generation, so $e_n= e, \mu,
\tau$, \etc.

\vspace*{0.2cm}\noindent
\underline{{\bf Higgs-Gauge-Gauge}}\par
\begin{eqnarray}
        {\cal L} &=&
  \left[ C^{W^+W^- h}h + C^{W^+W^- H}H \right] W_\mu^+W^{-\mu}
  +{1\over 2}\left[ C^{ZZh}h+C^{ZZH}H \right] Z_\mu Z^{\mu}, \nonumber \\
\end{eqnarray}
where
\begin{eqnarray}
   C^{W^+W^- h} &=& - g\,m_W \sin(\alpha-\beta),  \\
   C^{W^+W^- H} &=&  g\,m_W \cos(\alpha-\beta),   \\
   C^{ZZh} &=& - \frac{g m_Z}{ \cw} \sin(\alpha-\beta),  \\
   C^{ZZH} &=&  \frac{g m_Z}{ \cw} \cos(\alpha-\beta).
\end{eqnarray}

\vspace*{0.2cm}\noindent
\underline{{\bf Higgs-Neutralino-Neutralino}}\par
\begin{eqnarray}
    {\cal L}&=&
      {1\over 2}\sum_{i,j=1}^{4} \overline{\chi_i^0}
        \left[ C^{\chi_i^0\chi_j^0 H}_S H
              +C^{\chi_i^0\chi_j^0 h}_S h
              +C^{\chi_i^0\chi_j^0 A}_P \gamma_5A \right]\chi_j^0 \, ,
\end{eqnarray}
where
\begin{eqnarray}
     C^{\chi_i^0\chi_j^0 H}_S &=&-{g\over 2} \left\{
        [N_{i2}-N_{i1}\tan\theta_W]
        [\cos\alpha\,N_{j3}-\sin\alpha\,N_{j4}]
            +(i\leftrightarrow j) \right\}, \\
     C^{\chi_i^0\chi_j^0 h}_S &=& -{g\over 2} \left\{
        [N_{i2}-N_{i1}\tan\theta_W]
        [-\sin\alpha\,N_{j3}-\cos\alpha\,N_{j4}]
            +(i\leftrightarrow j) \right\},  \\
     C^{\chi_i^0\chi_j^0 A}_P &=& -{i g\over 2} \left\{
        [N_{i2}-N_{i1}\tan\theta_W]
        [\sin\beta\,N_{j3}-\cos\beta\,N_{j4}]
            +(i\leftrightarrow j) \right\}.
\end{eqnarray}
Note that we assume no CP violating phases in the
neutralino mass matrix and we use the convention that $N_{ij}$
is real. 
%
\vspace{1cm} \\
\underline{{\bf Gauge-Gauge-Gauge}}\par
\begin{eqnarray}
 {\cal L}&=& i C^{WWZ}\lbrack
         (\partial_\mu W_\nu^--\partial_\nu W_\mu^-) W^{+\mu}Z^\nu
        - (\partial_\mu W^+_\nu-\partial_\nu W^+_\mu) W^{-\mu} Z^\nu
            \nonumber \\
      & & ~~~~~~~~~~-(\partial_\mu Z_\nu-\partial_\nu Z_\mu)W^{+\mu} W^{-\nu}
            \rbrack,
\end{eqnarray}
where
\begin{eqnarray}
C^{WWZ} & = & g \cw.
\end{eqnarray}

\vspace*{0.2cm}\noindent
\underline{{\bf Neutral Higgs-Higgs-Higgs}}\par
\begin{eqnarray}
 {\cal L} &=&
   {1 \over {6}}C^{hhh} h^3+{1 \over {6}}C^{HHH} H^3
  +{1 \over {2}}C^{hhH} h^2 H+{1 \over {2}}C^{HHh} H^2 h \nonumber \\
  & & + \, {1 \over {2}}(C^{AAh} h+C^{AAH}  H) A^2,
\end{eqnarray}
where
\begin{eqnarray}
  C^{hhh} &=&   -\frac{3g}{2\,\cw } m_Z \cos 2\alpha \sin(\alpha+\beta), \\
  C^{HHH} &=&   -\frac{3g}{2\,\cw } m_Z \cos 2\alpha \cos(\alpha+\beta), \\
  C^{AAh} &=&   -\frac{g}{2\,\cw } m_Z \cos 2\beta \sin(\alpha+\beta), \\
  C^{AAH} &=&   \frac{g}{2\,\cw } m_Z \cos 2\beta \cos(\alpha+\beta), \\
  C^{hhH} &=&   -\frac{g}{2\,\cw } m_Z [2 \sin 2\alpha\sin(\alpha+\beta)
                 -\cos 2\alpha \cos(\alpha+\beta)], \\
  C^{HHh} &=&   \frac{g}{2\,\cw } m_Z  [2 \sin 2\alpha\cos(\alpha+\beta)
                 +\cos 2\alpha\sin(\alpha+\beta)].
\end{eqnarray}

\vspace*{0.2cm}\noindent
\underline{{\bf H$^+$-H$^-$-Higgs}}\par
\begin{eqnarray}
    {\cal L} &=& (C^{H^+H^- h}h+C^{H^+H^- H}H)H^+H^-,
\end{eqnarray}
where
\begin{eqnarray}
    C^{H^+H^- h}  &=& gm_Z \left[ \cw \sin(\alpha-\beta)
           -\frac{1}{2\cw}\sin(\alpha+\beta)\cos 2\beta \right], \\
    C^{H^+H^- H} &=& - \, gm_Z \left[ \cw \cos(\alpha-\beta)
           -\frac{1}{2\cw}\cos(\alpha+\beta)\cos 2\beta \right].
\end{eqnarray}

\vspace*{0.2cm}\noindent
\underline{{\bf Neutralino-Fermion-Sfermion}}\par
\begin{eqnarray}
   {\cal L}& =&
\sum_{i=1}^4 \sum_{n=1}^3 \left\{ \sum_{f=u,d,e}\sum_{a=1}^6
\overline{\chi^0_i} \left( \Lambda_{nai}^{(f)L} \frac{1-\gamma_5}{2}
+ \Lambda_{nai}^{(f)R} \frac{1+\gamma_5}{2}  \right) f_n \widetilde{f}_a^* \right.
\nonumber \\
 & & \hspace{4cm} + \left. \sum_{a=1}^3
\overline{\chi^0_i} \Lambda_{nai}^{(\nu)L} \frac{1-\gamma_5}{2}
\nu_n \tilde{\nu}_a^*  + h.c. \right\},   \
\label{eq:n-f-sf}
\end{eqnarray}
where
\begin{eqnarray}
\label{eq:lambdal}
\Lambda_{nai}^{(f)L} & = &
- \frac{g}{\sqrt{2}} \left( \Gamma_{(f)L}^{an} \ell^{\chi^0_i f_n L}
          + \Gamma_{(f)R}^{an} \ell^{\chi^0_i f_n R} \right),  \\
\label{eq:lambdar}
\Lambda_{nai}^{(f)R}
 & = &
- \frac{g}{\sqrt{2}} \left( \Gamma_{(f)L}^{an} r^{\chi^0_i f_n L}
          + \Gamma_{(f)R}^{an} r^{\chi^0_i f_n R} \right). 
\end{eqnarray}
Note that in eqs.~(\ref{eq:lambdal}) and~(\ref{eq:lambdar}), 
$f$ $=$ $u$, $d$, $e$, $\nu$ represents
the type of the fermion, while $f$ in the main text denotes each fermion
$f$ $=$ $u$, $c$, $t$, etc. 
The slepton and squark mass eigenstates $ \widetilde{f}_a$ ($\tilde{\nu}_a$
with $a=1,2,3$ and $\tilde{e}_a$,$\tilde{u}_a$ and $\tilde{d}_a$ with
$a=1,...,6$) are related to the
sfermion gauge eigenstates $ \widetilde{f}_{nL}$ and $ \widetilde{f}_{nR}$
($n=1,2,3$) via
\begin{eqnarray}
  \widetilde{f}_{nL}=
     \sum_{a} \widetilde{f}_a \Gamma^{an *}_{(f)L} \, ,  \\
  \widetilde{f}_{nR}=
     \sum_{a} \widetilde{f}_a \Gamma^{an *}_{(f)R} \, ,
\end{eqnarray}
where $\widetilde{V}_f$ $=$ $(\Gamma_{(f)L},\Gamma_{(f)R})$ 
denotes a $6 \times 6$ matrix which 
diagonalizes the sfermions mass matrix given by
eqs.~(\ref{eqn:ll})--(\ref{eqn:rr}):
$\widetilde{V}_f$ ${\cal M}^2_{\widetilde{f}}$ 
$\widetilde{V}_f^\dagger$ $=$ 
${\rm diag}(m_{\widetilde{f}_1}^2,\cdots,m_{\widetilde{f}_6}^2)$.
Note that in the case of squarks and charged sleptons the mixing matrices
$\Gamma_{(u)\, L,R},\Gamma_{(d)\, L,R}$ and $\Gamma_{(e)\,L,R}$ have
dimension $6\times3$, while for sneutrinos $\Gamma_{(\nu)\,L}$ is a mixing
matrix of order $3\times3$.
(We neglect here the CKM matrix for simplicity.)
Finally $\ell^{\chi_i^0 f_n L,R}$ and $r^{\chi_i^0 f_n L,R}$ are given
as follows:
\begin{eqnarray}
 \ell^{\chi_i^0 \nu_n L}& =& N_{i2} -\tw N_{i1} ,                \\
 \ell^{\chi_i^0 e_n L} &=& -N_{i2} -\tw N_{i1} ,                 \\
 \ell^{\chi_i^0 u_n L} &=& N_{i2} +\frac{1}{3}\tw N_{i1} ,       \\
 \ell^{\chi_i^0 d_n L} &=& -N_{i2} +\frac{1}{3}\tw N_{i1} ,      \\
 r^{\chi_i^0 \nu_n L} &=& 0\, ,                                  \\
 r^{\chi_i^0 e_n L} &=& \frac{m_{e_n}}{m_W\cos\beta} N_{i3}\, ,  \\
 r^{\chi_i^0 u_n L} &=& \frac{m_{u_n}}{m_W\sin\beta} N_{i4}\, ,  \\
 r^{\chi_i^0 d_n L} &=& \frac{m_{d_n}}{m_W\cos\beta} N_{i3}\, ,  \\
 \ell^{\chi_i^0 \nu_n R} &=& 0 \, ,                              \\
 \ell^{\chi_i^0 e_n R} &=& \frac{m_{e_n}}{m_W\cos\beta} N_{i3}\, ,
      ~~~~~~~~~~~~~~                                              \\
 \ell^{\chi_i^0 u_n R} &=& \frac{m_{u_n}}{m_W\sin\beta} N_{i4}\, ,\\
 \ell^{\chi_i^0 d_n R} &=& \frac{m_{d_n}}{m_W\cos\beta} N_{i3}\, ,\\
 r^{\chi_i^0 \nu_n  R} &=& 0\, ,                                  \\
 r^{\chi_i^0 e_n R} &=& 2\tw N_{i1}\, ,                           \\
 r^{\chi_i^0 u_n R} &=& -{4\over 3}\tw N_{i1}\, ,                 \\
 r^{\chi_i^0 d_n R} &=& {2\over 3}\tw N_{i1}\,.
\end{eqnarray}

\newpage
\section{Auxiliary Functions}\label{appxb}


Here we give expressions for the auxiliary functions used in the
text. First, we  define
\begin{eqnarray}
   D(s,x,y_1,y_2)&\equiv& x+\frac{y_{1}+y_{2}}{2}-\frac{s}{2} \, ,\\
   F(s,x,y_1,y_2)&\equiv& \frac{1}{2}\sqrt{s-4\,x}
\sqrt{s-(\sqrt{y_1}+\!\sqrt{y_2})^2}
\,\sqrt{1-\frac{(\sqrt{y_1}-\!\sqrt{y_2})^2}{s}} \, .
\end{eqnarray}
If we define $D\equiv D(s,x,y_{1},y_{2})$, $F\equiv F(s,x,y_{1},y_{2})$,
$t_{\pm}(s,x,y_1,y_2) \equiv D \pm F$
 and $({\mathcal T}_{i},{\mathcal Y}_{i})$ $\equiv$
$({\mathcal T}_{i},{\mathcal Y}_{i}) (s,x,y_{1},y_{2},z_{1},z_{2})$
($i$ $=$ $0,\cdots, 4$), then
\begin{eqnarray}
  {\mathcal F}(s,x,y_1,y_2,z) &=& \frac{1}{2\,F}\,
\ln\left|\frac{t_{+}(s,x,y_1,y_2)-z}{t_{-}(s,x,y_1,y_2)-z}\right|\, ,
\end{eqnarray}
and
\begin{eqnarray}
\!\!\!\!\!\!\!\!\! {\mathcal T}_{0} \! &=& \!
\frac{1}{z_1-z_2}[{\mathcal F}(s,x,y_1,y_2,z_1)
                      -{\mathcal F}(s,x,y_1,y_2,z_2)],              \\
\!\!\!\!\!\!\!\!\! {\mathcal T}_{1} \! &=& \!
\frac{1}{z_1-z_2}[z_1\,{\mathcal F}(s,x,y_1,y_2,z_1)
                     -z_2\,{\mathcal F}(s,x,y_1,y_2,z_2)],           \\
\!\!\!\!\!\!\!\!\! {\mathcal T}_{2} \! &=& \!
1+\frac{1}{z_1-z_2}[z_1^2\,{\mathcal F}(s,x,y_1,y_2,z_1)
                     -z_2^2\,{\mathcal F}(s,x,y_1,y_2,z_2)],         \\
\!\!\!\!\!\!\!\!\! {\mathcal T}_{3} \! &=& \!
D+z_1+z_2
     +\frac{1}{z_1-z_2}[z_1^3\,{\mathcal F}(s,x,y_1,y_2,z_1)
                     -z_2^3\,{\mathcal F}(s,x,y_1,y_2,z_2)],          \\
\!\!\!\!\!\!\!\!\! {\mathcal T}_{4} \! &=& \!
D^2+D\,(z_1+z_2)+(z^2_1+z^2_2+z_1 z_2)+\frac{1}{3}F^2 \nonumber \\
 & & \hspace{2cm} + \, \frac{1}{z_1-z_2}[z_1^4\,{\mathcal F}(s,x,y_1,y_2,z_1)
                      -z_2^4\,{\mathcal F}(s,x,y_1,y_2,z_2)],        \\
\!\!\!\!\!\!\!\!\! {\mathcal Y}_{0} \! &=& \!
\frac{1}{z_1+z_2-2\,D}[{\mathcal F}(s,x,y_1,y_2,z_1)
                        +{\mathcal F}(s,x,y_1,y_2,z_2)],            \\
\!\!\!\!\!\!\!\!\! {\mathcal Y}_{1} \! &=& \!
\frac{1}{z_1+z_2-2\,D}[2\,(z_1-D)\,{\mathcal F}(s,x,y_1,y_2,z_1)
                   -2\,(z_2-D)\,{\mathcal F}(s,x,y_1,y_2,z_2)], \nn  \\
                                                                    \\
\!\!\!\!\!\!\!\!\! {\mathcal Y}_{2} \! &=& \!
1 +\frac{1}{z_1+z_2-2\,D}
[z_1\,(z_1-2\,D){\mathcal F}(s,x,y_1,y_2,z_1) \nonumber \\
& & \hspace{3.cm} + z_2\,(z_2-2\,D)\,{\mathcal
  F}(s,x,y_1,y_2,z_2)],     \\
\!\!\!\!\!\!\!\!\! {\mathcal Y}_{3} \! &=& \!
2\,(z_1-z_2)+\frac{1}{z_1+z_2-2\,D}
[z_1\,(2\,z_1^2-6\,z_1\,D+4\,D^2){\mathcal F}(s,x,y_1,y_2,z_1) \nonumber \\
& & \hspace{4.5cm} -\,z_2\,(2\,z_2^2-6\,z_2\,D+4\,D^2)
                                {\mathcal F}(s,x,y_1,y_2,z_2)],   \\
\!\!\!\!\!\!\!\!\! {\mathcal Y}_{4} \! &=& \!
-D^2-D\,(z_1+z_2)+(z^2_1+z^2_2-z_1\,z_2)+\frac{1}{3} F^2 \nonumber \\
& & \hspace{2cm}  + \,\frac{1}{z_1+z_2-2\,D}
    [z_1^2\,(z_1^2-4\,z_1\,D+4\,D^2){\mathcal F}(s,x,y_1,y_2,z_1) \nonumber \\
& &   \hspace{4.7cm} + \, z_2^2\,(z_2^2-4\,z_2\,D+4\,D^2)
                                       {\mathcal F}(s,x,y_1,y_2,z_2)].
\end{eqnarray}


\newpage


\begin{thebibliography}{99}
\bibitem{kt90}
E.W. Kolb and M.S. Turner, {\em The Early Universe}, Addison-Wesley (1990).

\bibitem{jkg96}
G. Jungman, M. Kamionkowski and K. Griest, \prep{267}{1996}{195}.

\bibitem{susyreview}
For a review of supersymmetric models 
see, for instance, H.P. Nilles, \prep{110}{1984}{1} and
H.E.~Haber and G.~Kane, \prep{117}{1985}{75}.

\bibitem{ckr} 
L.~Covi, J.E.~Kim and L.~Roszkowski, \prl{82}{1999}{4180};
L.~Covi, H.B.~Kim, J.E.~Kim and L.~Roszkowski, \jhep{0105}{2001}{033}
(\hepph{0102308}).

\bibitem{gravitino}
H. Pagels and J.R. Primack, 
\prl{48}{1982}{223};
S. Weinberg, \prl{48}{1982}{1303}; 
J. Ellis, J.E. Kim, and D.V.  Nanopoulos, \plb{145}{1984}{181}. For a
recent re-examination, see M.~Bolz, W.~Buchm{\" u}ller, 
M. Pl{\" u}macher, \plb{443}{1998}{209}.

\bibitem{cmbrreview} For a recent review, see, \eg, A.~Melchiorri, 
talk given at COSMO-01, \astroph{0201237}.

\bibitem{goldberg83}
H. Goldberg, \prl{50}{1983}{1419}.

\bibitem{ehnos}
J. Ellis, J.S. Hagelin, D.V. Nanopoulos, K.A. Olive and M. Srednicki,
\plb{127}{1983}{233} and \npb{238}{1984}{453}.

\bibitem{krauss83}
L.M.~Krauss, \npb{227}{1983}{556}.

\bibitem{griest88}
K. Griest, \prd{38}{1988}{2357}; Erratum \prd{39}{1989}{3802}.

\bibitem{gkt90}
K. Griest, M. Kamionkowski, and M.S. Turner, \prd{41}{1990}{3565}.

\bibitem{erl90}
J. Ellis, L. Roszkowski and Z. Lalak, \plb{245}{1990}{545}.

\bibitem{os91}
K.A. Olive and M. Srednicki, \plb{230}{1989}{78} and \npb{355}{1991}{208}.

\bibitem{dn93}
M. Drees and M. Nojiri, \prd{47}{1993}{376}.

\bibitem{gs91}
K. Griest and D. Seckel, \prd{43}{1991}{3191}.

\bibitem{gg91}
P. Gondolo and G. Gelmini, \npb{360}{1991}{145}.

\bibitem{an93}
P.~Nath and R.~Arnowitt, \prl{69}{1992}{725}.

\bibitem{lny93}
J.L. Lopez, D.V. Nanopoulos and K. Yuan, \prd{48}{1993}{2766}.

\bibitem{nrr1}
T. Nihei, L. Roszkowski and R. Ruiz de Austri, \jhep{0105}{2001}{063}
(\hepph{0102308}).

\bibitem{dy96}
M.~Drees and A.~Yamada, \prd{53}{1996}{1586}.

\bibitem{swo88}
M. Srednicki, R. Watkins and K.A. Olive, \npb{310}{1988}{693}.

\bibitem{darksusy00}
P.~Gondolo {\it et.al.}, \astroph{0012234}.
The code DarkSUSY is available from
http://www.physto.se/\~{\mbox{}}edsjo/darksusy.

\bibitem{mizuta92}
S. Mizuta and M. Yamaguchi, \plb{298}{1993}{120}.

\bibitem{edsjo97}
J. Edsj\"o and P. Gondolo, \prd{56}{1997}{1879}.

\bibitem{efos98}
J. Ellis, T. Falk, K.A. Olive and M. Srednicki, 
\app{13}{2000}{181}.

\bibitem{bdd00}
C.~Boehm, A.~Djouadi and M.~Drees, \prd{62}{2000}{035012}, \hepph{9911496}.

\bibitem{chiasdm} 
L.~Roszkowski, \plb{262}{1991}{59}.

\bibitem{rr93}
R.G. Roberts and L. Roszkowski, \plb{309}{1993}{329}.

\bibitem{kkrw94}
G.L. Kane, C. Kolda, L. Roszkowski, and J.D. Wells,
\prd{49}{1994}{6173}.

\bibitem{annecy01}
G.~Belanger, F. Boudjema, A. Pukhov, and A. Semenov, \hepph{0112278}. 

\bibitem{reheatfactor}
K.A. Olive, D. Schramm and G. Steigman,
\npb{180}{1981}{497}.

\bibitem{gh}
J.F. Gunion and H.E. Haber, \npb{272}{1986}{1}.

\bibitem{ck98} Chen and M.~Kamionkowski, \jhep{9807}{1998}{001}.

\bibitem{simple:ref} 
See, for example, L.~Roszkowski, \prd{50}{1994}{4842}.


\end{thebibliography}
\end{document}